\documentclass[sigconf, nonacm]{acmart}
\usepackage{pvldb}
\renewcommand\vldbdoi{10.14778/3860016.3860020}
\renewcommand\vldbpages{510 - 523}
\renewcommand\vldbvolume{20}
\renewcommand\vldbissue{3}
\renewcommand\vldbyear{2026}
\renewcommand\vldbauthors{\authors}
\renewcommand\vldbtitle{\shorttitle} 
\renewcommand\vldbavailabilityurl{https://github.com/cslyuxu/ObliGraph}
\renewcommand\vldbpagestyle{empty}

\AtBeginDocument{%
  \providecommand\BibTeX{{%
    \normalfont B\kern-0.5em{\scshape i\kern-0.25em b}\kern-0.8em\TeX}}}

\usepackage{amsmath}
\usepackage{amsthm}

\usepackage{amssymb}
\usepackage{amsfonts}
\usepackage{epstopdf}
\usepackage{xcolor}
\usepackage{kantlipsum}
\usepackage{multicol}
\usepackage{multirow}
\usepackage{balance}
\usepackage{footmisc}
\usepackage{graphicx,caption}
\usepackage{utfsym}
\usepackage{epstopdf}
\usepackage{threeparttable}
\usepackage[labelformat=simple]{subcaption}

\usepackage{diagbox}

\usepackage[ruled,vlined,linesnumbered]{algorithm2e}
\SetKwProg{Fn}{Function}{}{}
\SetVlineSkip{0pt}
\SetAlCapNameFnt{\small}
\SetAlCapFnt{\small}
\SetAlFnt{\small}
\makeatletter
\usepackage{etoolbox}

\makeatletter
\patchcmd\algocf@Vline{\vrule}{\vrule \kern-0.4pt}{}{}
\patchcmd\algocf@Vsline{\vrule}{\vrule \kern-0.4pt}{}{}
\makeatother

\usepackage{url}
\usepackage{enumerate}
\usepackage{color}
\usepackage{caption}
\usepackage{enumitem}
\setlist[itemize]{leftmargin=10pt}
\usepackage{array}
\usepackage{tikz}
\usepackage{csquotes}

\newtheorem{example}{Example}
\newtheorem{lemma}{Lemma}
\newtheorem{definition}{Definition}
\newtheorem{theorem}{Theorem}

\newcommand{\shorteat}[1]{}

\definecolor{myBrown}{RGB}{128,23,31}
\definecolor{myGreen}{RGB}{29, 131, 72}
\definecolor{expcolor}{RGB}{0,140,0}
\newcommand{\two}[1]{\textcolor{myBrown}{#1}}
\newcommand{\three}[1]{\textcolor{myGreen}{#1}}
\newcommand{\four}[1]{\textcolor{blue}{#1}}
\renewcommand{\two}[1]{{#1}}
\renewcommand{\three}[1]{{#1}}
\renewcommand{\four}[1]{{#1}}
\newcommand{\TR}[1]{}

\newcommand{\stab}{\rule{0pt}{8pt}\\[-3ex]}
\newcommand{\eg}{{\it e.g.}, }
\newcommand{\ie}{{\it i.e.}, }
\newcommand{\st}{{\it s.t.} }

\newcommand{\resp}{{\it resp.} }

\let\oldnl\nl%
\newcommand{\nonl}{\renewcommand{\nl}{\let\nl\oldnl}}%

\newcommand{\User}{\mathsf{User}}
\newcommand{\SP}{\mathsf{SP}}

\newcommand{\wrt}{{\it w.r.t.} }

\newcommand{\access}{\mathsf{Access}}

\newcommand{\dis}[2]{\mathsf{ged}(#1, #2)}

\newcommand*\circled[1]{\tikz[baseline=(char.base)]{
            {\small \node[shape=circle,draw,inner sep=0.6pt] (char) {#1};}}}

\newcommand{\iso}{\mathsf{Sub\textnormal{-}Iso}}
\newcommand{\homo}{\mathsf{Hom}}

\newcommand{\ssim}{\mathsf{SSim}}
\newcommand{\gsim}{\mathsf{GSim}}
\newcommand{\ppdo}{\mathsf{DOS}}
\newcommand{\fullname}{\mathsf{ObliGraph}}
\newcommand{\DO}{\mathsf{DO}}
\newcommand{\FO}{\mathsf{FO}}

\newcommand{\gss}{\mathsf{GS}}
\newcommand{\ged}{\mathsf{GED}}
\newcommand{\base}{{{\mathsf{FOS}}}}
\newcommand{\opt}{{{\mathsf{Base}}}}
\newcommand{\cs}{\mathsf{CSet}}
\newcommand{\cseq}{\mathsf{CS}}
\newcommand{\cm}{\mathsf{CM}}
\newcommand{\aid}{\textsc{Aids}}
\newcommand{\chem}{\textsc{PubChem}}
\newcommand{\synchem}{\textsc{Chem1M}}
\newcommand{\syn}{\textsc{Syn}}
\newcommand{\slashdot}{\textsc{Slashdot}}
\newcommand{\dblp}{\textsc{DBLP}}

\newtheorem{manuallemmainner}{Lemma}
\newenvironment{manuallemma}[1]{%
  \renewcommand\themanuallemmainner{#1}%
  \manuallemmainner
}{\endmanualtheoreminner}

\newtheorem{manualtheoreminner}{Theorem}

\usepackage[linewidth=1pt]{mdframed}%
\usepackage{hyperref}

\mdfdefinestyle{@commentstyle}{
  backgroundcolor=black!15,
  hidealllines=true,
  innertopmargin=\topskip,
  innerbottommargin=\topskip,
  skipabove=0pt,
  skipbelow=0pt,
}

\newmdenv[style=@commentstyle]{reviewbox}

\usepackage{fancyhdr}

\makeatletter
\fancypagestyle{revisionstyle}{%
  \fancyhf{}
}
\makeatother

\definecolor{myBrown}{RGB}{128,23,31}
\definecolor{myGreen}{RGB}{29, 131, 72}

\apptocmd\normalsize{%
   \setlength{\abovedisplayskip}{2pt}
   \setlength{\belowdisplayskip}{2pt}
   \setlength{\abovedisplayshortskip}{2pt}
   \setlength{\belowdisplayshortskip}{2pt}
}{}{}

\begin{document}
\title{ObliGraph: Graph Pattern Query Processing with Full and Differential Obliviousness}

\author{Lyu Xu}
\affiliation{%
  \institution{Nanyang Technological University}
  \country{Singapore}
}
\email{lyu.xu@ntu.edu.sg}

\author{Xinyi Zhang}
\affiliation{%
  \institution{Hong Kong Baptist University}
  \city{Hong Kong SAR}
  \country{China}
}
\email{csxyzhang@comp.hkbu.edu.hk}

\author{Qichen Wang}
\affiliation{%
  \institution{Nanyang Technological University}
  \country{Singapore}
}
\email{qichen.wang@ntu.edu.sg}

\begin{abstract}
Graph pattern query is a fundamental primitive in graph databases for retrieving data graphs that match a query graph under various structural constraints. As graph data is increasingly outsourced to third-party cloud servers for scalable query processing, however, both the query graph and the data graphs may leak sensitive structural information through server-side access patterns. In this paper, we study privacy-preserving graph pattern query processing in the outsourced setting, aiming to protect the structures of both the query and data graphs from access-pattern leakage. To this end, we propose $\mathsf{ObliGraph}$, the first oblivious graph search framework that supports multiple classes of graph pattern queries under two privacy modes: \textsf{FOS} achieves full obliviousness by uniformly evaluating candidate mappings, making the server-side access pattern independent of the structures of the query and data graphs; \textsf{DOS} further improves query efficiency through pruning while retaining a provable privacy guarantee of differential obliviousness. This design enables a smooth trade-off between stronger privacy and higher efficiency within a single framework. Experiments on real and synthetic datasets demonstrate that $\mathsf{ObliGraph}$ efficiently supports a range of graph pattern queries, achieving substantial improvements while preserving strong privacy guarantees.
\end{abstract}

\maketitle

\vldbtopmatter

\section{Introduction}\label{sec:introduction}
Graph pattern query processing is a fundamental primitive for retrieving graph-structured data from graph databases. It has been widely used in applications such as bioinformatics, chemistry, circuit design, and social-network analysis~\cite{sprinzak2003reliable, DBLP:journals/bioinformatics/TianMSSP07, DBLP:conf/icde/TianP08, DBLP:conf/dac/OhlrichEGS93, DBLP:conf/sigmod/YanYH05, DBLP:journals/pvldb/ZouCO09, DBLP:conf/sigmod/ShangLZYW10, DBLP:journals/tods/MaCFHW14}. Given a query graph $Q$, a graph pattern query searches for data graphs whose structures match or resemble the structure of $Q$. Different query semantics impose different levels of structural rigidity. For example, \emph{subgraph isomorphism} ($\iso$)~\cite{DBLP:conf/stoc/Cook71} requires an exact one-to-one embedding of $Q$ into a data graph, while \emph{subgraph homomorphism} ($\homo$)~\cite{DBLP:journals/pvldb/KimSHHC15} allows many-to-one mappings. Simulation-based semantics, including \emph{graph simulation} ($\gsim$)~\cite{DBLP:conf/focs/HenzingerHK95} and \emph{strong simulation} ($\ssim$)~\cite{DBLP:journals/tods/MaCFHW14}, capture recursive preservation of the topology of $Q$. Similarity-based semantics, including \emph{graph similarity} ($\gss$) queries~\cite{DBLP:journals/prl/BunkeA83, DBLP:journals/tsmc/SanfeliuF83}, measure structural resemblance using criteria such as \emph{maximum common subgraph}~\cite{DBLP:journals/prl/BunkeS98, DBLP:conf/sigmod/YanYH05, DBLP:conf/sigmod/ShangLZYW10, DBLP:journals/tkde/ZhuQYC20}, \emph{graph alignment}~\cite{DBLP:journals/bioinformatics/TianMSSP07}, and missing edges or features~\cite{DBLP:journals/pvldb/ZhangYJ10, DBLP:conf/sigmod/ZhuLZZY12}.

Graph pattern queries are often computationally expensive, especially when the graph database is large, and the query semantics require complex structural matching. A natural deployment model is therefore data outsourcing: the data owner outsources the graph database to a service provider ($\SP$), and query users rely on $\SP$'s cloud resources to process graph pattern queries at scale. This model provides elastic computation and high availability, but it also raises serious privacy concerns. Both the outsourced data graphs and the query graphs may contain sensitive structural information. For example, in biological or social graph databases, the existence of a particular substructure may itself reveal private information. Even if the graph contents are protected, an untrusted $\SP$ may still infer structural information from the memory and storage access patterns generated during query processing.

\begin{table*}
\centering
\caption{Summary of state-of-the-art works on graph pattern queries with various privacy preservation targets}
\vspace*{-3ex}
\begin{threeparttable}
\resizebox{0.85\linewidth}{!}{
\begin{tabular}{|c|c|c|c|c|c|c|c|c|m{4cm}<{\centering}|}\hline
\multirow{2}{*}{\textbf{Methods}} & \multicolumn{3}{c|}{\textbf{Privacy to be preserved}} & \multicolumn{5}{c|}{\textbf{Query semantics}} &  \multirow{2}{*}{\textbf{Security basis}} \\ \cline{2-9}
& \textbf{Query graph} & \textbf{Data graph} & \textbf{Access pattern} & $\iso$ & $\homo$ & $\gss$ & $\ssim$ & $\gsim$ & \\\hline
$\mathsf{SPsubIso}$~\cite{DBLP:conf/icde/FanCXB15} & edge &\usym{2613} &  \emph{FO} & \usym{2713} & \usym{2613} & \usym{2613} & \usym{2613} & \usym{2613} & \emph{SHE}
\\ \hline
$\mathsf{All\_Eff}$~\cite{DBLP:conf/dasfaa/GaoXLCY018} & node label, edge & node label, edge  & \usym{2613} & \usym{2613} & \usym{2613} & \usym{2613} & \usym{2713} & \usym{2613} & $k$-automorphism \\ \hline
\four{$\mathsf{OblivGM}$~\cite{DBLP:journals/tifs/WangZJHW22}} & \four{node label, edge} & \four{node label, edge} & \four{\emph{FO}} & \four{\usym{2713}} & \four{\usym{2613}} & \four{\usym{2613}} & \four{\usym{2613}} & \four{\usym{2613}} & \four{secret sharing} \\ \hline
$\mathsf{Prilo}$~\cite{DBLP:journals/pacmmod/XuCPXB23} & edge & \usym{2613} & \emph{FO} & \usym{2713} & \usym{2713} & \usym{2613} & \usym{2713} & \usym{2613} & \emph{SHE} $+$ \emph{TEE}\\\hline
$\mathsf{PrigSim}$~\cite{DBLP:journals/tkde/WangZJHW23} & node label, edge & node label, edge & \emph{FO} + $k$-automorphism  & \usym{2613} & \usym{2613} & \usym{2713} & \usym{2613} & \usym{2613} & secret sharing \\ \hline
\four{$\mathsf{PGSim}$~\cite{DBLP:journals/tifs/ZhengZLGZWSL23}} & \four{node label, edge} & \four{node label, edge} & \four{\usym{2613}} & \four{\usym{2613}} & \four{\usym{2613}} & \four{\usym{2713}} & \four{\usym{2613}} & \four{\usym{2613}} & \four{encrypted index} \\ \hline
$\mathsf{eGrass}$~\cite{DBLP:journals/tifs/WangZJW24} & node label & node label, edge & \emph{FO} & \usym{2713} & \usym{2613} & \usym{2613} & \usym{2613} & \usym{2613} & secret sharing \\ \hline
\four{$\mathsf{GraphGuard}$~\cite{DBLP:conf/uss/WangZJ24}} & \four{node label, edge} & \four{node label, edge} & \four{\emph{FO}} & \four{\usym{2713}} & \four{\usym{2613}} & \four{\usym{2613}} & \four{\usym{2613}} & \four{\usym{2613}} & \four{secret sharing $+$ edge \emph{DP}} \\ \hline
$\mathsf{Fresh}$~\cite{DBLP:conf/icde/HuangL0TZC0024} & \usym{2613} & edge  & \usym{2613} & \usym{2713} & \usym{2613} & \usym{2613} & \usym{2613} & \usym{2613} & $k$-automorphism \\ \hline
scheme in \cite{DBLP:journals/tdsc/GeYH24} & edge & edge & \usym{2613} & \usym{2713} & \usym{2613} & \usym{2613} & \usym{2613} & \usym{2613} & \emph{ORE}, secret sharing \\ \hline
$\mathsf{PPGSQ}$~\cite{DBLP:journals/tkde/Gexinrui26} & trapdoor token & index & \usym{2613} & \usym{2613} & \usym{2613} & \usym{2713} & \usym{2613} & \usym{2613} & Order-revealing encryption \\ \hline
$\fullname$ (ours) & node label, edge & node label, edge & \emph{FO} \& \emph{DO} & \usym{2713} & \usym{2713} & \usym{2713} & \usym{2713} & \usym{2713} & \four{\emph{TEE} $+$ oblivious operators} \\ \hline
\end{tabular}
}
\begin{tablenotes}
\scriptsize
\item \hspace{9ex} \emph{FO}: full obliviousness \hspace{4ex} \emph{DO}: differential obliviousness \hspace{4ex} \emph{SHE}: somewhat homomorphic encryption \hspace{4ex} \emph{TEE}: trusted execution environment
\end{tablenotes}
\end{threeparttable}
\label{tab:solutionsummary}
\end{table*}

\stab
\noindent\textbf{Limitations of existing works.}
Recent studies have considered privacy-preserving graph pattern query processing under different protection targets, protecting the query graph~\cite{DBLP:conf/icde/XuJCXB21, DBLP:journals/pacmmod/XuCPXB23}, the data graph~\cite{DBLP:conf/ccs/LaiYSLLL19, DBLP:journals/vldb/HuangHZGYZ22, DBLP:conf/ccs/KotiKPG24, DBLP:conf/icde/HuangL0TZC0024}, or both~\cite{DBLP:journals/tifs/WangZJHW22, DBLP:journals/tkde/WangZJHW23, DBLP:journals/tifs/ZhengZLGZWSL23, DBLP:conf/uss/WangZJ24, DBLP:journals/tifs/WangZJW24, DBLP:journals/tdsc/GeYH24, DBLP:journals/tkde/Gexinrui26}. However, only a few of them~\cite{DBLP:conf/icde/FanCXB15, DBLP:journals/pacmmod/XuCPXB23, DBLP:journals/tkde/WangZJHW23, DBLP:journals/tifs/WangZJW24, DBLP:journals/tifs/WangZJHW22, DBLP:conf/uss/WangZJ24} also protect the \emph{access pattern} of the search process. This is a crucial gap because graph pattern query processing is highly data-dependent: pruning rules, candidate generation orders, and early termination conditions can all reveal information about the structures of the query and data graphs. \four{Moreover, those that do protect the access pattern either leave the data graph unprotected~\cite{DBLP:conf/icde/FanCXB15, DBLP:journals/pacmmod/XuCPXB23}, rely on \emph{multiple non-colluding} servers~\cite{DBLP:journals/tifs/WangZJHW22, DBLP:journals/tkde/WangZJHW23, DBLP:conf/uss/WangZJ24, DBLP:journals/tifs/WangZJW24}, or target different queries over attributed or streaming graphs~\cite{DBLP:journals/tifs/WangZJHW22, DBLP:conf/uss/WangZJ24, DBLP:journals/tifs/WangZJW24}; and where privacy is traded for efficiency, the relaxation is applied to the \emph{data} being searched~\cite{DBLP:journals/tkde/WangZJHW23, DBLP:conf/uss/WangZJ24} rather than to the access pattern.} \two{$\mathsf{GraphOS}$~\cite{DBLP:journals/pvldb/ChamaniD0PJ23} is the closest peer, as it also runs inside a TEE at a single service provider, but it is designed for graph queries that take \emph{no} input query graph.} \four{Data-dependent optimizations from plaintext graph mining systems~\cite{DBLP:conf/sosp/MawhirterW19, DBLP:conf/sc/ShiZXZ20} are likewise inapplicable, since they produce precisely the access-pattern leakage that we aim to eliminate.}

Existing 
{methods} also face a scalability challenge {that has}
two main sources. First, many privacy-preserving designs rely on expensive cryptographic operations on encrypted graph data, making large-scale graph search difficult. Second, when access-pattern leakage is considered, existing designs often pursue \emph{full obliviousness} (\emph{FO}), which requires the access pattern during query processing to be independent of the structures of both the query and data graphs. Although \emph{FO} provides strong privacy guarantees, it is in tension with efficient graph search. In ordinary graph pattern query processing, efficiency usually comes from data-dependent pruning: the algorithm avoids exploring candidates that cannot lead to answers. Under \emph{FO}, such structure-aware pruning is infeasible, as different pruning decisions may create distinguishable access patterns.

The relaxation principle behind differential privacy (\emph{DP})~\cite{DBLP:conf/tcc/DworkMNS06} has been widely used in privacy-preserving data analysis~\cite{DBLP:conf/sigmod/ZhangCPSX15, DBLP:journals/csur/LiPRSRN23, DBLP:conf/icde/Liu0MLY24, DBLP:journals/tkde/LiuWLCL24}. Classical \emph{DP} mechanisms usually randomize the input, the computation, or the output, thereby trading accuracy for a bounded privacy loss. Accordingly, the \emph{DP} variant applies to only connected graphs and for two graphs differing in one edge or one node. This is not ideal for processing exact graph pattern queries, where users expect correct answers under the specified query semantics.  This motivates the central question in this paper:

\enquote{Can we answer graph pattern queries exactly while allowing useful search optimizations under a formal privacy guarantee?}

\stab
\noindent\textbf{Our approach.} \four{Recent advances in \emph{trusted execution environment} (\emph{TEE}) technologies, including \emph{Intel TDX}~\cite{IntelTDX2022}, \emph{AMD SEV}~\cite{AMDMemoryEncryption2021}, and \emph{NVIDIA}'s confidential computing~\cite{NVIDIAConfidentialComputing2023}, provide a promising foundation for confidential graph query processing. With a TEE, graph computations can be executed in plaintext inside a hardware-based memory-protected environment, thereby avoiding the high cost of processing graphs purely in the ciphertext domain.} However, a TEE alone does not eliminate leakage through access patterns. Therefore, we study how to build a privacy-preserving graph search framework that uses a TEE for confidential computation while controlling what can be inferred from the access patterns generated during query processing.

We propose $\fullname$, a privacy-preserving search framework for graph pattern queries. The goal of $\fullname$ is twofold: (i) to protect the structures and labels of both the query graph and the data graphs against access-pattern leakage, and (ii) to support search optimizations with a formal privacy guarantee. Achieving these goals raises two main challenges.

\stab
\noindent\textbf{Challenge 1.} \emph{Supporting 
{various} graph pattern 
{queries} under a~strong privacy guarantee.} 
Table~\ref{tab:solutionsummary} summarizes existing privacy-preserving systems and their %
{privacy} targets. To the best of our knowledge, no prior work supports all graph pattern query semantics considered in this paper while also protecting the structures and labels of both the query and data graphs against access-pattern leakage.

To address this challenge, we design $\base$, the fully oblivious search protocol in $\fullname$. $\base$ supports three representative query semantics $\mathcal{F}$, $\iso$, $\gss$, and $\gsim$. It also extends naturally to $\homo$ and $\ssim$ through the corresponding extensions of $\iso$ and $\gsim$, respectively. The key idea is to separate graph pattern query processing into two parts. The first part is a generic candidate-enumeration procedure that obliviously enumerates candidate mappings between the vertices of the query graph $Q$ and a data graph $G$. The second part is a query-specific verification or scoring procedure that evaluates these candidate mappings based on the target query semantics. By making the candidate enumeration and the following computation oblivious, $\base$ ensures that the access pattern is independent of the structures and labels of the input graphs, thereby satisfying \emph{FO}.

\stab
\noindent\textbf{Challenge 2.} \emph{Improving search efficiency with a provable privacy guarantee.}
Although $\base$ provides a strong privacy guarantee, full obliviousness also limits efficiency by forcing the algorithm to explore candidates without data-dependent pruning. To improve efficiency, we need to allow pruning while still bounding the information leaked by the resulting access patterns.

We address this challenge by using \emph{differential obliviousness} (\emph{DO})~\cite{DBLP:conf/soda/ChanCMS19}, which has recently been used for several query-processing tasks~\cite{DBLP:journals/pvldb/QinJS0ZC22, DBLP:journals/pvldb/QiuKMNK23, DBLP:journals/pacmmod/ZhangWXPX24, DBLP:journals/pacmmod/WuDH26}. \two{These \emph{DO} designs are all developed for \emph{relational} operators; even $\mathsf{GraphOS}$~\cite{DBLP:journals/pvldb/ChamaniD0PJ23}, the state-of-the-art system for oblivious graph processing, adopts the stronger notion of \emph{FO} and reports that it is \enquote{\emph{still far from practical}}.} Compared with \emph{FO}, \emph{DO} requires access patterns on any two \emph{neighboring} instances to be statistically indistinguishable up to a bounded privacy loss. This gives a principled way to trade privacy for efficiency.

Based on this idea, we design $\ppdo$, an optimized search approach of $\fullname$. Instead of pruning the query-specific cost computation, which depends heavily on the concrete 
{query}
semantics, $\ppdo$ prunes the candidate-enumeration process of $\base$. A set of pruning parameters controls the pruning behavior. We analyze how these parameters affect the access patterns on neighboring inputs and derive thresholds that bound the sensitivity of the pruning traces. We then add appropriate \emph{non-negative} noise to these thresholds based on the given privacy budget. The one-sided nature of the noise is important: it may cause $\ppdo$ to retain additional candidates and perform some redundant computation, but it never removes candidates that are needed for correctness. Therefore, $\ppdo$ preserves exact query answers while satisfying the privacy guarantee of \emph{DO} for the access patterns generated by candidate pruning.

\noindent\textbf{Contributions.} Our main contributions are summarized as follows:

\begin{itemize}[wide, labelwidth=!, labelindent=0pt]

\item We propose $\fullname$, the first privacy-preserving search framework for graph pattern query processing that protects the structures and labels of both the query and data graphs, and controls the access-pattern leakage generated during query processing.

\item We design $\base$, a fully oblivious search protocol for $\fullname$. $\base$ supports $\iso$, $\gss$, and $\gsim$ queries over confidential graph data, and it further supports $\homo$ and $\ssim$ through the corresponding extensions of $\iso$ and $\gsim$. Its access pattern satisfies the privacy guarantee of \emph{FO}.

\item We develop $\ppdo$, an optimized search protocol that improves the efficiency of $\base$ through candidate pruning. We quantify the access-pattern leakage caused by pruning across neighboring inputs and show how to bound it under the privacy guarantee of 
{\emph{DO}.}

\item We implement a prototype of $\fullname$ and evaluate it on \two{five} real-world datasets and one synthetic dataset. The experimental results show that $\fullname$ supports multiple graph pattern query semantics with practical performance under 
{\emph{FO} and \emph{DO}.}
\end{itemize}
\section{Preliminaries}
\label{sec:preliminaries}

\subsection{Graph Pattern Query}
\label{subsec:query}
In this paper, we consider \emph{connected}, \emph{directed}, and \emph{labeled} graphs. For simplicity, we focus only on node labels. Our techniques can be extended to undirected graphs and edge-labeled graphs with minor modifications.
Some frequently used notations are summarized in 
Appendix~A
of the technical report~\cite{repo:dos}.

\stab
\noindent\textbf{Graph.} A \emph{graph} is denoted by $G = (V_G, E_G, \Sigma_G, \ell_G)$, where (i) $V_G$ is the node set; (ii) $E_G: V_G\times V_G \to \{0, 1\}$ is the edge function, such that $E_G(u, v) = 1$ if there exists an edge from node $u$ to node $v$, and $E_G(u, v) = 0$ otherwise; (iii) $\Sigma_G$ is the label set; and (iv) $\ell_G$: $ V_G \to \Sigma_G$ is the node-labeling function. When no confusion arises, we omit the subscript of $G$.

\stab
\noindent\textbf{Graph edit distance ($\ged$).} We adopt \emph{graph edit distance}~\cite{DBLP:conf/kdd/QinBS20, DBLP:conf/sigmod/Kim21, DBLP:conf/icde/PengCCX22, DBLP:journals/tkde/ChangFYQZ23, DBLP:journals/tkde/WangZJHW23} to measure the similarity between graphs. Given two graphs $G_1$ and $G_2$, the \emph{graph edit distance} between $G_1$ and $G_2$, denoted by $\dis{G_1}{G_2}$, is the \emph{minimum} number of edit operations required to transform $G_1$ into $G_2$.  We consider the following edit operations, which align with the literature~\cite{DBLP:conf/kdd/QinBS20, DBLP:conf/icde/PengCCX22, DBLP:journals/pacmmod/ChengYWHZ25}: 
\begin{itemize}
\item insertion or deletion of an isolated node with label $l$;
\item insertion or deletion of an edge; and
\item replacement of the label (\ie relabeling) of a node with label $l$.
\end{itemize}
Here, node insertion and deletion are defined on isolated nodes, while changes to incident edges are handled separately by edge insertions and deletions.

\begin{figure}[t]
  \setlength{\abovecaptionskip}{0pt}
  \setlength{\belowcaptionskip}{0pt}
  \centering
  \graphicspath{{figures/}}
  \includegraphics[width=0.48\textwidth]{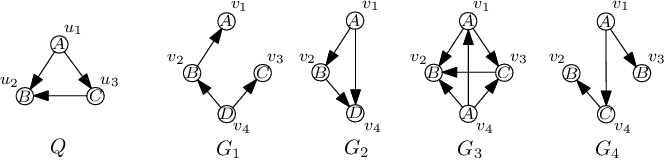}
  \caption{An example of query $Q$ and data graphs $G_i$ ($1\le i \le 4$)}
  \label {fig:definition}
\end{figure}

\begin{example}
\label{example:two}
Figure~\ref{fig:definition} shows a query graph $Q = (V_Q, E_Q, \Sigma_Q, \ell_Q)$ and four data graphs $G_i$, $1\le i\le 4$. For $Q$, we have $V_Q = \{u_1, u_2, u_3\}$, $E_Q(u_1, u_2) = E_Q(u_1, u_3) = E_Q(u_3, u_2) = 1$, $\Sigma_Q = \{A, B, C\}$, and $\ell_Q(u_1) = A$, $\ell_Q(u_2) = B$, and $\ell_Q(u_3) = C$.
Consider the edit distance between $Q$ and $G_1$. One minimum edit sequence that transforms $Q$ into $G_1$ is:
(i) relabel $u_1$ from $A$ to $D$ $\to$
(ii) delete the edge from $u_3$ to $u_2$ $\to$
(iii) insert an isolated node $u_4$ with label $A$ $\to$
(iv) insert an edge from $u_2$ to $u_4$.
Hence, $\dis{Q}{G_1} = 4$. By symmetry, the same distance is obtained when transforming $G_1$ into $Q$. Similarly, we have $\dis{Q}{G_2} = 2$, $\dis{Q}{G_3} = 4$, and $\dis{Q}{G_4} = 3$.
\end{example}

We next present three representative query semantics studied in depth in this paper: 
\emph{subgraph isomorphism}, \emph{graph similarity}, and \emph{graph simulation}.

\begin{definition}[\textbf{Subgraph Isomorphism \textnormal{(}$\iso$\textnormal{)}}]
\label{def:iso}
Given a query graph $Q$ and a data graph $G$, a \emph{subgraph isomorphism} of $Q$ in $G$ is an \emph{injective} function $f: V_Q \to V_G$ satisfying the conditions:
\begin{itemize}[leftmargin=4ex]
\item[(i)] $\forall u \in V_Q$, $\ell_Q(u) = \ell_G\big(f(u)\big)$; and
\item[(ii)] $\forall u, v\in V_Q$ \st $E_Q(u, v) = 1$, $E_G\big(f(u), f(v)\big) = 1$.
\end{itemize}

If~$f$~exists, we say that $G$ matches $Q$ via $\iso$.
\end{definition}

\begin{definition}[\textbf{Graph Similarity \textnormal{(}$\gss$\textnormal{)}}]
\label{def:gss}
Given a query~graph~$Q$, a data graph $G$, and a similarity parameter $\tau$, $G$ matches $Q$ via \emph{graph similarity} if $\dis{Q}{G} \le \tau$.
\end{definition}

\begin{definition}[\textbf{Graph Simulation \textnormal{(}$\gsim$\textnormal{)}}]
\label{def:gsim}
Given a query graph $Q$ and a data graph $G$, $G$ matches $Q$ via \emph{graph simulation} if there exists a binary relation $S\subseteq V_Q \times V_G$ satisfying the conditions:
\begin{itemize}[leftmargin=4ex]
\item[(i)] $\forall \langle u, v\rangle \in S$, $\ell_Q(u) = \ell_G(v)$; and
\item[(ii)] for any node $u$ in $V_Q$, there exists $v$ in $V_G$ satisfying:
\begin{itemize}
\item $\langle u, v\rangle \in S$; and
\item $\forall u' \in V_Q$ \st $E_Q(u, u') = 1$, $\exists v' \in V_G$ such that $E_G(v, v') = 1$ and $\langle u', v'\rangle \in S$.
\end{itemize}
\end{itemize}
\end{definition}

\begin{example}
Consider the example in Figure~\ref{fig:definition} again.
\begin{itemize}
\item \textbf{$\iso$ semantics.} $G_3$ matches $Q$ via $\iso$ since there exists an injective function $f: V_Q \to V_{G_3}$ with
$f(u_1) = v_1$, $f(u_2) = v_2$, and $f(u_3) = v_3$, satisfying the conditions in Definition~\ref{def:iso}.
\item \textbf{$\gss$ semantics with $\tau = 2$.} As shown in Example~\ref{example:two}, we have $\dis{Q}{G_2} = 2 \le \tau$. Hence, $G_2$ matches $Q$ via $\gss$.
\item \textbf{$\gsim$ semantics.} There exist binary relations $S_3 \subseteq V_Q \times V_{G_3} = \big\{\langle u_1, v_1\rangle$, $\langle u_1, v_4\rangle$, $\langle u_2, v_2\rangle$, $\langle u_3, v_3\rangle\big\}$ and $S_4 \subseteq V_Q \times V_{G_4} = \big\{\langle u_1, v_1\rangle$, $\langle u_2, v_2\rangle$, $\langle u_2, v_3\rangle$, $\langle u_3, v_4\rangle\big\}$ satisfying the conditions in Definition~\ref{def:gsim}. Thus, both $G_3$ and $G_4$ match $Q$ via $\gsim$.
\end{itemize}
\end{example}

We are now ready to define the 
graph query studied in this paper.

\stab
\noindent\textbf{Graph pattern query.}
Given a query graph $Q$, a query semantics $\mathcal{F}$ $\in$ $\{\iso$, $\gss$, $\gsim\}$ with similarity parameter $\tau$, and a graph database $\mathcal{D}$ $=$ $\{G_1, ..., G_n\}$, the \emph{graph pattern query} $(Q, \mathcal{F}, \tau)$ returns the answer set $\mathcal{R}(Q,  \mathcal{F}, \tau, \mathcal{D})$ of graphs in $\mathcal{D}$ that match $Q$ via $\mathcal{F}$.

\stab
\noindent\underline{\textbf{Remark.}}
With minor modifications, our methods can be extended to support \emph{graph homomorphism} ($\homo$) and \emph{strong simulation} ($\ssim$), based on $\iso$ and $\gsim$, respectively. We omit their definitions due to space limitations. For $\mathcal{F} \in \{\iso, \homo, \gsim, \ssim\}$, the similarity parameter $\tau$ can~be taken to be \emph{zero}. 

\subsection{Data Outsourcing and Privacy Target}
\label{subsec:system}
\begin{figure}[t]
    \centering
    \graphicspath{{figures/}}
    \includegraphics[width=0.5\textwidth]{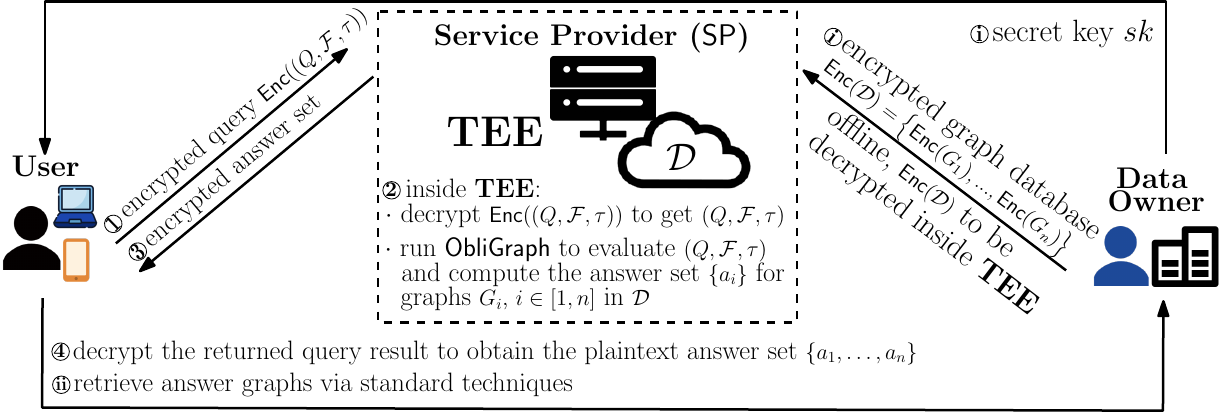}
    \vspace{-5.5ex}
    \caption{\four{The system model}}
    \label{fig:system}
\end{figure}

\noindent\textbf{System model.}
\four{Following a standard outsourced-database model~\cite{DBLP:conf/icdcs/CaoYWRL11, DBLP:conf/icde/WuLLYYW19, DBLP:conf/icde/LeiLLT19, DBLP:conf/icde/Cui0WLW20, DBLP:conf/icde/XuJCXB21, DBLP:journals/pacmmod/XuCPXB23, DBLP:journals/tkde/WangZJHW23, DBLP:journals/tdsc/GeYH24}, we illustrate in Figure~\ref{fig:system} the system model used in this work, which involves a data owner, a user, and a service provider equipped with a \emph{trusted execution environment} (\emph{TEE}).}

\begin{itemize}[wide, labelwidth=!, labelindent=0pt]
\item \textbf{Data owner}. The data owner holds a graph database $\mathcal{D}$. \four{In step \circled{i}, the data owner generates a secret key $sk$,
encrypts $\mathcal{D}$ offline using a standard symmetric encryption scheme (\eg \emph{AES-256}), and sends the encrypted $\mathcal{D}$, namely $\mathsf{Enc}(\mathcal{D})$, 
to~the~TEE deployed at the service provider through a secure channel. $\mathsf{Enc}(\mathcal{D})$ is decrypted inside the TEE.} Only authorized users can obtain~$sk$.

\item \textbf{User}. \four{In step \circled{1}, through a secure channel, $\User$ sends a query $(Q, \mathcal{F}, \tau)$ encrypted via \emph{AES-256} to the service provider's TEE.} After receiving the encrypted query answers in step \circled{3}, $\User$ decrypts them in step \circled{4} to obtain the identifiers of the answer graphs. Using the secret key $sk$, $\User$ then retrieves the answer graphs in step \circled{ii} via standard techniques, \eg \emph{private information retrieval}~\cite{DBLP:journals/jacm/ChorKGS98}.

\item \textbf{Service provider ($\SP$)}. After receiving the encrypted query from $\User$ in step \circled{1}, \four{inside $\SP$'s TEE, $\SP$ decrypts the query and executes our proposed $\fullname$ framework in step \circled{2} for query evaluation. For each data graph $G_i$, $i \in [1, n]$, $\fullname$ computes inside the TEE an answer value $a_i$ to indicate if $G_i$ is an answer graph. $\SP$ encrypts the set $\{a_i\}$ of query answers via \emph{AES-256}, and sends the encrypted answers back to $\User$ through a secure channel in step~\circled{3}.}
\end{itemize}

The overall workflow of $\fullname$ is
\circled{i} $\to$ \circled{1} $\to$ \circled{2} $\to$ \circled{3} $\to$ \circled{4} $\to$ \circled{ii}.
In this paper, we focus on the key step \circled{2}, namely privacy-preserving graph pattern query processing at the $\SP$ side, as highlighted by the dashed box in Figure~\ref{fig:system}. The other steps rely on standard techniques and are therefore omitted.

\stab
\noindent\textbf{Privacy target.}
As in prior privacy-preserving query processing work~\cite{DBLP:journals/pvldb/BaterEEGKR17, DBLP:journals/tkde/DuWWCJM22, DBLP:journals/tkde/WangZJHW23, DBLP:journals/pacmmod/ZhangWXPX24, DBLP:journals/vldb/TongZSPFXZZCXXL25}, we consider a \emph{semi-honest} adversary model. That is, $\SP$ follows the prescribed protocol but attempts to infer sensitive information about both $\User$'s query graph and the data graphs in $\mathcal{D}$ from what it observes during query processing.

Under this model, we consider two types of leakage. First, $\SP$ should not obtain the raw, sensitive data during the processing, such as the query graph or the data graphs. Second, even if the data are protected, $\SP$ may still infer structural information from the \emph{access pattern} generated by the computation; for example, by observing memory accesses, the adversary can infer the size or order of intermediate results and further the graph structure.

Accordingly, we protect the following two kinds of privacy during query processing at $\SP$:
\begin{itemize}[labelwidth=!,labelindent=0pt]
\item {\bf Data privacy.} Data privacy concerns the structural information of $\User$'s query graph $Q$ and each data graph $G_i$, including their edge {functions}, label sets, and labeling functions. More precisely, during query processing, $\SP$ should learn nothing about $Q$ and each $G_i$ beyond an explicitly allowed leakage profile, including the numbers of nodes (\ie $|V_Q|$ and $|V_{G_i}|$), the query semantics $\mathcal{F}$, and the similarity threshold $\tau$.

\item {\bf Access pattern privacy.} Access pattern privacy concerns the access pattern observed by $\SP$ during query processing, including the order, location, and size of accesses. Depending on the instantiated mechanism, we protect this leakage under either the strong notion of \emph{full obliviousness} or the relaxed \emph{differential obliviousness}. The formal definitions will be given in \S~\ref{subsec:access}.

\end{itemize}

\section{Security Basis}
\label{sec:access}

\subsection{Trusted Execution Environment}
\label{subsec:tee}
\four{A straightforward way to protect the sensitive contents of graphs (\ie the data privacy in \S~\ref{subsec:system}) is to keep all data encrypted and execute queries in the ciphertext domain. However, such ciphertext-domain computation is typically expensive for graph pattern queries due to their combinatorial nature. Recent \emph{trusted execution environment} (\emph{TEE}) and confidential-computing technologies 
provide a more practical alternative by enabling computation inside hardware-protected execution environments (\S~\ref{subsec:system}).}

A TEE improves confidentiality and efficiency, but it does not, by itself, hide access patterns, such as memory accesses or execution traces. Such leakage may allow an adversarial service provider to infer sensitive structural information about the query graph or the outsourced data graphs~\cite{DBLP:conf/sp/XuCP15}. Therefore, the goal of $\fullname$ is not only to protect graph contents, but also to control access-pattern leakage during graph pattern query processing.

\subsection{Obliviousness}
\label{subsec:access}
\emph{Obliviousness}~\cite{DBLP:conf/stoc/Goldreich87} is a standard security notion for protecting against leakage through access patterns. To formalize it in our setting, we first define the \emph{leakage function} and \emph{access pattern} of a protocol~$\Pi$.

\stab
\noindent\textbf{Leakage function.}
Let $\mathcal{L}\big((Q, \mathcal{F}, \tau), \mathcal{D}\big)$ be a leakage function that captures the information revealed during the execution of $\Pi$ on a database $\mathcal{D} = \{G_1, ..., G_n\}$ for a graph pattern query $(Q, \mathcal{F}, \tau)$. We assume the node set size $|V_Q|$ of $Q$, the node set size $|V_{G_i}|$ of each data graph $G_i$, the query semantics $\mathcal{F}$, and the similarity parameter $\tau$ are not sensitive. With this assumption, we have $\mathcal{L}\big((Q, \mathcal{F}, \tau), \mathcal{D}\big) = \big\{|V_Q|$, $|V_{G_1}|, ..., |V_{G_n}|$, $\mathcal{F}, \tau\big\}$. Then, the access~pattern during the execution of $\Pi$ for the query is defined as follows.

\stab
\noindent\textbf{Access pattern.}
\four{$\fullname$ protects the sequence of \emph{logical memory accesses} issued by the query processing logic inside the TEE, \ie the order, the logical address, and the size of every read and write performed on its data structures. Micro-architectural side channels that are observable \emph{below} the logical-address level, such as cache-timing, page-fault, and speculative-execution channels, are \emph{beyond the scope} of this work.} Let $\access^\Pi\big((Q, \mathcal{F}, \tau), \mathcal{D}\big) := (o_1, o_2, ..., o_l)$ denote the \emph{access pattern} of length $l$, which defines the adversary's observation during the execution of protocol $\Pi$ on database $\mathcal{D}$ for query $(Q, \mathcal{F}, \tau)$. Each access $o$ $\in$  $\access^\Pi\big((Q, \mathcal{F}, \tau), \mathcal{D}\big)$ can be represented as a 3-tuple $\langle op$, $addr$, $data\rangle$, where (i) $op \in \{read, write\}$; (ii) $addr$ is the logical address accessed by $\Pi$; and (iii) $data$ is the data structure to be written/read.

Depending on the strength of the privacy guarantee, we consider two notions of access-pattern protection: \emph{full obliviousness}~\cite{DBLP:conf/stoc/Goldreich87}, and \emph{differential obliviousness}~\cite{DBLP:conf/soda/ChanCMS19}.

\begin{definition}[\textbf{Full Obliviousness}]
Consider a database $\mathcal{D}$,~a graph pattern query $(Q, \mathcal{F}, \tau)$, and any graph pattern query $(\hat{Q}, \mathcal{F}, \tau)$ such that $\hat{Q}$ has the same node set size as that of $Q$, \ie $|V_Q|$ $=$~$|V_{\hat{Q}}|$. We say a protocol $\Pi$ is \emph{fully oblivious (or simply $\FO$) \wrt query graph} if a polynomial-time adversary $\mathcal{A}$ only has a negligible advantage in distinguishing $Q$~and~$\hat{Q}$:
\begin{align}
\label{eqa:fo}
&Pr\Big[Sim^\Pi\Big(\mathcal{L}, \access^\Pi\big((Q, \mathcal{F}, \tau), \mathcal{D}\big)\Big) = 1\Big] \le \\
\nonumber
&Pr\Big[Sim^\Pi\Big(\mathcal{L}, \access^\Pi\big((\hat{Q}, \mathcal{F}, \tau), \mathcal{D}\big)\Big) = 1\Big] + \delta,
\end{align}
where (i) the simulator $Sim^\Pi\Big(\mathcal{L}, \access^\Pi\big((\hat{Q}, \mathcal{F}, \tau), \mathcal{D}\big)\Big)$ equals $1$~if~$\mathcal{A}$ determines that $\hat{Q}$ is the same as $Q$ based on $\mathcal{L}$ and $\access^\Pi\big((\hat{Q}, \mathcal{F}, \tau)\big)$; and (ii) $\delta$ is a security parameter to control the negligible probability that the adversary can learn additional information beyond $\mathcal{L}$.
\end{definition}

Although full obliviousness provides strong privacy guarantees, it may significantly degrade performance by ruling out data-dependent access patterns. To improve efficiency, a relaxed notion called \emph{($\epsilon, \delta$)-differential obliviousness} has been proposed~\cite{DBLP:conf/soda/ChanCMS19, DBLP:journals/pvldb/QinJS0ZC22, DBLP:journals/tdsc/WuLNHW22, DBLP:journals/pacmmod/ZhangWXPX24}. Intuitively, instead of requiring indistinguishability for all admissible inputs, differential obliviousness only requires that neighboring inputs induce statistically close access-pattern distributions.

\begin{definition}[\textbf{($\epsilon$, $\delta$)-Differential Obliviousness}]
\label{def:dp}
Consider a database $\mathcal{D}$, and 
two neighboring graph pattern queries $(Q_1$, $\mathcal{F}$, $\tau)$ and $(Q_2$, $\mathcal{F}$, $\tau)$
satisfying that (i) $Q_1$ and $Q_2$ have the same node set size, \ie $|V_{Q_1}| = |V_{Q_2}|$; and (ii) $Q_1$ and $Q_2$ can be edited to become each other with one edit operation in the $\ged$ metric, \ie $\dis{Q_1}{Q_2} = 1$.\footnote{\label{fn:neighbor}As $Q_1$ and $Q_2$ are assumed to be connected graphs, the edit operation between $Q_1$ and $Q_2$ \emph{cannot} be the insertion or deletion of an \emph{isolated} node.}
We say a protocol $\Pi$ is \emph{($\epsilon$, $\delta$)-differentially oblivious} (or simply $\DO$) \wrt \emph{query graph} if it holds that:
\begin{align}
\label{eqa:do}
&Pr\Big[Sim^\Pi\Big(\mathcal{L}, \access^\Pi\big((Q_1, \mathcal{F}, \tau), \mathcal{D}\big) \Big) = 1\Big] \le \\
\nonumber
&e^\epsilon \cdot Pr\Big[Sim^\Pi\Big(\mathcal{L}, \access^\Pi\big((Q_2, \mathcal{F}, \tau), \mathcal{D}\big)\Big) = 1\Big] + \delta,
\end{align}
where $\epsilon$ is the privacy budget and $\delta$ is a negligible probability with which the $\epsilon$-bound may fail.
\end{definition}

Compared with full obliviousness, differential obliviousness allows limited input-dependent variation in the access pattern, as long as this variation is quantitatively bounded for neighboring inputs. This relaxation enables more efficient privacy-preserving query-processing algorithms.

\stab
\noindent\underline{\textbf{Remark}.}
As the sensitive data of the query graph (excluding node set sizes) is independent of that of each data graph in $\mathcal{D}$, the notions~of full obliviousness and differential obliviousness can be equivalently extended to graph databases {in the context of their data graphs}.~Under our security model, a protocol $\Pi$ is fully oblivious (\resp differentially oblivious) if it satisfies full obliviousness $\big($\resp ($\epsilon$, $\delta$)-differential obliviousness$\big)$ \wrt both the query~graph and the database. For brevity, we simply say that $\Pi$ is $\FO$ $\big($\resp~$\DO$$\big)$. We also assume in this paper that the arithmetic operations, including bitwise operators, addition, and multiplication, satisfy full obliviousness. These oblivious arithmetic operations have been extensively studied in prior works~\cite{DBLP:conf/ccs/KellerOS16, DBLP:conf/crypto/ApplebaumDI0Z17, DBLP:conf/ndss/LeeJFG21}.

\section{Fully Oblivious Search}
\label{sec:base}
In this section, we present 
$\fullname$'s $\FO$ search base $\base$.
$\base$ consists of two 
components: (i)~an $\FO$ \emph{candidate mapping~enumeration} that enumerates all possible candidates {of mapping relations from the query graph to the data graph;}
and (ii) {query-specific} $\FO$ \emph{cost computation} that computes a cost of each candidate {for finding the query answer.} 
In the~following, we first present the 
candidate mapping enumeration in \S~\ref{subsec:canenumerate}, and then detail the cost computation for various query semantics in \S~\ref{subsec:editcost}.
\three{Due to space limitations, all proofs are presented in 
Appendix~B
of the technical report~\cite{repo:dos}.}

\subsection{Candidate Mapping Enumeration of $\base$}
\label{subsec:canenumerate}
Although candidate enumeration for graph pattern queries has~been well studied, existing methods fail to support various query semantics simultaneously while satisfying full obliviousness (with the leakage function $\mathcal{L}$ introduced in \S~\ref{subsec:access}).
To achieve this,~we explore \four{the basis of existing candidate enumeration paradigms for various graph pattern query semantics~\cite{DBLP:journals/pami/CordellaFSV04, DBLP:conf/sigmod/Kim21, DBLP:journals/tods/MaCFHW14}, based on which we design a general enumeration procedure.}

\subsubsection{Enumeration for various query semantics}
\label{subsubsec:canenumerate}
Given a query graph $Q$ 
and a data graph $G$,
we first design a general candidate mapping ($\cm$) enumeration for various query semantics, including $\iso$, $\gss$, and $\gsim$.
Given a node $u$ of  $Q$, we consider the candidate set of $u$, denoted by $\cs(u)$, for mapping $G$ to $Q$ as follows:

\begin{itemize} [wide, labelwidth=!, labelindent=0pt]
\item\textbf{$\iso$.} 
For each node $u$ of $Q$, $\cs(u)$ is initialized as the set of the nodes of $G$ having the same label as $u$'s label. Then, the injective function for $\iso$ (Definition~\ref{def:iso}) can be constructed as $\cm$s by choosing any node from $\cs(u)$ to match $u$.
In this way, the $\cm$ enumeration for $\iso$ constructs \emph{all possible} injective functions, and computes the cost for each function to verify whether there exist functions satisfying $\iso$. As there are $|V_Q|$ nodes to be matched in each function, the enumeration depth for $\iso$ is $|V_Q|$, and the enumeration space complexity is $O\big(|V_G|^{|V_Q|}\big)$.  
\three{For example, Figure~\ref{fig:enumeration-iso}(a) shows a simple illustration for enumerating the functions for mapping $G_3$ to $Q$ (from Figure~\ref{fig:definition}) via $\iso$, where $\cs(u_1)$ $=$ $\{v_1, v_4\}$, $\cs(u_2)$ $=$ $\{v_2\}$, and $\cs(u_3)$ $=$ $\{v_3\}$.~The two enumerated functions $f_1$, $f_2$ will be further verified for~$\iso$.}

\begin{figure}[t]
    \centering
    \graphicspath{{figures/}}
    \includegraphics[width=0.48\textwidth]{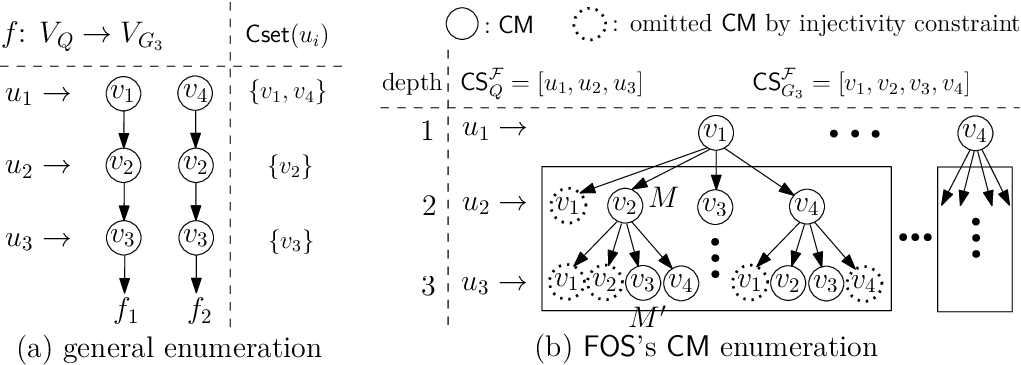}
    \vspace{-5.5ex}
    \caption{\three{Illustration of the candidate mapping enumerations for mapping $G_3$ to $Q$ (from Figure~\ref{fig:definition}) via $\iso$}}
    \label{fig:enumeration-iso}
\end{figure}

\item \textbf{$\gss$.} To evaluate the $\ged$ between $Q$ and $G$ for the $\gss$ query, $Q$ needs to be edited to become $G$, such that the edited $Q$ has the same number of nodes as $G$. For ease of $\ged$ cost computation, $\big||V_G|-|V_Q|\big|$ virtual nodes are inserted into $V_Q$ if $Q$ has fewer nodes than $G$; otherwise, virtual nodes are inserted into $V_G$.  
 
Then, for each node $u$ of $Q$, $\cs(u)$ is initialized as the set of all nodes of $G$. Same as $\iso$, the $\cm$ enumeration constructs \emph{all possible} injective functions from $V_Q$ to $V_G$ as $\cm$s, and computes the cost for each function to verify whether there exist functions satisfying $\gss$. Let $d$ $=$ $\max\big\{|V_Q|, |V_G|\big\}$. As there are $d$ nodes to be matched in each function, the enumeration depth for $\gss$ is $d$, and the enumeration space complexity is $O(d^d)$.

\item \textbf{$\gsim$.} A baseline method~\cite{DBLP:conf/focs/HenzingerHK95} answers the $\gsim$ query as follows: (i) for each node $u$ of $Q$, assume that all the nodes of $G$ having the same label as $u$'s label initially match $u$; (ii) iteratively remove the matches that cannot satisfy $\gsim$'s structural restriction $\big($\ie condition (ii) of Definition~\ref{def:gsim}$\big)$ until there are no more removals; and (iii) check whether there exists at least one match for each node of $Q$. If yes, $G$ matches $Q$ via $\gsim$. 

Based on this baseline, the candidate set of each node $u$ of $Q$~is initialized to be
$\cs(u)$ $=$ $\big\{\langle u, v\rangle~|~v\in V_G, \ell_Q(u)=\ell_G(v)\big\}$,~and~hence, 
the set $\big\{\langle u, v\rangle~|~\langle u,v\rangle\in \cs(u), u\in V_Q\big\}$ of all the elements in the candidate sets of $Q$'s nodes serves as the initial $\cm$.
Then, to obtain the final \emph{one} $\cm$, the $\cm$ enumeration iteratively traverses all the elements in the current $\cm$ $M$ to check the removal. The cost of $M$ is the number of removals in $M$.  
With {$|V_Q|\cdot |V_G|$} such traversals, it is guaranteed that there are no more removals since each traversal should remove at least one element if any removal exists. Hence, the $\cm$ enumeration processes $|V_Q|\cdot |V_G|$ $\cm$s, so the enumeration depth is $|V_Q|\cdot |V_G|$.
\end{itemize}

With the analyses above, a general enumeration paradigm consists of (i) candidate set initialization for nodes of $Q$, (ii) candidate set-based $\cm$ enumeration up to a specific depth, and (iii) cost computation for $\cm$s. Next, we 
{introduce the general $\FO$ $\cm$ enumeration of $\base$ by leveraging such an enumeration paradigm.}

\subsubsection{Overall {search process} of $\base$}
\label{subsubsec:overall}
\four{Based on the enumeration paradigm in \S~\ref{subsubsec:canenumerate}, we design $\base$'s $\cm$ enumeration by enumerating redundant $\cm$s to satisfy full obliviousness. Such an enumeration can be realized in either a \emph{BFS} or a \emph{DFS} manner, which are interchangeable in $\fullname$: 
the $\cm$ set $\mathcal{M}_i$ enumerated by the \emph{BFS}-based dynamic programming table at depth $i$ is identical to the one enumerated by the \emph{DFS}-based recursion at level $i$, and the costs of the corresponding $\cm$s are the same. To protect the access pattern, we aim to enumerate the $\cm$s in an $\FO$ or $\DO$ manner, together with $\FO$ cost computation for each $\cm$.
 For ease of presentation, we adopt the \emph{BFS} enumeration via \emph{dynamic programming}, whereas our prototype adopts the \emph{DFS} one to avoid materializing $\mathcal{M}_i$ (\S~\ref{sec:exp}).}
Algorithm~\ref{alg:base} shows the overall {search process} of $\base$.

\begin{algorithm}[tb]
	\caption{{The $\FO$ search base  $\base$ of $\fullname$}} 
	\label{alg:base}
	\small
	\SetKwFunction{canenum}{CanEnum}
  \SetKwFunction{funccost}{EditCost}
  \SetKwFunction{oblimax}{Obli\_Max}
    \SetKwFunction{oblieq}{Obli\_NonEqual}
	\SetKwProg{int}{int}{\string:}{}
	\SetKwProg{void}{void}{\string:}{}
	\SetKwInOut{Input}{Input}
	\SetKwInOut{Output}{Output}
	\SetKwData{Result}{result}
	\SetKwData{True}{true}
	\SetKwData{False}{false}
	\Indentp{-1em}
	\Input{The graph pattern query $(Q, \mathcal{F}, \tau)$ and the graph database $\mathcal{D}$}
	\Output{The answer set $\mathcal{R}$}
	\Indentp{1em}

	\tcp*[h]{$\base$'s main procedure}
  
	 $\mathcal{R}$ $\gets$ $\emptyset$\;

    \ForEach{\textnormal{graph $G$ in $\mathcal{D}$}}{

    $r$ $\gets$ \canenum{$G$}\;

      $\mathcal{R}$ $\gets$ $\mathcal{R}\cup \{r\}$\;
    }

    	\Return $\mathcal{R}$\;

	\BlankLine
  \nonl \textbf{Procedure~}{\canenum{$G$}:}
  
    \tcp*[h]{Initialization of enumeration}
  
  $r$ $\gets$ $1$, $d$ $\gets$ $\mathsf{depth}_\mathcal{F}$; \hfill\tcp*[h]{\S~\ref{subsubsec:canenumerate}}

\If{\textnormal{$\mathcal{F}$ $=$ $\gss$}}{

            virtual node insertion for ease of $\ged$ cost computation;

			}

  generate candidate sequences $\cseq_Q^{\mathcal{F}}$ for $Q$ and $\cseq_G^{\mathcal{F}}$ for $G$,  initialize  $H$: $V_Q\times V_G\to\mathbb{R}$ as a null function; \hfill\tcp*[h]{\S~\ref{subsubsec:overall}} 

\BlankLine
\tcp*[h]{Dynamic programming: initial basis}

  $u$ $\gets$ the first element in $\cseq_Q^{\mathcal{F}}$, $\mathcal{M}_1$ $\gets$ $\emptyset$\;

    \ForEach{\textnormal{element} $v \in \cseq_G^{\mathcal{F}}$}{

        $M(u)$ $\gets$ $v$; \hfill\tcp*[h]{enumeration at depth $1$} 

        $\mathcal{M}_1$ $\gets$ $\mathcal{M}_1\cup\{M\}$, $\mathsf{dp}[M]$ $\gets$ $\mathsf{Cost}_\mathcal{F}(1, v, -,  H)$\; 
    }

\BlankLine
\tcp*[h]{Dynamic programming: iterative expansion}

\For{$i \gets 2$ \textnormal{to} $d$}{

    $u$ $\gets$ the $i^\textnormal{th}$ element in $\cseq_Q^{\mathcal{F}}$, $\mathcal{M}_i$ $\gets$ $\emptyset$\;
        \ForEach{\textnormal{candidate mapping $M$ $\in$ $\mathcal{M}_{i-1}$}}{
          \ForEach{\textnormal{element $v \in \cseq_G^{\mathcal{F}}$}}{

           $M'$ $\gets$ $M$, $M'(u)$ $\gets$ $v$\;

           $\mathcal{M}_i$ $\gets$ $\mathcal{M}_i\cup\{M'\}$\;

           $\mathsf{dp}[M'] \gets \mathsf{dp}[M] + \mathsf{Cost}_\mathcal{F}(i, v, M, H)$\;         
          }
        }

}

\BlankLine
\tcp*[h]{Dynamic programming: final extraction}

 \ForEach{\textnormal{candidate mapping $M$ $\in$ $\mathcal{M}_d$}}{
    $r$ $\gets$ $r$ $\times$ \oblimax{$0$, $\mathsf{dp}[M]-\tau$}\;
 }

 \Return $r$\;

	\BlankLine
  \nonl \textbf{{Operation}~}{\oblimax{$a$, $b$}:}
	
      mask $\gets$ the first bit of $a-b$,   \Return mask $\times$ $b$ + ($1-$mask$)$ $\times a$\;

	\BlankLine
  \nonl \textbf{{Operation}~}{\oblieq{$a$, $b$}:} \hfill\tcp*[h]{$a\neq b$: 1; $a=b$: 0}

      $c$ $\gets$   \oblimax{$a-b$, $b-a$},
      \Return the first bit of $0-c$\;

\end{algorithm}

Taking a graph pattern query $(Q, \mathcal{F}, \tau)$ and a graph database $\mathcal{D}$ as inputs,  Algorithm~\ref{alg:base} outputs the answer set $\mathcal{R}(Q, \mathcal{F}, \tau, \mathcal{D})$ for querying $(Q, \mathcal{F}, \tau)$ on $\mathcal{D}$. 
Specifically, Line 1 first initializes an empty answer set $\mathcal{R}$. For each data graph $G$ in $\mathcal{D}$, Line 3 invokes the {\tt\small CanEnum} procedure, which enumerates all the $\cm$s for the cost computation, and then returns an answer value $r$ indicating whether $G$ matches $Q$ via $\mathcal{F}$. $r$ is inserted into $\mathcal{R}$ as the answer for $G$ (Line 4). After all the data graphs in $\mathcal{D}$ have been evaluated, Line 5 returns $\mathcal{R}$ as the output answer set.
Regarding the dynamic programming-based {\tt\small CanEnum} procedure:
\begin{itemize} [wide, labelwidth=!, labelindent=0pt]
\item[1.]\textbf{Initialization.} 
Line 6 first initializes two parameters, including the answer value $r$ and the enumeration depth $d$ for semantic $\mathcal{F}$. According to the analyses in \S~\ref{subsubsec:canenumerate}, $d$ equals $|V_Q|$,  $\max\big\{|V_Q$, $|V_G|\big\}$, and {$|V_Q|\cdot |V_G|$} for $\iso$, $\gss$, and $\gsim$, respectively. Then, Line~9 initializes two candidate sequences $\cseq_Q^{\mathcal{F}}$ for $Q$ and $\cseq_G^{\mathcal{F}}$ for $G$, which will be used to iteratively generate the \four{\emph{BFS}} $\cm$s at each depth. 
As illustrated in \S~\ref{subsubsec:canenumerate}, the enumeration spaces for $\iso$ and $\gss$ queries are exponential, while for $\gsim$ queries are polynomial.
Hence, it is crucial to design specific initializations of $\cseq_Q^{\mathcal{F}}$ and  $\cseq_G^{\mathcal{F}}$ for different query semantics, as follows:
\begin{itemize}
\item $\iso$ and $\gss$.  $\cseq_Q^{\mathcal{F}}$ (\resp $\cseq_G^{\mathcal{F}}$) is initialized to \emph{any} sequence of \three{\emph{all}} the nodes in $V_Q$ (\resp $V_G$); and
\vspace{-0.5ex}
\item $\gsim$. $\cseq_Q^{\mathcal{F}}$  is initialized to \emph{any} sequence of \three{\emph{all}} the elements in $\big\{\langle u, v\rangle\big\}$, $u\in V_Q, v\in V_G$, and $\cseq_G^{\mathcal{F}}$ is initialized to a sequence of any single element.
\end{itemize}
In particular, for ease of the cost computation for $\gss$ and $\gsim$ queries, Lines~7-8 conduct virtual node insertion and Line~9  initializes a function $H$: $V_Q\times V_G\to\mathbb{R}$ to indicate whether the elements in a $\cm$ have been removed, respectively.

Note that the candidate sequence $\cseq_Q^{\mathcal{F}}$ $\big($\resp $\cseq_G^{\mathcal{F}}$$\big)$ is initialized to contain \emph{all} the nodes of $Q$ (\resp $G$), so $\base$'s $\cm$ enumeration is independent of the label sets and the labeling functions. The label information is used only inside the $\FO$ cost computation for each enumerated $\cm$ (\S~\ref{subsec:editcost}). Likewise,~the numbers of edges of $Q$ and $G$ are protected, as $\base$ accesses the edge functions only through $\FO$ arithmetic operations for $\cm$~cost computation inside the TEE.

\item[2.]\textbf{$\cm$ enumeration.} After the initialization, Lines 10-22 %
enumerate all the $\cm$s by iteratively constructing $\cm$ at each depth without label-based constraints, which enumerates redundant $\cm$s to achieve full obliviousness. 

First, Lines 10-12 generate the $\cm$s at depth $1$ by mapping the first element in $\cseq_Q^{\mathcal{F}}$ to each element in $\cseq_G^{\mathcal{F}}$. These $\cm$s are inserted into the $\cm$ set $\mathcal{M}_1$ for depth $1$ (Line~13). For each partial mapping $M$ at depth $1$ (\ie the mapping in Line~12), Line 13 invokes the {\tt\small Cost} procedure to compute the cost for $M$'s mapping at depth $1$ (\ie the mapping in Line~12)
and stores this cost in the dynamic programming table $\mathsf{dp}[\cdot]$ as $M$'s cost.

Then, Lines 14-18 iteratively expand the $\cm$s at depth $i$ from $i$ $=$ $2$ to $i$ $=$ $d$ by mapping the $i^\textnormal{th}$ element in $\cseq_Q^{\mathcal{F}}$ to each element~in $\cseq_G^{\mathcal{F}}$ 
based on each partial mapping $M$ in $\mathcal{M}_{i-1}$. These constructed $\cm$s are inserted into the $\cm$ set $\mathcal{M}_i$ for depth $i$ (Line~19). 
For each mapping $M'$ in $\mathcal{M}_i$ enumerated by expanding $M$, Line 20 invokes the {\tt\small Cost} procedure to compute the cost for $M'$'s mapping at depth $i$ (\ie the mapping in Line~18), and then
computes the sum of this cost and $\mathsf{dp}[M]$ as the cost of $M'$. $M'$'s cost is also stored in~$\mathsf{dp}[\cdot]$.

Finally, Lines 21-22 compute the answer value $r$ for $G$ by aggregating the evaluation result for each $\cm$ at depth $d$ via multiplication. Specifically, 
if the cost of a $\cm$ $M$ in $\mathcal{M}_d$ is not larger than the similarity parameter $\tau$, $r$ is updated as $0$, indicating that $G$ matches $Q$ through $M$; otherwise, $r$ is multiplied by a positive value, which has no effect on the query result for mapping $G$ to $Q$. To achieve such an update of $r$, the {\tt\small Obli\_Max} {operation} (Line 24) is invoked to obtain the larger one between $0$ and $\mathsf{dp}[M]-\tau$.
After the aggregation, Line 23 returns $r$ as the output for $G$. 
\end{itemize}

\begin{example}
\label{example:base}
Consider the query graph $Q$ and data graph $G_3$  in Figure~\ref{fig:definition}. To map $G_3$ to $Q$ via $\mathcal{F}$, the enumeration of $\base$ is as follows:
\begin{itemize} [wide, labelwidth=!, labelindent=0pt]
\item \textbf{$\mathcal{F}$ $=$ $\iso$.}  The enumeration depth is $d$ $=$ $|V_Q|$ $=$ $3$, and the candidate sequences are $\cseq_Q^{\mathcal{F}}$ $=$ $[u_1, u_2, u_3]$ and $\cseq_G^{\mathcal{F}}$ $=$ $[v_1, v_2, v_3, v_4]$. 
After $\base$'s $\cm$ enumeration, there are $4^3$ $=$ $64$ $\cm$s in the $\cm$ set $\mathcal{M}_3$.
\three{However, as shown in Figure~\ref{fig:enumeration-iso}(b), there are actually $|\cseq_G^{\mathcal{F}}|!/\big(|\cseq_G^{\mathcal{F}}|-d\big)!$ $=$ $4!/1!$ $=$ $24$ $\cm$s in $\mathcal{M}_3$ due to the injectivity constraint.\footnote{\three{Such a constraint can be achieved with minor modifications on Algorithm~\ref{alg:base} with \emph{no} effect on the correctness of the proofs for various obliviousness.}} As graphs $G_1$ and $G_4$ from Figure~\ref{fig:definition} have the same number of nodes as $G_3$, $\base$ will enumerate the same $\cm$ set $\mathcal{M}_3$ for mapping $G_1$ or $G_4$ to $Q$ via $\iso$.}

\item \textbf{$\mathcal{F}$ $=$ $\gss$.} The enumeration depth is $d$ $=$ $\max\big\{|V_Q|, |V_G|\big\}$ $=$ $4$, and the candidate sequences are $\cseq_Q^{\mathcal{F}}$ $=$  $[u_1, u_2, u_3, u_4]$ ($u_4$ is the virtual node inserted into $V_Q$) and $\cseq_G^{\mathcal{F}}$ $=$ $[v_1, v_2, v_3, v_4]$. 
\three{With the injectivity constraint, $\base$ enumerates $|\cseq_G^{\mathcal{F}}|!/\big(|\cseq_G^{\mathcal{F}}|-d\big)!$ $=$ $24$ $\cm$s in $\mathcal{M}_4$.}

\item \textbf{$\mathcal{F}$ $=$ $\gsim$.} The enumeration depth is {$d$ $=$ $|V_Q|\cdot |V_G|$ $=$ $3\cdot 4$ $=$ $12$}, and the candidate sequences are $\cseq_Q^{\mathcal{F}}$ $=$  $\big[\langle u_1, v_1\rangle$, $\langle u_1, v_2\rangle$, $\langle u_1, v_3\rangle$, $\langle u_1, v_4\rangle$, $\langle u_2, v_1\rangle$, $\langle u_2, v_2\rangle$, $\langle u_2, v_3\rangle$, $\langle u_2, v_4\rangle$, $\langle u_3, v_1\rangle$, $\langle u_3, v_2\rangle$, $\langle u_3, v_3\rangle$, $\langle u_3, v_4\rangle\big]$ and $\cseq_G^{\mathcal{F}}$ $=$ $\big[\langle u_1, v_1\rangle\big]$. During the $\cm$ enumeration, there exists only one $\cm$ in each set $\mathcal{M}_{i}$, $1\le i \le {12}$.
\end{itemize}
\end{example}

\noindent\textbf{Analysis.}
In $\base$, the {\tt\small Obli\_NonEqual} {operation} (Line 25 in Algorithm~\ref{alg:base}) is to compute the non-equality of two values, which will be used later in \S~\ref{subsec:editcost} for the cost computations. 
{Note that both of these operations are composed solely of oblivious arithmetic operations and are therefore $\FO$.}
\three{With simulations on Algorithm~\ref{alg:base} for any two inputs of the same size,}
we have the following lemma to show that the $\cm$ enumeration in $\base$ satisfies full obliviousness. 

\begin{lemma}
\label{lemma:base}
The $\base$ search is $\FO$ if 
$\mathsf{Cost}_\mathcal{F}(\cdot)$ is $\FO$. 
\end{lemma}

Regarding $\base$'s space complexity, there are in total $O\big(|V_G|^{|V_Q|}\big)$, $O(d^d)$, and {$|V_Q|\cdot |V_G|$} $\cm$s enumerated  for $\iso$, $\gss$, and $\gsim$ queries, respectively, where $d$ $=$ $\max\big\{|V_Q|, |V_G|\big\}$.

\stab
\noindent\underline{\textbf{Remark.}} 
With Lemma~\ref{lemma:base}, the full obliviousness of $\base$ can be guaranteed by designing $\FO$ $\mathsf{Cost}_\mathcal{F}(\cdot)$. For better understanding, $\base$'s guarantees of full obliviousness and its correctness of outputting exact query answers will be presented in \S~\ref{subsec:editcost} along with the specific designs of $\mathsf{Cost}_\mathcal{F}(\cdot)$ for different query semantics~$\mathcal{F}$.

\subsection{Cost Computation of $\base$}
\label{subsec:editcost}
In this subsection, we present the designs of $\cm$ cost computation $\mathsf{Cost}_\mathcal{F}(\cdot)$ (Lines 13 and 20 in Algorithm~\ref{alg:base}), which enables $\base$ to output exact query answers for different query semantics~$\mathcal{F}$ and to satisfy full obliviousness. Given a query graph $Q$, a data graph $G$, and a $\cm$ $M$ for mapping $G$ to $Q$,  $\mathsf{Cost}_\mathcal{F}(\cdot)$ is defined as follows.

\subsubsection{$\iso$ semantics}
\label{subsubsec:iso}
Assume $G$ does not match $Q$ through $M$. Then, there must exist at least one condition in Definition~\ref{def:iso} that cannot be satisfied:
\begin{itemize}
\item \textbf{Case 1.} there exists node $u$ of $Q$ such that $u$ and $M(u)$ have different labels; and 
\item \textbf{Case 2.} there exist two nodes $u, v$ of $Q$ such that $E_Q(u, v) = 1$ and $E_G\big(M(u), M(v)\big) = 0$.
\end{itemize}

By considering the cost of $M$ for $\iso$ as the number of \emph{distinct} existences of the above cases, we can easily yield Lemma~\ref{lemma:iso}.

\begin{lemma}
\label{lemma:iso}
$G$ matches $Q$ via $\iso$ through $M$ if and only if the cost of $M$ for $\iso$ is $0$.
\end{lemma}

\begin{algorithm}[tb]
	\caption{{$\mathsf{Cost}_\mathcal{F}(\cdot)$ for $\iso$}} 
	\label{alg:base-iso}
	\small
	
	\SetKwFunction{OGHA}{AMatch}
	\SetKwFunction{query}{QueryMatch}
	\SetKwFunction{editcost}{EditCost}
  \SetKwFunction{funccost}{Cost}
  \SetKwFunction{oblimax}{Obli\_Max}
  \SetKwFunction{oblieq}{Obli\_NonEqual}
	\SetKwProg{int}{int}{\string:}{}
	\SetKwProg{void}{void}{\string:}{}
	\SetKwInOut{Input}{Input}
	\SetKwInOut{Output}{Output}
	\SetKwData{Result}{result}
	\SetKwData{Flag}{CandidateSet}
	\SetKwData{True}{true}
	\SetKwData{False}{false}
	\Indentp{-1em}
	\Input{The query graph $Q$,
  the data graph $G$,
  the candidate sequence $\cseq_Q^{\mathcal{F}}$,  current depth $i$, node $v$ of $G$, and a $\cm$ $M$ at depth $i-1$}
	\Output{The cost  for mapping  $v$ to the $i^{\textnormal{th}}$ node in $\cseq_Q^{\mathcal{F}}$  \wrt $M$}
	\Indentp{1em}

	$u$ $\gets$ the $i^{\textnormal{th}}$ element in $S_{Q}$, $S$ $\gets$ the set of $\cseq_Q^{\mathcal{F}}$'s first $i-1$ elements\; %

      $c_1$ $\gets$ \oblieq{$\ell_Q(u)$, $\ell_G(v)$}\;
      
      $c_2^\textnormal{out}$ $\gets$ $0$, $c_2^\textnormal{in}$ $\gets$ $0$\;

    \ForEach{\textnormal{node} $u' \in S$}{

    $c_2^\textnormal{out}$ $\gets$ $c_2^\textnormal{out}$ $+$ $E_Q(u, u')\times \Big(1-E_G\big(v, M(u')\big)\Big)$\;

    $c_2^\textnormal{in}$ $\gets$ $c_2^\textnormal{in}$ $+$ $E_Q(u', u)\times \Big(1-E_G\big(M(u'), v\big)\Big)$\;

    }

        \Return $c_1$ $+$ $c_2^\textnormal{out}$ $+$ $c_2^\textnormal{in}$\;

\end{algorithm}

Based on Lemma~\ref{lemma:iso}, the design of $\mathsf{Cost}_\mathcal{F}(\cdot)$ for $\iso$  is shown in Algorithm~\ref{alg:base-iso}. Taking as inputs the query graph $Q$, the data graph $G$, the candidate sequence $\cseq_Q^{\mathcal{F}}$ for $Q$, the current depth $i$, node $v$ of $G$,  and the candidate mapping $M$ at depth $i-1$, Algorithm~\ref{alg:base-iso} outputs 
the cost for mapping  $v$ to the $i^{\textnormal{th}}$ node in $\cseq_Q^{\mathcal{F}}$ \wrt $M$ (\ie the mapping at depth $i$ as shown in Line 18 of Algorithm~\ref{alg:base}). 
Specifically,
Line 1 first initializes (i) node $u$ of $Q$ as the node to be mapped to $v$,
and (ii) set $S$ as the set of the first $i-1$ nodes in $\cseq_Q^{\mathcal{F}}$. For mapping $v$ to $u$, Line 2 computes the number $c_1$ of the existence of Case 1
by invoking the {\tt\small Obli\_NonEqual} {operation} to check whether $u$ and $v$ have the same label; 
Line 5 (\resp Line 6) evaluates a polynomial expression to compute
the total number $c_2^\textnormal{out}$ (\resp $c_2^\textnormal{in}$) of the existence of Case 2 for $u$'s outgoing (\resp incoming) edges to (\resp from) nodes in $S$. 
Finally, Line 7 returns the sum of $c_1$, $c_2^\textnormal{out}$, and $c_2^\textnormal{in}$ as the output. Note that, as $\tau$ equals $0$ for $\iso$ queries, Line 23 of Algorithm~\ref{alg:base} returns $r$ $=$ $0$ if $G$ matches $Q$ via $\iso$; otherwise, a positive integer is returned.

\begin{example}
\label{example:base-iso}
\three{Following Example~\ref{example:base}, we consider mapping data graph $G_3$ to query graph $Q$ via $\iso$ as shown in Figure~\ref{fig:enumeration-iso}(b).}
Assume the inputs: (i) the current depth is $3$, (ii) node $v_4$ in $G_3$ to be mapped to the $3^{\textnormal{rd}}$ node in $\cseq_Q^{\mathcal{F}}$ (\ie $u_3$), and (iii) a $\cm$ $M$ at depth $2$ such that $M(u_1)$ $=$ $v_1$, $M(u_2)$ $=$ $v_2$. 
\three{According to the equations in Algorithm~\ref{alg:base-iso}, we have
$c_1$ $=$ $1$, $c_2^\textnormal{out}$ $=$ $0$, and $c_2^\textnormal{in}$ $=$~$1$. Hence, the cost for mapping $v_4$ to $u_3$ \wrt $M$ equals $c_1$ $+$ $c_2^\textnormal{out}$ $+$ $c_2^\textnormal{in}$ $=$ $2$.} 
Then, for the $\cm$ $M'$ at depth $3$ with $M'(u_1)$ $=$ $v_1$, $M'(u_2)$ $=$ $v_2$,
and $M(u_3)$ $=$ $v_4$, 
the cost of $M'$ for $\iso$ equals $\mathsf{dp}[M]$ $+$ $c_1$ $+$ $c_2^\textnormal{out}$ $+$ $c_2^\textnormal{in}$ $>$ $0$. With Lemma~\ref{lemma:iso}, $G_3$ does not match $Q$ via $\iso$ through~$M'$.
\end{example}

\noindent\textbf{Correctness.} The correctness of $\base$ with $\mathsf{Cost}_\mathcal{F}(\cdot)$ 
for answering $\iso$ queries can be shown by (i) analyzing on the truth table for the polynomial computations in Lines 5-6 of Algorithm~\ref{alg:base-iso}, ensuring a correct number computation for new distinct existences of Cases $1$ and $2$ led by the mapping at depth $i$,
and (ii) conducting a simple mathematical induction with a hypothesis: for any $\cm$ $M$ in the $\cm$ set $\mathcal{M}_i$ for depth $i$ ($i\ge 1$), $\mathsf{dp}[M]$ equals the cost of $M$.

\subsubsection{$\gss$ semantics} As both the $\iso$ and $\gss$ queries involve an exponentially large $\cm$ enumeration space, the design of $\mathsf{Cost}_\mathcal{F}(\cdot)$ for $\gss$ is similar to the one for $\iso$. According to Definition~\ref{def:gss}, the cost of $M$ for $\gss$ is the graph edit distance ($\ged$) between $Q$ and $G$ \wrt $M$. Then, we can compute the cost for $\gss$ by replacing the computation of $c_1$, $c_2^\textnormal{out}$, and $c_2^\textnormal{in}$ in Algorithm~\ref{alg:base-iso} with computations to count the number of edit operations for $\ged$ defined in \S~\ref{subsec:query}. Specifically,   there are two cases of the 
edit-operation computation:

\begin{itemize}
\item \textbf{Case 1: insertion/deletion of an isolated node, and label replacement of a node.}  %
In this case, the cost increases by $1$ if a (virtual) node $u$ of $Q$ has a label different from that of $M(u)$ in $G$; otherwise, the cost remains unchanged. Hence, the computation for Case 1 is the same as that of $c_1$ in Algorithm~\ref{alg:base-iso}.

\item \textbf{Case 2: insertion/deletion of an edge.}
In this case, the cost increases by $1$ if the existence of an outgoing edge of node $u$ to node $v$ in $Q$ is different from the edge existence from $M(u)$ to $M(v)$ in $G$; otherwise, the cost remains unchanged. To compute such a cost, we update the computation of $c_2^\textnormal{out}$ as follows:
\begin{equation}
\nonumber
c_2^\textnormal{out} = c_2^\textnormal{out} + \textnormal{$E_Q(u, u') + E_G\big(v, M(u')\big) - 2 \cdot E_Q(u, u')\cdot E_G\big(v, M(u')\big)$}
\end{equation}
Similarly, the cost for the existence of $u$'s incoming edges is:
\begin{equation}
\nonumber
c_2^\textnormal{in} = c_2^\textnormal{in} + \textnormal{$E_Q(u', u) + E_G\big(M(u'), v\big) - 2 \cdot E_Q(u', u)\cdot E_G\big(M(u'), v\big)$}
\end{equation}
\end{itemize}

\three{With the above cost computation for $\gss$, $G$ matches $Q$ via $\gss$ through an $\cm$ $M$ if and only if the cost of $M$ is not larger than the similarity parameter $\tau$.}
We omit the detailed correctness analysis 
since 
{it is similar to the one for $\iso$.}

\subsubsection{$\gsim$ semantics}
\label{subsubsec:gsim}

\begin{algorithm}[tb]
	\caption{{$\mathsf{Cost}_\mathcal{F}(\cdot)$ for $\gsim$}} 
	\label{alg:base-gsim}
	\small
	
	\SetKwFunction{OGHA}{AMatch}
	\SetKwFunction{query}{QueryMatch}
	\SetKwFunction{editcost}{EditCost}
  \SetKwFunction{funccost}{Cost}
  \SetKwFunction{oblimax}{Obli\_Max}
  \SetKwFunction{oblieq}{Obli\_NonEqual}
	\SetKwProg{int}{int}{\string:}{}
	\SetKwProg{void}{void}{\string:}{}
	\SetKwInOut{Input}{Input}
	\SetKwInOut{Output}{Output}
	\SetKwData{Result}{result}
	\SetKwData{Flag}{CandidateSet}
	\SetKwData{True}{true}
	\SetKwData{False}{false}
	\Indentp{-1em}
	\Input{The query graph $Q$,
  the data graph $G$, 
  the candidate sequence $\cseq_Q^{\mathcal{F}}$,  current depth $i$, the single element in  $\cseq_G^{\mathcal{F}}$, the $\cm$ $M$ at depth $i-1$, and function $H$ at depth $i-1$}
	\Output{The cost of $M$'s mapping at depth $i$ for $\gsim$}
	\Indentp{1em}

  \tcp*[h]{Initialization}

  count $\gets$ $0$\;

  \If{$i=1$}{

    \ForEach{\textnormal{element} $e =\langle u, v\rangle \in \cseq_Q^{\mathcal{F}}$}{

      $H(e)$ $\gets$ $1-$ \oblieq{$\ell_Q(u)$, $\ell_G(v)$}\;

    }

  }

  \BlankLine

  \tcp*[h]{update of $H$ according to (ii) of Definition~\ref{def:gsim}}

  \If{$i<{|V_Q|\cdot |V_G|}$}{

  \ForEach{$e_i =\langle u_i, v_i\rangle \in \cseq_Q^{\mathcal{F}}$}{

      $c_1 \gets 1$; %

    \ForEach{\textnormal{node} $u \in V_Q-\{u_i\}$}{

      $c_2$ $\gets$ $1-E_Q(u_i, u)$\;

        \ForEach{$e_j =\langle u_j, v_j\rangle$$\in$$\cseq_Q^{\mathcal{F}}$ \textnormal{having} $u_j=u$}{

            $c_2$ $\gets$ $c_2 + E_G(v_i, v_j)\times H(e_j)$\;
        }

      $c_1$ $\gets$ $c_1 \times c_2$\;

    }

    $H(e_i)$ $\gets$ $H(e_i) \times c_1$\;

  }

    \BlankLine

      \Return $0$\;   
    }

  \BlankLine

    \tcp*[h]{final cost computation for (ii) of Definition~\ref{def:gsim}}

    \ForEach{\textnormal{node} $u \in V_Q$}{

      $c \gets 0$\;

            \ForEach{$e_i =\langle u_i, v_i\rangle \in \cseq_Q^{\mathcal{F}}$ \textnormal{such that} $u_i=u$}{

            $c$ $\gets$ $c$ $+$ $H(e_i)$\;
        }

      count $\gets$ count $+$ \oblieq{$c$, $0$}\;

    }

        \Return $|V_Q|$ $-$ count\;
	
\end{algorithm}

As introduced in \S~\ref{subsubsec:canenumerate}, 
to answer $\gsim$ queries, we need to conduct {$|V_Q|\cdot |V_G|$} traversals on the initial $\cm$ to iteratively remove its elements that violate $\gsim$'s structural restriction $\big($\ie condition (ii) in Definition~\ref{def:gsim}$\big)$. The number of traversals can ensure that all violations are removed. To achieve $\FO$ computations for such an iterative removal with $\base$'s $\cm$ enumeration, we adopt a function $H$ (Line 9 of Algorithm~\ref{alg:base}) to 
record and update the state of each element $\langle u,v\rangle$
indicating whether $\langle u,v\rangle$ has been removed. 

For the final check on condition (ii) of Definition~\ref{def:gsim} after the {$|V_Q|\cdot |V_G|$} traversals, we can simply compute the number $n$ of $Q$'s nodes having at least one match, and check whether $n$ equals $|V_Q|$.  Accordingly, $|V_Q|-n$ is the cost of the $\cm$ at depth {$|V_Q|\cdot |V_G|$}.

With the analysis above, the design of $\mathsf{Cost}_\mathcal{F}(\cdot)$ for $\gsim$ is shown in Algorithm~\ref{alg:base-gsim}.
Taking the same inputs as those for Algorithm~\ref{alg:base-iso} with an additional function $H$ at depth $i-1$, Algorithm~\ref{alg:base-gsim} outputs the cost of $M$'s mapping at depth $i$ for $\gsim$. Specifically,  Lines 2-4  initialize $H$ for the $\cm$ at depth $1$ as follows: for each element $e$ $=$ $\langle u, v\rangle$ in $\cseq_Q^{\mathcal{F}}$, if $u$ and $v$ have the same label, the {\tt\small Obli\_NonEqual} operation is invoked in Line~4 to set $H(e)$ to $1$; otherwise, $H(e)$ is set to $0$. Then, for the ${i}^\textnormal{th}$ $\big($$i$ $<$ $|V_Q|\cdot |V_G|$$\big)$ iteration of the removal traversals,
Lines 6-13 conduct polynomial computations for each element $e_i$ in $\cseq_Q^{\mathcal{F}}$ to update $H(e_i)$, and then a cost of $0$ is returned in Line~14 to avoid the accumulation of unnecessary costs in the dynamic programming table (Line 20 of Algorithm~\ref{alg:base}). Finally, for the $\cm$ at depth {$|V_Q|\cdot |V_G|$}, Lines 15-19 compute the number $n$ of  $Q$'s nodes having at least one match, and Line 20 returns $|V_Q|$ $-$ $n$ as the output. Note that, as $\tau$ equals $0$ for $\gsim$ queries, Line 23 of Algorithm~\ref{alg:base} returns $r$ $=$ $0$ if $G$ matches $Q$ via $\gsim$; otherwise, a positive integer is returned.

\begin{example}
Consider mapping data graph $G_3$ to query graph $Q$ via $\gsim$ as shown in Example~\ref{example:base}. For depth $1$, function $H$ is initialized as: $H\big(\langle u_1, v_1\rangle\big)$ $=$ $H\big(\langle u_1, v_4\rangle\big)$ $=$ $H\big(\langle u_2, v_2\rangle\big)$ $=$ $H\big(\langle u_3, v_3\rangle\big)$ $=$ $1$; for the remaining elements $e$ in $\cseq_Q^{\mathcal{F}}$, $H(e)$ $=$ $0$. Then, for 
$\langle u_1, v_1\rangle$, Lines~7-12 of Algorithm~\ref{alg:base-gsim} compute  $c_1$ $=$ $2$ such that Line~13 updates $H\big(\langle u_1, v_1\rangle\big)$ to remain positive instead of $0$.  
$H\big(\langle u_1, v_4\rangle\big)$, $H\big(\langle u_2, v_2\rangle\big)$, and $H\big(\langle u_3, v_3\rangle\big)$ are similarly updated to be positive integers, while
the other values remain $0$. 
After {$12$} iterations of such updates, Line~20 returns $0${, so the final $\mathsf{dp}[M]$ equals $0$}, indicating that $G_3$ matches $Q$ via $\gsim$. 
\end{example}

\noindent\textbf{Correctness.} Following the idea of iterative element removals, we can show that $\base$ with $\mathsf{Cost}_\mathcal{F}(\cdot)$ in Algorithm~\ref{alg:base-gsim} answers $\gsim$ queries correctly by conducting truth table-based analyses on the correctness of the following claims: (i) in Line 12 of Algorithm~\ref{alg:base-gsim}, $c_2$ $=$ $0$ is the only case that leads to the element removal, and (ii)  in Line 20 of Algorithm~\ref{alg:base-gsim}, count $=$ $|V_Q|$ is the only case that satisfies condition (ii) of Definition~\ref{def:gsim}.

\stab
\noindent\textbf{Analysis of $\base$.}
The full obliviousness of $\base$'s cost computation (Algorithms~\ref{alg:base-iso}-\ref{alg:base-gsim}) can be shown by conducting simulations in a way similar to the proof of Lemma~\ref{lemma:base}. Then, we have Lemma~\ref{lemma:cost}.

\begin{lemma}
\label{lemma:cost}
The cost computation $\mathsf{Cost}_\mathcal{F}(\cdot)$ of $\base$
is $\FO$.
\end{lemma}

Regarding the complexity of $\base$'s cost computation, there are at most $O\big(|V_Q|\big)$,  $O\big(|V_Q|\big)$, and $O\big(|V_Q|^2\cdot|V_G|^2\big)$  arithmetic operations at each depth for $\iso$, $\gss$, and $\gsim$, respectively.

With Lemmas~\ref{lemma:base} and \ref{lemma:cost}, and the correctness analyses in \S~\ref{subsubsec:iso}-\ref{subsubsec:gsim}, we yield the following theorem.

\begin{theorem}
The $\base$ {search of $\fullname$}  is $\FO$ and outputs exact query answers for the $\iso$, $\gss$, and $\gsim$ queries.
\end{theorem}

\section{Differentially Oblivious Search}
\label{sec:op}
To optimize $\base$'s query efficiency, we propose in this section to prune $\base$'s candidate mapping enumeration, while preserving access pattern privacy via differential obliviousness. We first introduce in \S~\ref{subsec:dopruning} our pruning strategies for various query semantics, together with sensitivity analyses on the corresponding access patterns for neighboring inputs. Then, in \S~\ref{subsec:doanalysis}, we present 
$\fullname$'s $\DO$ search approach $\ppdo$ 
with privacy analysis.

\subsection{
{Pruning Strategy and Sensitivity Analysis}}
\label{subsec:dopruning}

According to Definition~\ref{def:dp} in \S~\ref{subsec:access}, 
{to achieve the privacy guarantee of differential obliviousness},
we need to 
{quantify the relationship between the pruning access patterns for neighboring inputs $\big($\ie Equation~(\ref{eqa:do})$\big)$}. 
In the following, we present 
our pruning strategy for each query semantics $\mathcal{F}$, with bounded sensitivity to access patterns for neighboring inputs.

\stab
\noindent\textbf{Pruning strategy for query semantics $\mathcal{F}$.} 
Based on $\base$, we adopt {some parameters of the $\cm$ enumeration} as the \emph{pruning parameters} to control 
$\ppdo$'s pruning computations for various $\mathcal{F}$, as follows:

\begin{itemize}[wide, labelwidth=!, labelindent=0pt]
\item \textbf{$\mathcal{F}$ $=$ $\iso$.} Consider a $\cm$ $M_i$ enumerated by $\base$ at depth $i$ for mapping a data graph $G$ to a query graph $Q$.
{$M_i$'s cost is $\mathsf{dp}[M_i]$.}
According to Lemma~\ref{lemma:iso}, if $\mathsf{dp}[M_i]$ $>$ $0$, $G$ cannot match $Q$ through $M_i$ or any $\cm$s $M_j$ ($j>i$) enumerated  by expanding $M_i$. 
However, if we adopt such a condition to prune the candidates enumerated in $\base$, $\base$'s access patterns for $Q$ and $Q$'s neighboring query $Q'$ may either be the same or differ significantly, making the sensitivity either $0$ or unbounded. As the access pattern of a $\cm$ enumeration with pruning is too complicated to be quantified precisely by a single parameter, we propose to design a pruning condition such that the corresponding access patterns for neighboring queries are \emph{always the same}, leading to a sensitivity of $0$.

\begin{figure}[t]
  \setlength{\abovecaptionskip}{0pt}
  \setlength{\belowcaptionskip}{0pt}
  \centering
  \graphicspath{{figures/}}
     \includegraphics[width=0.48\textwidth]{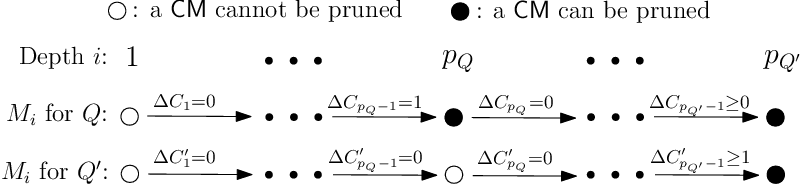}
    \vspace{-1.5ex}
  \caption{An illustration of sensitivity analysis for $\iso$, where (i) $Q$ and $Q'$ are neighboring input query graphs; (ii) $M_i$ is a $\cm$ at depth $i$ for $\iso$ enumerated by $\base$;
  (iii) $\Delta C_i$ and $\Delta C'_i$  are the costs of the mappings at depth $i$ (\ie $\mathsf{Cost}_\mathcal{F}$ computed in Line 20 of Algorithm~\ref{alg:base}) for $Q$ and $Q'$, respectively; and (iv) $p_Q$ and $p_{Q'}$ are the smallest depths to prune $M_i$ for $Q$ and $Q'$, respectively}
  \label {fig:iso-illustration}
  \end{figure}

We analyze the case that the access patterns of pruned $\cm$ enumeration for $Q$ and $Q'$ differ from each other. 
Let $p_Q$ (\resp  $p_{Q'}$) denote the smallest depth that $M_i$ can be pruned for (mapping $G$ to) $Q$ (\resp $Q'$), and let $\Delta C_i$ (\resp $\Delta C'_i$) denote the cost of the mapping at depth $i$ 
for $Q$ (\resp $Q'$). W.l.o.g.,  we assume that $p_Q < p_{Q'}$ as illustrated in Figure~\ref{fig:iso-illustration}. 
Then, we have the following deductions:
\begin{itemize}[leftmargin=0ex,wide, labelwidth=!, labelindent=0pt]
\item[(i)] $\Delta C_{i}$ $=$ $\Delta C'_{i}$ $=$ $0$, $1$ $\le$ $i$ $\le$ $p_Q-2$. Otherwise, $M_i$ for $Q$ and $Q'$ can be pruned at a depth smaller than $p_Q$;
\item[(ii)] $\mathsf{dp}[M_{p_Q}]$ $=$ $\sum_{1\le i \le p_Q-1}{\Delta C_i}$ $=$ $\Delta C_{p_Q-1}$ $=$ $1$. This can be proved by contradiction: assume that $\Delta C_{p_Q-1}$ $\ge$ $2$. As $\dis{Q}{Q'}$ $=$ $1$ (Definition~\ref{def:dp} in \S~\ref{subsec:access}), according to $\mathsf{Cost}_\mathcal{F}(\cdot)$ for $\iso$ (Lines 2-7 of Algorithm~\ref{alg:base-iso}),  we have $\Delta C'_{p_Q-1}$ $\ge$ $\Delta C_{p_Q-1}$ $-$ $1$ $\ge$ $1$. This indicates that $M_i$ for $Q'$ can also be pruned at depth $p_Q$, which contradicts the assumption that $p_{Q'}$ $>$ $p_Q$;
\item[(iii)] $\Delta C_{i}$ $=$ $\Delta C'_{i}$ $=$ $0$, $p_Q$ $\le$ $i$ $\le$ $p_{Q'}-2$. This can also be proved by contradiction in similar ways as those for deductions (i) and (ii).
\end{itemize}

With the above deductions, we propose to prune $M_i$ by using the pruning condition of "if $\mathsf{dp}[M_i]$ $>$ $\tau$ $+$ $1$ $=$ $1$ and $\Delta C_{i-1}$ $\ge$ $2$", which guarantees that $M_i$ for $Q$ and $Q'$ must be pruned at the same depth not smaller than $p_{Q'}$. Compared with the pruning condition of "if $\mathsf{dp}[M_i]$ $>$ $0$", such a pruning condition (i) relaxes the pruning power to achieve the same access pattern for neighboring input queries, leading to a sensitivity of $0$, and (ii) has no effect on the correctness of query answers.  We remark that this pruning condition can also be applied to achieve the same access pattern for neighboring input databases with no effect on correctness.

\item \textbf{$\mathcal{F}$ $=$ $\gss$.}  Based on the analysis for $\iso$, we can 
similarly set the pruning condition to be "if $\mathsf{dp}[M_i]$ $>$ $\tau$ $+$ $1$ and $\Delta C_{i-1}$ $\ge$ $2$", achieving a sensitivity of $0$ for the pruned $\cm$ enumeration. %

\item \textbf{$\mathcal{F}$ $=$ $\gsim$.}  Consider mapping a data graph $G$ to a query graph~$Q$.  As introduced in \S~\ref{subsec:canenumerate}, $\base$'s $\cm$ enumeration depends on the worst-case depth $\mathsf{depth}_{\mathsf{\gsim}}$ $=$ $|V_Q|\cdot|V_G|$, which ensures 
no more element removals after $\mathsf{depth}_{\mathsf{\gsim}}$ traversals.
Accordingly, the access pattern of $\base$'s $\cm$ enumeration for $\gsim$ is uniquely determined by $\mathsf{depth}_{\mathsf{\gsim}}$. To prune the $\cm$ enumeration, we propose to replace $\mathsf{depth}_{\mathsf{\gsim}}$ by a smaller value with bounded sensitivity for neighboring inputs.

Given a node $u$ of $Q$, let $N_G(u)$ denote the set of $G$'s nodes of  having label $\ell_Q(u)$. Then, in Algorithm~\ref{alg:base-gsim}, after %
$\mathsf{Cost}_\mathcal{F}(\cdot)$'s initialization for $\gsim$
(Lines~2-4), there are $\sum_{u\in V_Q}{\big|N_G(u)\big|}$ elements $e$ in the candidate sequence $\cseq_Q^{\mathcal{F}}$ such that $H(e)$ $=$ $1$ (Line~4). This indicates that $\sum_{u\in V_Q}{\big|N_G(u)\big|}$ traversals are enough to compute the query answer since each traversal should remove at least one element if any removal exists. 
Based on this observation, we propose to 
set $\mathsf{depth}_{\mathsf{\gsim}}$ to $d^*$ $=$ $|V_Q|\times \max_{u\in V_Q} \big\{|N_G(u)|\big\}$ for pruning the $\cm$ enumeration. This is because, as $d^*$ $\ge$ $\sum_{u\in V_Q}{\big|N_G(u)\big|}$, such a pruning may lead to slightly more traversals with no effect on the correctness of query answers. However, the sensitivity of $d^*$ for neighboring inputs can be bounded as shown by Lemma~\ref{lemma:depth3}. 
\begin{lemma}
\label{lemma:depth3}
The sensitivity of $d^*$ is bounded by $\max{\big\{|V_Q|, |V_G|\big\}}$.
\end{lemma}
\end{itemize}

With the above pruning strategies and sensitivity analyses, we are now ready to present 
$\fullname$'s $\DO$ search approach $\ppdo$.

\begin{algorithm}[tb]
	\caption{{$\DO$ search approach $\ppdo$ of $\fullname$}} 
	\label{alg:dos}
	\small
	\SetKwFunction{canenum}{CanEnum}
  \SetKwFunction{funccost}{EditCost}
  \SetKwFunction{oblimax}{Obli\_Max}
    \SetKwFunction{oblieq}{Obli\_NonEqual}
	\SetKwProg{int}{int}{\string:}{}
	\SetKwProg{void}{void}{\string:}{}
	\SetKwInOut{Input}{Input}
	\SetKwInOut{Output}{Output}
	\SetKwData{Result}{result}
	\SetKwData{True}{true}
	\SetKwData{False}{false}
	\Indentp{-1em}
\Input{The graph pattern query $(Q, \mathcal{F}, \tau)$, the graph database $\mathcal{D}$, \textcolor{blue}{privacy budget $\epsilon$, and maximum error probability $\lambda$}}
	\Output{The answer set $\mathcal{R}$}
	\Indentp{1em}

  	\tcp*[h]{$\base$'s main procedure}

  \nonl \textbf{Procedure~}{\canenum{$G$}:}
  
    \tcp*[h]{Initialization of enumeration}

\tcp*[h]{Dynamic programming: initial basis}

\tcp*[h]{Dynamic programming: iterative expansion}

\For{$i \gets 2$ \textnormal{to} $d$}{

    $u$ $\gets$ the $i^\textnormal{th}$ element in $\cseq_Q^{\mathcal{F}}$, $\mathcal{M}_i$ $\gets$ $\emptyset$\;
        \ForEach{\textnormal{candidate mapping $M$ $\in$ $\mathcal{M}_{i-1}$}}{
          \ForEach{\textnormal{element $v \in \cseq_G^{\mathcal{F}}$}}{

          $c$ $\gets$ $\mathsf{Cost}_\mathcal{F}(i, v, M, H, \textcolor{blue}{d})$; \hfill\tcp*[h]{\textcolor{blue}{for $\gsim$}}

            $M'$ $\gets$ $M$, $M'(u)$ $\gets$ $v$\;
          
            $\mathsf{dp}[M'] \gets \mathsf{dp}[M] + c$\;

            \If{\textcolor{blue}{$\mathsf{dp}[M']$ $>$ $\tau+1$ \textnormal{and} $c$ $\ge$ $\textnormal{cost}_{\mathcal{F}}$}}{

              \textbf{{continue}}; \hfill\tcp*[h]{\textcolor{blue}{$\DO$ pruning}}
          }
          
          $\mathcal{M}_i$ $\gets$ $\mathcal{M}_i\cup\{M'\}$\;
          }           
          }

}

\BlankLine

\tcp*[h]{Dynamic programming: final extraction}

\end{algorithm}

\subsection{$\ppdo$'s {Search Process} and Privacy Analysis}
\label{subsec:doanalysis}

$\ppdo$'s overall search process  is shown in Algorithm~\ref{alg:dos}. Taking the same inputs as those of $\base$ (Algorithm~\ref{alg:base}) together with the privacy budget $\epsilon$ and maximum error probability $\lambda$, $\ppdo$ returns the same output as that of $\base$. Different from $\base$, $\ppdo$ 
further prunes the dynamic programming-based $\cm$ enumeration via the pruning condition in {Line 8}, where $\textnormal{cost}_\mathcal{F}$ is  the cost parameter of the pruning for query semantics $\mathcal{F}$. Specifically,
\begin{itemize}[leftmargin=0ex,wide, labelwidth=!, labelindent=0pt]
\item $\mathcal{F}$ $=$ $\iso$ or $\gss$. Based on the analysis in \S~\ref{subsec:dopruning}, we can simply set $\textnormal{cost}_\mathcal{F}$ in Algorithm~\ref{alg:dos} to $2$,  \four{so that the pruning in Lines~8-9 of Algorithm~\ref{alg:dos} achieves the \emph{same} access pattern for neighboring inputs. According to Equation~\ref{eqa:do}, such a pruning of $\ppdo$ 
achieves a privacy guarantee of \emph{($0$,~$\delta$)-differential obliviousness}.}  %

\item $\mathcal{F}$ $=$ $\gsim$.
With Lemma~\ref{lemma:depth3}, $\ppdo$ prunes the $\cm$ enumeration for $\gsim$ by (i) setting $\textnormal{cost}_{\mathcal{F}}$ in Algorithm~\ref{alg:dos} to $-1$; and (ii) updating the enumeration depth $d$ to a smaller one via the optimized cost computation for $\gsim$ as shown in Algorithm~\ref{alg:do-gsim}.
Specifically, in Algorithm~\ref{alg:do-gsim}, Lines~4-7 compute $|N_G(u)|$ for each node $u$ of $Q$, which enables Line 8 to update m\_size as $ \max_{u\in V_Q} \big\{|N_G(u)|\big\}$. Finally, \four{Line~9 updates $d$ to be the sum of $d^*$ and a non-negative noise, making $\ppdo$ satisfy ($\epsilon$, $\delta$)-differential obliviousness. Note that the non-negative noise \emph{does not affect the correctness} of query answers due to redundant traversals.}
\end{itemize}

\stab
\noindent\underline{\textbf{Remark.}} With simulations 
for neighboring inputs,
the computational operators in Algorithm~\ref{alg:do-gsim} are $\FO$, and hence,
$\ppdo$'s access pattern for $\gsim$ (Algorithms~\ref{alg:dos}-\ref{alg:do-gsim}) is uniquely determined by $d$ updated in Line~9 of Algorithm~\ref{alg:do-gsim}.
{Regarding the complexity, $\ppdo$ has the same worst-case complexities as $\base$.} 

In the following, we present how to insert a {proper} non-negative noise
to achieve ($\epsilon$, $\delta$)-differential obliviousness for $\gsim$.

\begin{algorithm}[tb]
	\caption{{$\DO$ $\mathsf{Cost}_\mathcal{F}(\cdot)$ for $\gsim$}} 
	\label{alg:do-gsim}
	\small
	
	\SetKwFunction{OGHA}{AMatch}
	\SetKwFunction{query}{QueryMatch}
	\SetKwFunction{editcost}{EditCost}
  \SetKwFunction{funccost}{Cost}
  \SetKwFunction{oblimax}{Obli\_Max}
  \SetKwFunction{oblieq}{Obli\_NonEqual}
	\SetKwProg{int}{int}{\string:}{}
	\SetKwProg{void}{void}{\string:}{}
	\SetKwInOut{Input}{Input}
	\SetKwInOut{Output}{Output}
	\SetKwData{Result}{result}
	\SetKwData{Flag}{CandidateSet}
	\SetKwData{True}{true}
	\SetKwData{False}{false}
	\Indentp{-1em}
  \Input{The query graph $Q$,
  the data graph $G$, 
  the candidate sequence $\cseq_Q^{\mathcal{F}}$,  current depth $i$, the single element in  $\cseq_G^{\mathcal{F}}$, the $\cm$ $M$ at depth $i-1$, function $H$ at depth $i-1$, and \textcolor{blue}{depth $d$}}
	\Output{The cost value $c$ for $M$ at depth $i$}
	\Indentp{1em}

 \tcp*[h]{Initialization}

   count $\gets$ $0$,  \textcolor{blue}{m\_size $\gets$ $0$}\;

  \If{$i=1$}{

    \ForEach{\textcolor{blue}{\textnormal{node $u$ $\in$ $V_Q$}}}{
      
    \textcolor{blue}{t\_size $\gets$ $0$}\;

    \ForEach{\textnormal{element} $e =\langle \textcolor{blue}{u}, v\rangle \in \cseq_Q^{\mathcal{F}}$}{

      $H(e)$ $\gets$ $1-$ \oblieq{$\ell_Q(u)$, $\ell_G(v)$}\;

      \textcolor{blue}{t\_size $\gets$ t\_size $+$ $H(e)$}\;

    }

    \textcolor{blue}{m\_size $\gets$ Obli\_Max(m\_size, t\_size)}\;

    }
    \textcolor{blue}{$d \gets$ $|V_Q|\times \textnormal{m\_size}+$ $\lceil\xi\rceil$}; \hfill\tcp*[h]{\textcolor{blue}{noise injection in \S~\ref{subsec:doanalysis}}}

  }

  \BlankLine
  
  \tcp*[h]{update of $H$ according to (ii) of Definition~\ref{def:gsim}}

    \If{$i<\textcolor{blue}{d}$}{
      $\dots$
    }

    \BlankLine

          \tcp*[h]{final cost computation for (ii) of Definition~\ref{def:gsim}}

\end{algorithm}

\stab
\noindent\textbf{Noise injection.} 
We adopt the \emph{Laplace} mechanism to inject noise on the quantification $d^*$ of $\ppdo$'s access pattern.
\begin{definition}[\textbf{Laplace mechanism $\mathcal{F}(\mathcal{X}, f,\epsilon)$}]
Given a function $f$: $\mathcal{X}\to\mathbb{R}_{\ge 0}$, $\mathcal{X} \in\{Q,\mathcal{D}\}$,
we add a non-negative noise $\xi$ drawn from the Laplace distribution $Lap(\frac{\Delta f}{\epsilon})$ to the output $f(\mathcal{X})$, where $\Delta f$ is the sensitivity of $f$ and $\epsilon$ is the privacy budget. Formally, the Laplace mechanism outputs:
\begin{equation}
\nonumber
\mathcal{F}(\mathcal{X}, f,\epsilon) = f(\mathcal{X}) + \xi,~~\xi\sim Lap(\frac{\Delta f}{\epsilon}).
\end{equation}
\end{definition}

With the Laplace mechanism, we set the non-negative noise $\xi$ 
inserted into $d^*$ 
in Line~9 of Algorithm~\ref{alg:do-gsim} via the following equation:
\begin{equation}
\nonumber
\xi = Lap(\frac{\Delta f}{\epsilon}) - \frac{\Delta f}{\epsilon}\cdot \ln(2\lambda),
\end{equation}
where 
(i) $\Delta f$ $=$ $\max{\big\{|V_Q|, |V_G|\big\}}$ according to Lemma~\ref{lemma:depth3}, (ii) 
$\epsilon$ and $\lambda$ are $\ppdo$'s input privacy budget and maximum error probability, respectively, and (iii) the offset $\frac{\Delta f}{\epsilon}\cdot \ln(2\lambda)$ is to ensure that $\xi$ is \four{a non-negative noise (without affecting the correctness of query answers)} with a probability of at least $1-\lambda$.

\stab
\noindent\textbf{Privacy analysis.} 
For the Laplace mechanism, we have Lemma~\ref{lemma:laplace}.
\begin{lemma}
\label{lemma:laplace}
For any neighboring inputs $\mathcal{X}_1$ and $\mathcal{X}_2$, the Laplace mechanism satisfies $(\epsilon, 0)$-differential privacy.
\end{lemma}

As $\ppdo$'s access pattern for $\gsim$ is uniquely determined by the depth of $\mathcal{F}(\mathcal{X}, f, \epsilon)$ updated in Line 9 of Algorithm~\ref{alg:do-gsim}, with Lemma~\ref{lemma:laplace} and Definition~\ref{def:dp}, we yield the following theorem.

\begin{theorem}
\label{theorem:do}
For any neighboring inputs $\mathcal{X}_1$ and $\mathcal{X}_2$, $\ppdo$ for answering $\gsim$ queries achieves $(\epsilon, 0)$-differential obliviousness.
\end{theorem}

\section{Experimental Evaluation}
\label{sec:exp}

\subsection{Experimental Settings}
\label{subsec:exp-settings}
\begin{table}
\caption{Dataset statistics}
\vspace{-3ex}
\centering
\resizebox{1.0\linewidth}{!}{
\begin{tabular}{|l|c|c|c|c|c|c|}\hline
Dataset  & \# of graphs $G$ & avg. $|V_G|$ & stddev. of $|V_G|$ & avg. $|E_G|$ & \# of node labels\\\hline
\hline
\textsc{Aids} & 42.7K &  25.6 &  12.2  & 27.6  & 62 \\\hline
\textsc{PubChem} & 22.8K & 48.1 &  9.4  &  50.6  & 10 \\\hline
\textsc{Chem1M} & 1M & 24.0 &  8.8  & 25.8  & 94 \\\hline
\textsc{Syn} & 1M & 14.3 &  1.8  & 20.5 & 1 \\\hline
\end{tabular}
}
\label{tab:datastat}
\end{table}

\noindent
{\bf Platform.} We implemented a prototype of $\fullname$ in C++ \four{with a \emph{DFS}-based $\cm$ enumeration (\S~\ref{subsubsec:overall})}, and conducted experiments on a server equipped with $2$TB RAM and dual $64$-core \emph{AMD EPYC} 7742 processors ($2.25$GHz), which supports a \emph{TEE} via the \emph{AMD SEV} technology. We used $160$ worker threads for all methods. \four{Across all evaluations, the peak memory usage was $12.6$GB.}

\stab
\noindent
{\bf Datasets.} Following prior work~\cite{DBLP:conf/sigmod/Kim21, DBLP:conf/icde/PengCCX22, DBLP:journals/pacmmod/XuCPXB23, DBLP:journals/tkde/ChangFYQZ23, DBLP:journals/tkde/WangZJHW23}, we conducted the experiments on two widely adopted real-world datasets, {\aid} and {\chem}, and two large-scale datasets, {\synchem} and {\syn}. Specifically, {\aid}~\cite{data:aid} is a set of antiviral compound graphs; {\chem} and {\synchem} are datasets of chemical compound graphs~\cite{data:pubchem}; and {\syn} is a synthetic graph dataset generated by using GraphGen~\cite{data:graphgen}.  Table~\ref{tab:datastat} summarizes the statistics of these datasets.

\stab
\noindent{\bf Queries and parameter settings.} For the query workload, we randomly generated $20$ graphs as query graphs $Q$ for a given node set size $|V_Q|$. We set the experimental parameters as follows:
\begin{itemize}
\item \textbf{Similarity parameter $\tau$.} For $\iso$ and $\gsim$ queries, $\tau$ is always set to $0$; for $\gss$ queries, we varied $\tau$ among $2$, $4$, $6$, $8$, and $10$, where $\tau = 8$ is the default value.
\item \textbf{Query (node set) size $|V_Q|$.} We varied $|V_Q|$ from $6$ to $10$. For $\iso$ and $\gsim$ queries, the default value of $|V_Q|$ is $6$; for $\gss$ queries, we remark that the evaluation on a data graph $G$ can be skipped directly without affecting the privacy guarantees if $\big||V_G| - |V_Q|\big| > \tau$. Therefore, to maximize the evaluation coverage, we set $|V_Q| = 10$ by default.
\item \textbf{Security parameters.} 
{For $\iso$ and $\gss$ queries, the $\ppdo$ approach of $\fullname$ achieves $(0, \delta)$-differential obliviousness, and hence, the privacy budget  $\epsilon$ is set to $0$.} 
For $\gsim$, we varied the privacy budget $\epsilon$ over $1$, $2$, $4$, $8$, and $16$, with a default value of $4$; we also set the maximum error probability $\lambda$ to $0.05$.
\item \textbf{\four{Restriction on graph size for $\base$.}} \four{As $\base$ requires  \emph{exponential} $\cm$ enumeration for $\iso$ and $\gss$ queries to ensure full obliviousness, we restricted $\base$ to evaluate the data graphs with fewer than $30$ (\resp $13$) nodes for $\iso$ (\resp $\gss$) queries.}
\end{itemize}

\noindent{\bf Benchmark methods.} 
We evaluate the following approaches:
\begin{itemize}
\item $\base$. \two{$\fullname$'s $\FO$ search approach inside the \emph{TEE}.}
\item $\ppdo$. \two{$\fullname$'s $\DO$ search approach inside the \emph{TEE}.}
\item $\ppdo^*$. \four{$\fullname$'s $\DO$ search approach outside the \emph{TEE}.}
\item $\ppdo^-$. \two{The $\ppdo$ approach inside the \emph{TEE} without the additional {optimization strategies} 
for differential obliviousness guarantee:} for $\iso$ and $\gss$ queries, $\ppdo^-$ prunes the $\cm$s {earlier than $\ppdo$ does} when the cost is larger than $\tau$; for $\gsim$ queries, $\ppdo^-$ is the same as $\ppdo$ without the noise injection.
\item $\opt$. {Efficient plaintext online search algorithms outside the \emph{TEE} without any privacy guarantee for the access pattern, namely (i) 
\two{the state-of-the-art 
open-source $\mathsf{DAF}$ algorithm~\cite{DBLP:conf/sigmod/HanKGPH19} for $\iso$,}
(ii) the state-of-the-art online $\mathsf{A^*}$ algorithm~\cite{DBLP:journals/tkde/ChangFYQZ23} for $\gss$, and (iii) a partition-refinement-free version of \cite{DBLP:conf/focs/HenzingerHK95} for $\gsim$.}  %

\end{itemize}

{As $\base$ is $\FO$, we report its runtime for one query. For the other approaches, we report the average runtimes over $20$ queries.}

\begin{figure}[t]
  \setlength{\abovecaptionskip}{0pt}
  \setlength{\belowcaptionskip}{0pt}
  \captionsetup{width=1\linewidth}
  \centering
  \vspace{-4ex}
  \graphicspath{{script/overall/}}
        \begin{subfigure}[t]{1\columnwidth}
        \centering
        \includegraphics[width=1.2\linewidth]{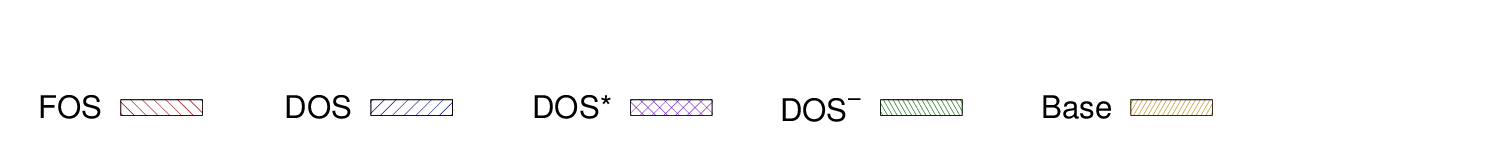}
    \end{subfigure}
      \begin{subfigure}[t]{1\columnwidth}
        \centering
        \vspace{-4.5ex}
        \includegraphics[width=\linewidth]{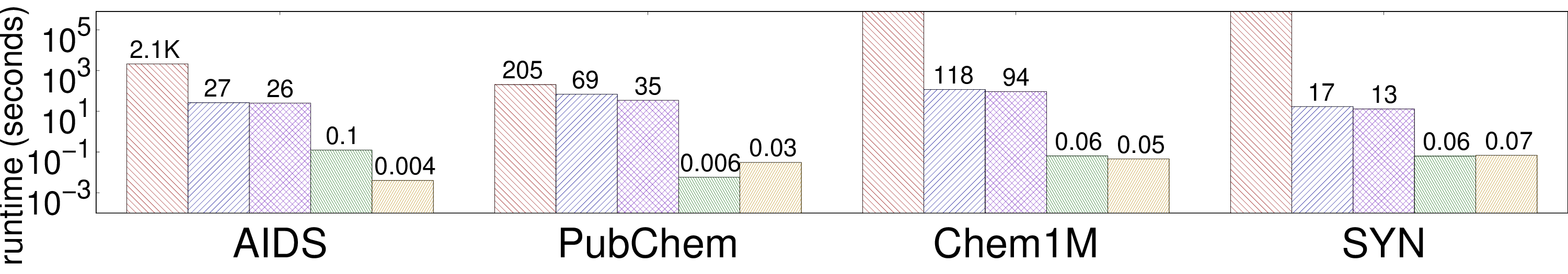}
        \vspace{-5ex}
        \caption{$\mathcal{F}$ $=$ $\iso$}
        \label{fig:overall-iso}
    \end{subfigure}
    \hfill
    \begin{subfigure}[t]{1\columnwidth}
        \centering
        \includegraphics[width=\linewidth]{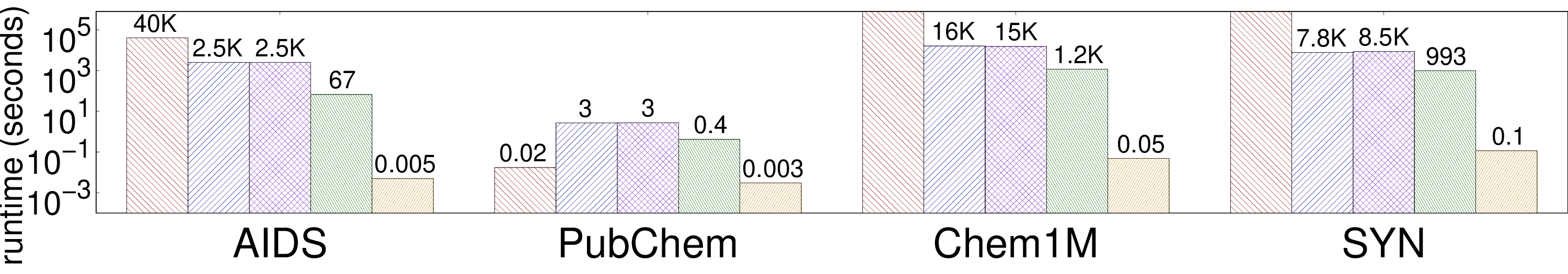}
        \vspace{-5ex}
        \caption{$\mathcal{F}$ $=$ $\gss$}
        \label{fig:overall-gss}
    \end{subfigure}
    \begin{subfigure}[t]{1\columnwidth}
        \centering
        \includegraphics[width=\linewidth]{gsim.eps}
        \vspace{-5ex}
        \caption{$\mathcal{F}$ $=$ $\gsim$}
        \label{fig:overall-gsim}
    \end{subfigure}
  \caption{Overall runtime performance of baselines for graph pattern queries under various $\mathcal{F}$}
  \label {fig:exp-overall}
  \end{figure}

\begin{figure}[t]
  \setlength{\abovecaptionskip}{0pt}
  \setlength{\belowcaptionskip}{0pt}
  \captionsetup{width=1.0\linewidth}
  \centering
  \graphicspath{{script/}}
      \begin{subfigure}[t]{0.485\columnwidth}
        \centering
        \includegraphics[width=\linewidth]{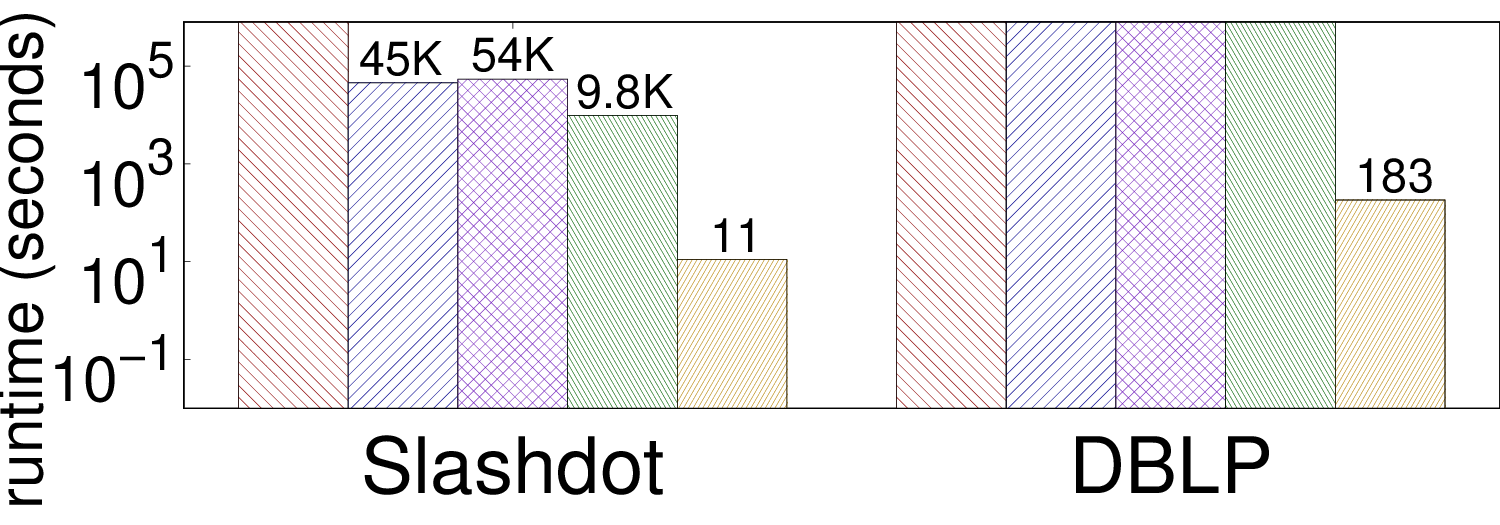}
        \vspace{-4.5ex}
        \caption{\scriptsize $\mathcal{F}$ $=$ $\gsim$; {\slashdot} and {\dblp}}
        \label{fig:exp-comparison-gsim}
    \end{subfigure}
    \hfill
    \begin{subfigure}[t]{0.485\columnwidth}
        \centering
        \includegraphics[width=\linewidth]{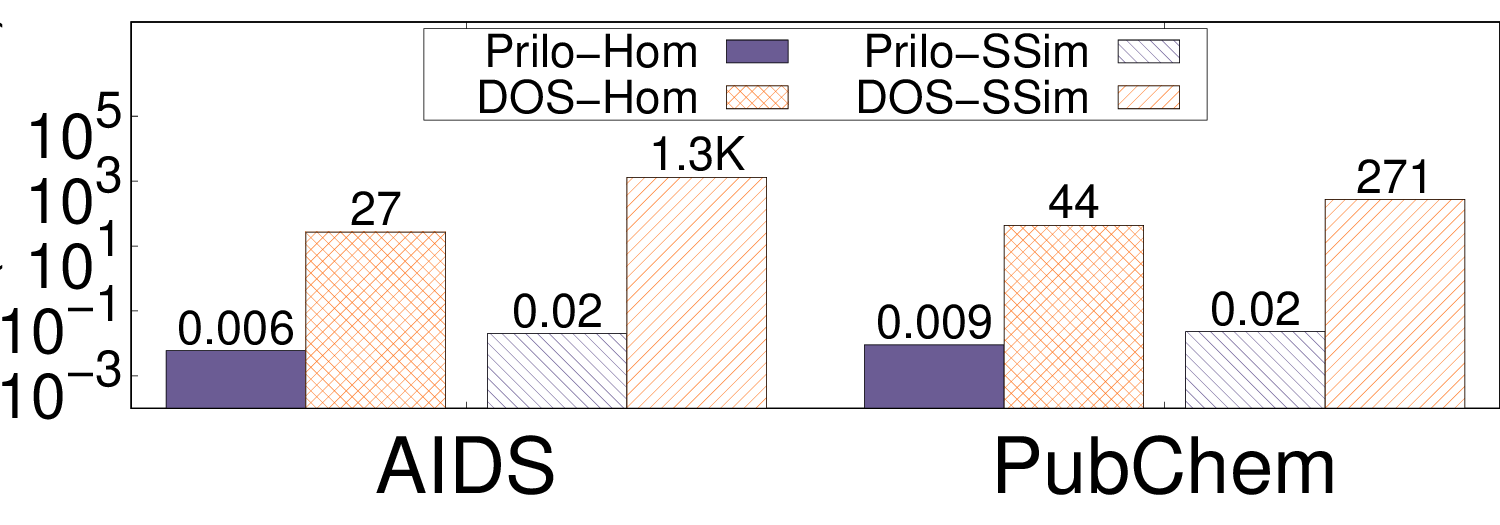}
        \vspace{-4.5ex}
        \caption{\scriptsize $\fullname$ vs. $\mathsf{Prilo}$}
        \label{fig:exp-comparison-prilo}
    \end{subfigure}
  \caption{\two{(a) 
  baseline runtimes for $\gsim$ queries on {\slashdot} and {\dblp}; (b) runtime comparison between $\ppdo$ and $\mathsf{Prilo}$}}
  \label {fig:exp-comparison}
  \end{figure}

\subsection{Overall Performance}
\label{subsec:overall}
We first evaluate the overall efficiency of the baselines on the four datasets under all three query semantics $\mathcal{F}$.
\begin{itemize}[wide, labelwidth=!, labelindent=0pt]
\item $\mathcal{F} = \iso$ \two{$\big($Figure~\ref{fig:overall-iso}$\big)$. As $\base$ enumerates exponentially many $\cm$s, $\base$ can handle small data graphs in \textsc{Aids} and \textsc{PubChem}  with 
$\le 30$ nodes under our default setting,
but could not finish the search on \textsc{Chem1M} and \textsc{Syn} within $24$ hours. 
As expected, $\base$ is the slowest among the five approaches due to its strongest privacy guarantee; $\ppdo$ ran slightly slower than $\ppdo^*$ due to the TEE's overhead for data protection; and the runtimes decrease from $\ppdo$ (or $\ppdo^*$) to  $\ppdo^-$ to $\opt$ as the information leakage from the pruning access pattern increases.}
\four{To quantify this overhead, we further measured $\ppdo$ against $\ppdo^*$ in single-threaded microbenchmarks: the overhead of \emph{AMD SEV} amounts to $0.68$\%, $2.53$\%, and $0.59$\% of the runtime for $\iso$, $\gss$, and $\gsim$ queries, respectively. This confirms that a TEE is a practical execution substrate for this workload.}

\item $\mathcal{F} = \gss$ $\big($Figure~\ref{fig:overall-gss}$\big)$.
Overall, the results follow the same pattern as for $\iso$ queries.\footnote{\two{The graphs in \textsc{PubChem} have at least $17$ nodes, so $\base$ skipped all of them.}} In particular, $\opt$ is consistently faster than $\ppdo^-$, showing the efficiency of $\mathsf{A^*}$ optimizations. 
\two{However, $\ppdo$ ran faster than $\ppdo^*$ on \textsc{Syn},
which reflects default configuration-level differences between the \emph{AMD SEV} guest and the bare-metal host (such as guest-kernel scheduling and isolation from OS noise) that offset SEV's added cost for memory encryption.
}

\item $\mathcal{F} = \gsim$.
\two{In addition to the four multi-graph datasets, we also evaluated $\gsim$ queries on {\slashdot}~\cite{data:snap} and {\dblp}~\cite{data:snap}, which are two single-graph datasets having $82$K and $317$K nodes and $948$K and $1$M edges, respectively. Their node labels were randomly assigned from a set of $128$ distinct labels.
As shown in Figures~\ref{fig:overall-gsim} and \ref{fig:exp-comparison-gsim}, the runtimes for the multi-graph datasets decrease from $\base$ to $\ppdo$ (or $\ppdo^*$) to $\ppdo^-$ to $\opt$; on {\slashdot}, $\base$ cannot finish within $24$ hours, while the remaining approaches follow the same decreasing order.} This is because the access pattern for $\gsim$ queries depends on the depth of $\cm$ enumeration. As the depth decreases from $\base$ to $\ppdo$ (or $\ppdo^*$) to $\ppdo^-$ to $\opt$, the results are consistent with the designs of these approaches.
\two{However, as {\dblp} has a large node set size $|V_G|$, neither $\base$ nor the $\ppdo$ variants can finish the query processing on {\dblp} within $24$ hours due to their time complexity of $O\big(|V_G|^3\big)$.}
\end{itemize}

\noindent\textbf{Summary.} %
$\fullname$'s $\base$ serves as an $\FO$ search baseline, which can handle searches on small graphs. Compared with $\base$,
$\ppdo$ (and $\ppdo^*$)
can search larger data graphs with improved efficiency,
while retaining the privacy guarantee of differential obliviousness.
Although $\ppdo^-$ and $\opt$ are more efficient than $\ppdo$, they provide no theoretical privacy guarantees for the access pattern. 

\subsection{Detailed Evaluation}
\label{exp:detail}

\noindent\textbf{\two{Comparison to SOTAs.}} 
\two{We compared $\ppdo$ to $\mathsf{Prilo}$~\cite{DBLP:journals/pacmmod/XuCPXB23} for subgraph homomorphism ($\homo$) and strong simulation ($\ssim$) queries. Figure~\ref{fig:exp-comparison-prilo} shows their average runtimes on \textsc{Aids} and \textsc{PubChem}. As expected, Prilo achieves significantly higher query efficiency than $\fullname$, as its topology-based pruning leverages the unprotected data-graph structure, whereas $\fullname$ protects both the query and data graphs.}

\two{Regarding the comparison to $\mathsf{PrigSim}$ for $\gss$ queries, \cite{DBLP:journals/tkde/WangZJHW23}~reports that $\mathsf{PrigSim}$ took approximately $400$ seconds on \textsc{Aids} to answer a $\gss$ query with $|V_Q| = 10$ and $\tau = 5$, using $16$ \emph{Intel I7-10700K} cores ($\sim$$4.7$GHz turbo) and $64$GB RAM. For a like-for-like comparison, we restricted $\ppdo$ to $16$ threads of our \emph{AMD EPYC 7742} processors ($2.25$GHz), \ie the same number of threads at a lower clock rate, with peak memory within the $12.6$GB reported above. Under this setting, $\ppdo$ took an average runtime of $9.57$ seconds to answer $20$ random $\gss$ queries on \textsc{Aids},} 
while evaluating an average of $5{,}396$ out of $42{,}689$ data graphs.

\begin{figure}[t]
  \setlength{\abovecaptionskip}{0pt}
  \setlength{\belowcaptionskip}{0pt}
  \captionsetup{width=1.1\linewidth}
  \centering
  \vspace{-4.5ex}
  \graphicspath{{script/VQ/}}
        \hspace{-12ex}
        \begin{subfigure}[t]{1\columnwidth}
        \centering
        \includegraphics[width=1.2\linewidth]{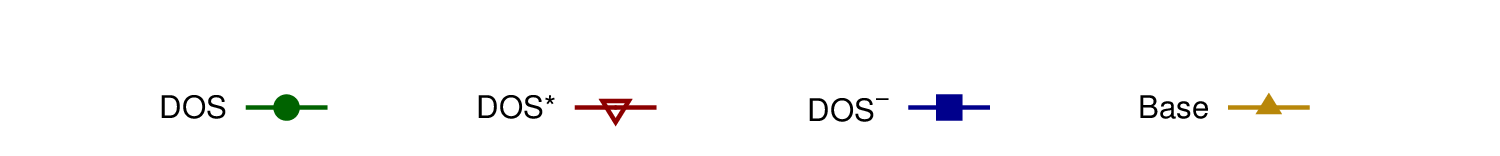}
    \end{subfigure}
      \begin{subfigure}[t]{0.325\columnwidth}
        \centering
        \vspace{-4ex}
        \includegraphics[width=\linewidth]{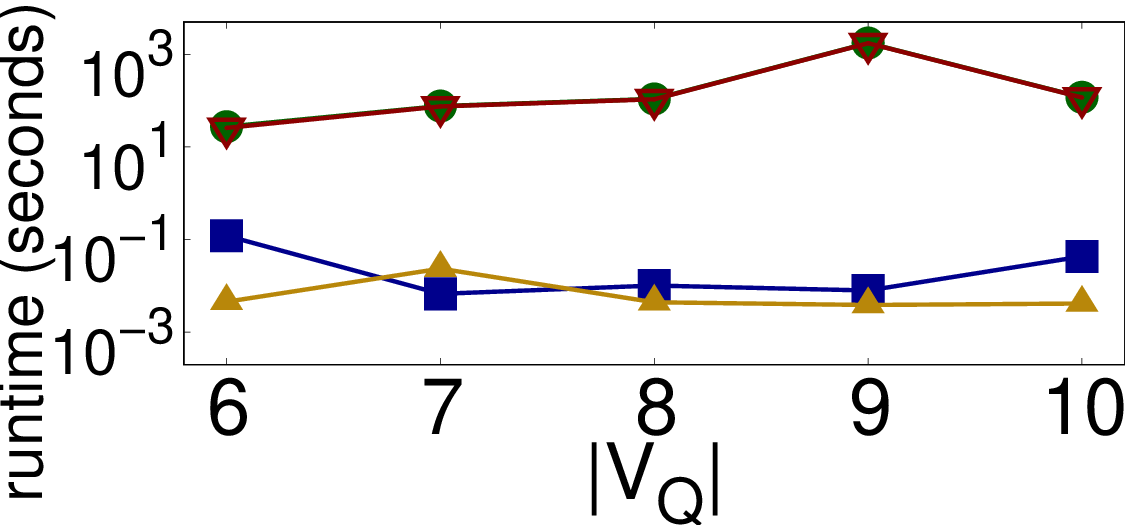}
        \vspace{-5ex}
        \caption{\scriptsize $\mathcal{F}$ $=$ $\iso$; \textsc{Aids}}
        \label{fig:VQ-aids-iso}
    \end{subfigure}
    \begin{subfigure}[t]{0.325\columnwidth}
        \centering
        \vspace{-4ex}
        \includegraphics[width=\linewidth]{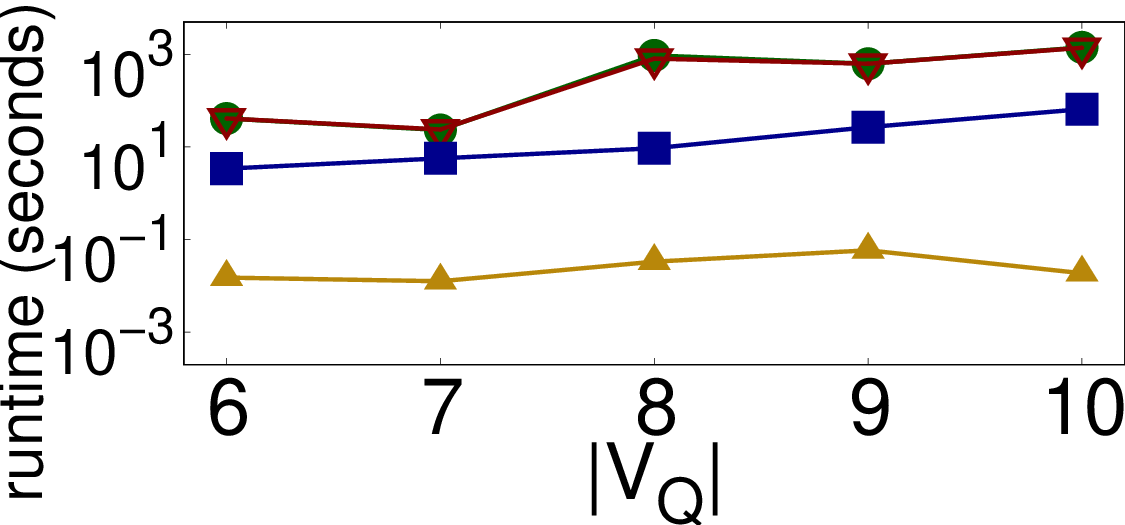}
        \vspace{-5ex}
        \caption{\scriptsize $\mathcal{F}$ $=$ $\gss$; \textsc{Aids}}
        \label{fig:VQ-aids-gss}
    \end{subfigure}
    \begin{subfigure}[t]{0.325\columnwidth}
        \centering
        \vspace{-4ex}
        \includegraphics[width=\linewidth]{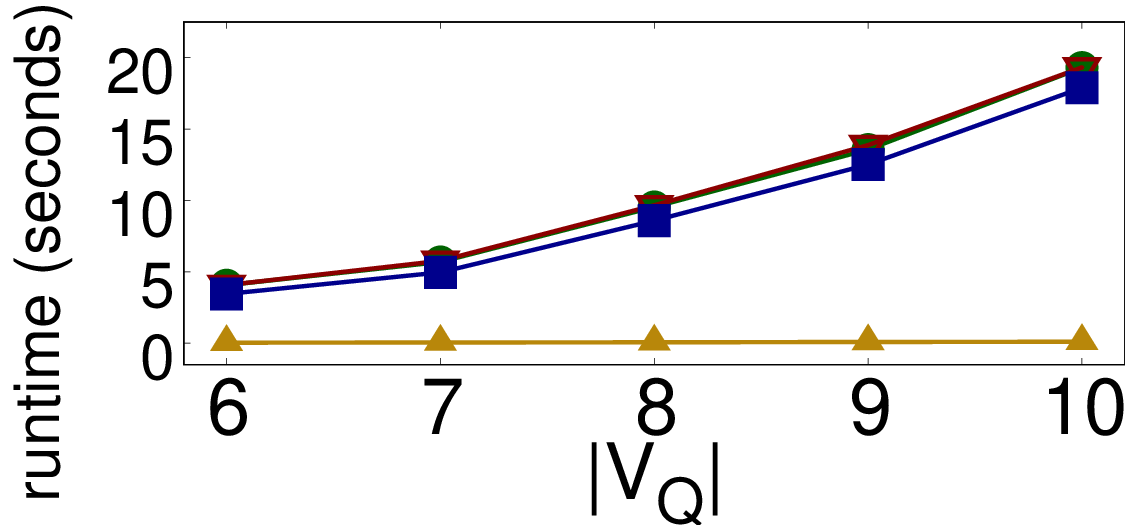}
        \vspace{-5ex}
        \caption{\scriptsize $\mathcal{F}$ $=$ $\gsim$; \textsc{Aids}}
        \label{fig:VQ-aids-gsim}
    \end{subfigure}
    \begin{subfigure}[t]{0.325\columnwidth}
        \centering
        \vspace{0.5ex}
        \includegraphics[width=\linewidth]{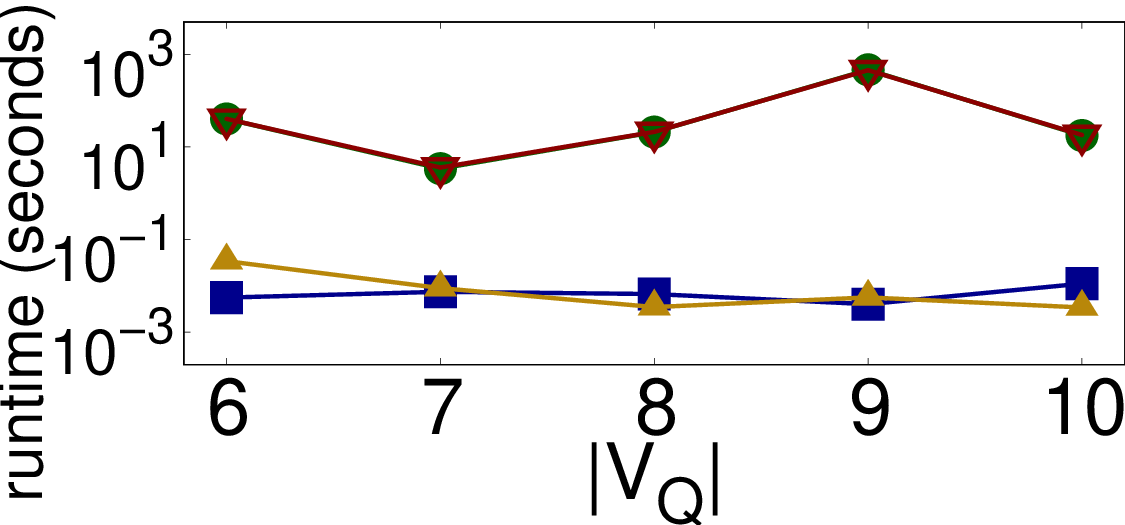}
        \vspace{-5ex}
        \caption{\scriptsize $\mathcal{F}$ $=$ $\iso$; \textsc{PubChem}}
        \label{fig:VQ-pubchem-iso}
    \end{subfigure}
    \begin{subfigure}[t]{0.325\columnwidth}
        \centering
        \vspace{0.5ex}
        \includegraphics[width=\linewidth]{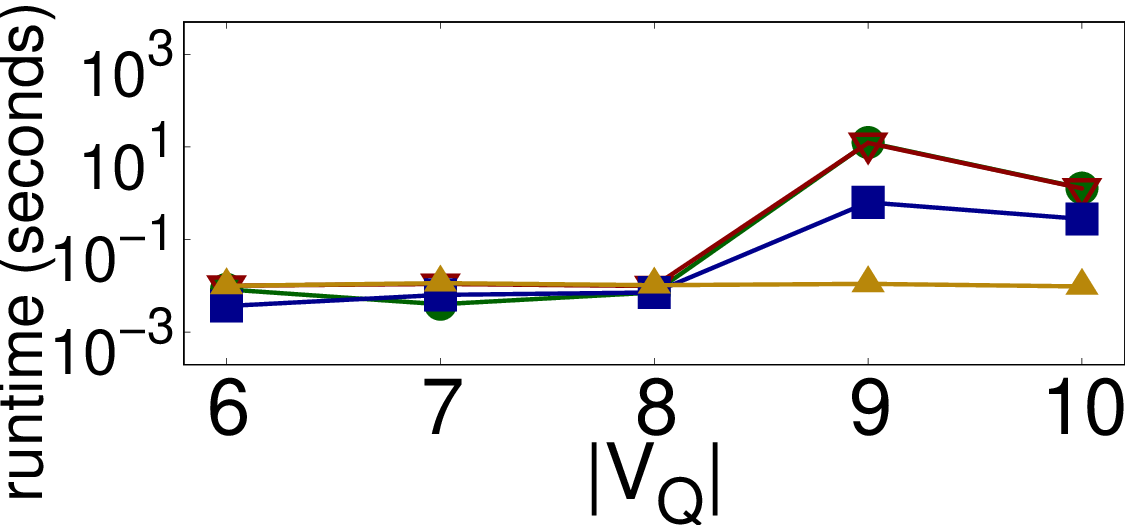}
        \vspace{-5ex}
        \caption{\scriptsize $\mathcal{F}$ $=$ $\gss$; \textsc{PubChem}}
        \label{fig:VQ-pubchem-gss}
    \end{subfigure}
    \begin{subfigure}[t]{0.325\columnwidth}
        \centering
        \vspace{0.5ex}
        \includegraphics[width=\linewidth]{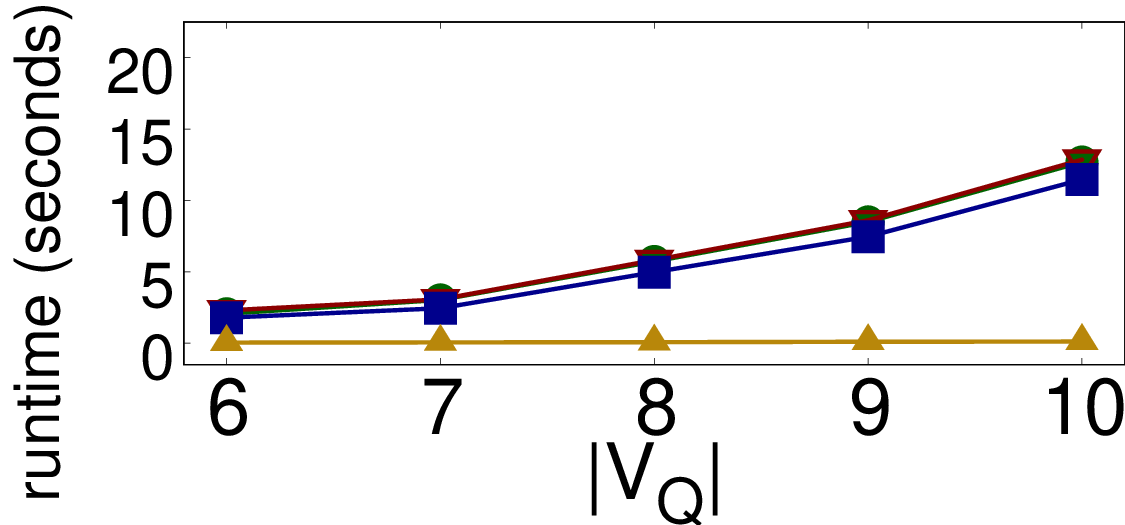}
        \vspace{-5ex}
        \caption{\scriptsize $\mathcal{F}$ $=$ $\gsim$; \textsc{PubChem}}
        \label{fig:VQ-pubchem-gsim}
    \end{subfigure}
  \caption{\four{Runtimes of $\ppdo$ variants, $\opt$ when varying $|V_Q|$}}
  \label {fig:exp-VQ}
  \end{figure}

\stab
\noindent\textbf{Varying $|V_Q|$ for different query semantics $\big($Figure~\ref{fig:exp-VQ}$\big)$.} We evaluated the runtimes of $\ppdo$
variants and $\opt$ on \textsc{Aids} and \textsc{PubChem} by varying the node set size of the query graphs from $6$ to $10$. 
Figure~\ref{fig:VQ-aids-iso} shows that the runtimes of $\ppdo$ 
variants
for $\iso$ queries exhibit an overall upward trend as $|V_Q|$ increases. This is consistent with their worst-case time complexity of $O\big(|V_G|^{|V_Q|}\big)$. For $\gss$ queries, Figure~\ref{fig:VQ-aids-gss} shows similar trends.
Here, the runtime increases because a larger $|V_Q|$ with a fixed $\tau$ leads to larger data graphs being evaluated.  The variance is due to differences in pruning power for specific topological structures of the randomly generated queries. Figure~\ref{fig:VQ-aids-gsim} shows that the runtimes of 
$\ppdo$ variants for $\gsim$ queries increase as $|V_Q|$ increases since a larger $|V_Q|$ leads to a larger candidate sequence $\cseq_Q^{\mathcal{F}}$ and a greater $\cm$ enumeration depth.  Regarding $\opt$, its runtimes for all three query types are insensitive to $|V_Q|$ because it leverages strong 
pruning strategies that are infeasible under the privacy target. Similar results for \textsc{PubChem} appear in Figures~\ref{fig:VQ-pubchem-iso}-\ref{fig:VQ-pubchem-gsim}.

\begin{figure}[t]
  \setlength{\abovecaptionskip}{0pt}
  \setlength{\belowcaptionskip}{0pt}
  \captionsetup{width=1\linewidth}
  \centering
  \graphicspath{{script/}}
      \begin{subfigure}[t]{0.485\columnwidth}
        \centering
        \includegraphics[width=0.97\linewidth]{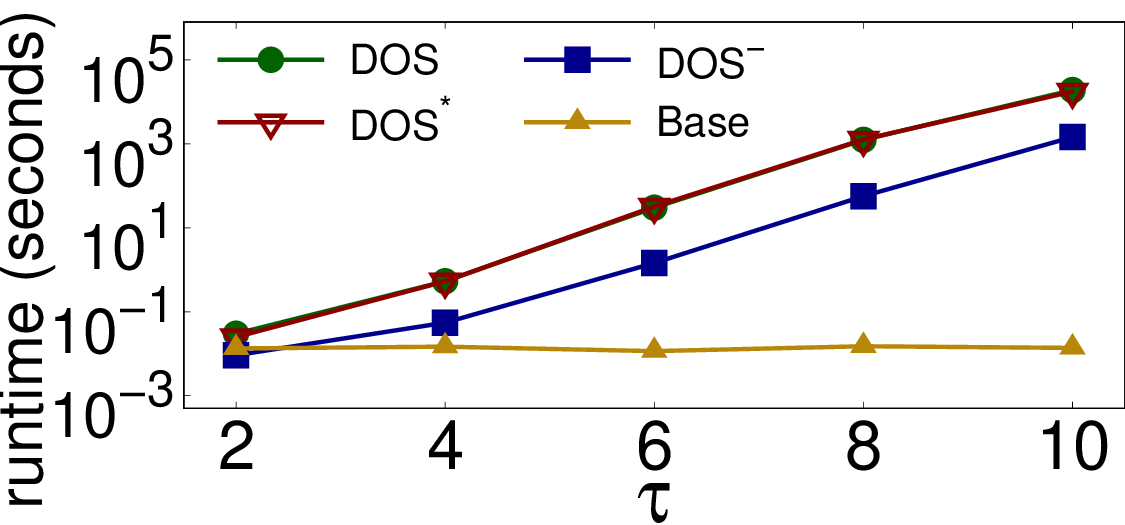}
        \vspace{-2ex}
        \caption{\scriptsize \four{$\ppdo$ variants, and $\opt$; $\mathcal{F}$ $=$ $\gss$; \textsc{Aids}}}
        \label{fig:Tau-aids-gss}
    \end{subfigure}
    \hfill
    \begin{subfigure}[t]{0.485\columnwidth}
        \centering
        \includegraphics[width=0.97\linewidth]{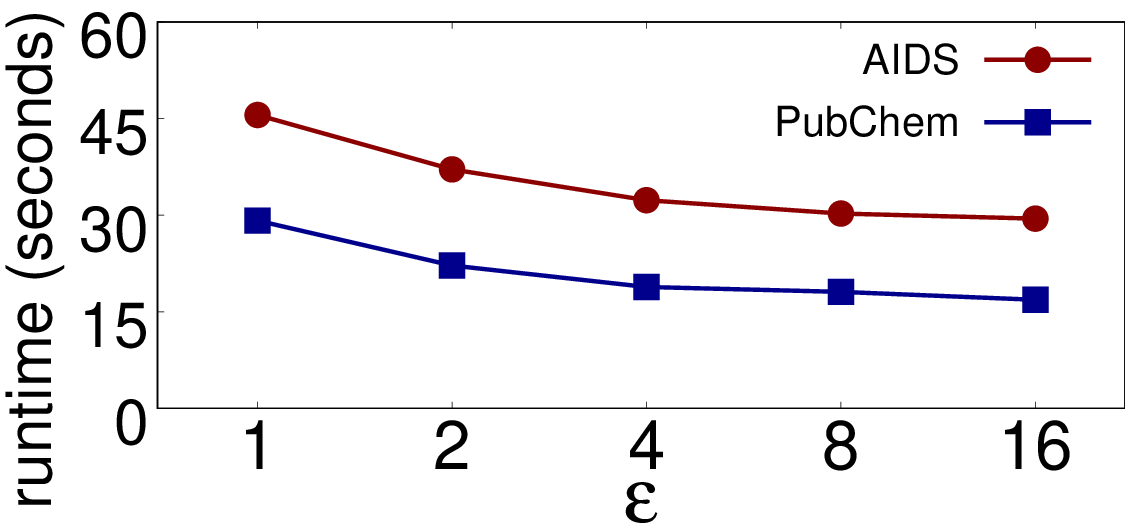}
        \vspace{-2ex}
        \caption{\scriptsize $\ppdo$; $\mathcal{F}$ $=$ $\gsim$; \textsc{Aids} and \textsc{PubChem}}
        \label{fig:Epsilon-aids-gsim}
    \end{subfigure}
  \caption{\four{Runtimes of baselines when varying $\tau$ and $\epsilon$ for $\gss$ and $\gsim$ queries, respectively}}
  \label {fig:exp-Tau}
  \end{figure}

\stab
\noindent\textbf{Varying $\tau$ for $\gss$ queries.} We evaluated the effect of $\tau$ on the $\gss$ query efficiency of $\ppdo$
variants and $\opt$. 
The results in Figure~\ref{fig:Tau-aids-gss} show that the runtimes of $\ppdo$ 
variants increase exponentially as $\tau$ increases. This is as expected. With a larger $\tau$, \four{more graphs having $\le \big(|V_Q|+\tau\big)$ nodes are evaluated, and these three approaches} use higher cost values to prune the $\cm$ enumeration, leading to a deeper $\cm$ enumeration. The runtime of $\opt$ increases slightly as $\tau$ increases due to $\mathsf{A^*}$'s lower bound pruning strategy.

\stab
\noindent\textbf{Varying $\epsilon$ for $\gsim$ queries.} Theorem~\ref{theorem:do} shows that $\ppdo$ achieves $(\epsilon, 0)$-differential obliviousness for 
$\gsim$ queries. 
Hence, we evaluated how the privacy budget $\epsilon$ affects $\ppdo$'s efficiency. Figure~\ref{fig:Epsilon-aids-gsim} shows $\ppdo$'s runtimes on \textsc{Aids} and \textsc{PubChem} as $\epsilon$ varies. As expected, 
$\ppdo$'s runtime decreases as $\epsilon$ increases 
since a larger $\epsilon$~requires less injected noise and 
hence a smaller enumeration depth.

\section{Conclusion}
\label{sec:con}
In this paper, we study 
privacy-preserving graph pattern query processing with the privacy target of the structural information of both the query graph and the data graphs. To handle various query semantics, we propose a query processing framework $\fullname$ that consists of an $\FO$ search base $\base$ and an optimized pruning search approach $\ppdo$ achieving the privacy guarantees of full obliviousness and differential obliviousness, respectively. 
The privacy analysis and experimental evaluation on real-world graph datasets demonstrate 
{$\fullname$'s promising performance.}
For future work, we plan to explore the application of differential obliviousness to improve query efficiency for other databases, 
\eg relational and vector databases.

\begin{acks}
    Qichen Wang was supported by the Singapore Ministry of Education under Grant No. RS32/25.  Xinyi Zhang conducted this research work while visiting Nanyang Technological University.
\end{acks}

\section*{AI Usage DECLARATION}
Generative AI tools were used only to polish the writing and assist with code implementation.

\clearpage

\balance

\bibliographystyle{ACM-Reference-Format}
\bibliography{main}

@inproceedings{DBLP:conf/sosp/MawhirterW19,
  author       = {Daniel Mawhirter and
                  Bo Wu},
  title        = {AutoMine: Harmonizing High-Level Abstraction and High Performance
                  for Graph Mining},
  booktitle    = {{ACM} {SOSP}},
  pages        = {509--523},
  year         = {2019}
}

@inproceedings{DBLP:conf/sc/ShiZXZ20,
  author       = {Tianhui Shi and
                  Mingshu Zhai and
                  Yi Xu and
                  Jidong Zhai},
  title        = {GraphPi: High Performance Graph Pattern Matching through Effective
                  Redundancy Elimination},
  booktitle    = {{IEEE/ACM} {SC}},
  pages        = {100:1--100:14},
  year         = {2020}
}

@article{DBLP:journals/tkde/ChangFYQZ23,
  author       = {Lijun Chang and
                  Xing Feng and
                  Kai Yao and
                  Lu Qin and
                  Wenjie Zhang},
  title        = {Accelerating Graph Similarity Search via Efficient {GED} Computation},
  journal      = {{IEEE} {TKDE}},
  volume       = {35},
  number       = {5},
  pages        = {4485--4498},
  year         = {2023},
  @url          = {https://doi.org/10.1109/TKDE.2022.3153523},
  @doi          = {10.1109/TKDE.2022.3153523},
  @timestamp    = {Sat, 29 Apr 2023 19:26:24 +0200},
  @biburl       = {https://dblp.org/rec/journals/tkde/ChangFYQZ23.bib},
  @bibsource    = {dblp computer science bibliography, https://dblp.org}
}

@article{DBLP:journals/tkde/WangZJHW23,
  author       = {Songlei Wang and
                  Yifeng Zheng and
                  Xiaohua Jia and
                  Hejiao Huang and
                  Cong Wang},
  title        = {PrigSim: Towards Privacy-Preserving Graph Similarity Search as a Cloud
                  Service},
  journal      = {{IEEE} {TKDE}},
  volume       = {35},
  number       = {10},
  pages        = {10478--10496},
  year         = {2023},
  @url          = {https://doi.org/10.1109/TKDE.2023.3266449},
  @doi          = {10.1109/TKDE.2023.3266449},
  @timestamp    = {Sat, 14 Oct 2023 20:13:37 +0200},
  @biburl       = {https://dblp.org/rec/journals/tkde/WangZJHW23.bib},
  @bibsource    = {dblp computer science bibliography, https://dblp.org}
}

@inproceedings{DBLP:conf/sigmod/Kim21,
  author       = {Jongik Kim},
  @editor       = {Guoliang Li and
                  Zhanhuai Li and
                  Stratos Idreos and
                  Divesh Srivastava},
  title        = {Boosting Graph Similarity Search through Pre-Computation},
  booktitle    = {{ACM} {SIGMOD}},
  pages        = {951--963},
  @publisher    = {{ACM}},
  year         = {2021},
  @url          = {https://doi.org/10.1145/3448016.3452780},
  @doi          = {10.1145/3448016.3452780},
  @timestamp    = {Mon, 21 Jun 2021 11:48:44 +0200},
  @biburl       = {https://dblp.org/rec/conf/sigmod/Kim21.bib},
  @bibsource    = {dblp computer science bibliography, https://dblp.org}
}

@inproceedings{DBLP:conf/kdd/QinBS20,
  author       = {Zongyue Qin and
                  Yunsheng Bai and
                  Yizhou Sun},
  @editor       = {Rajesh Gupta and
                  Yan Liu and
                  Jiliang Tang and
                  B. Aditya Prakash},
  title        = {GHashing: Semantic Graph Hashing for Approximate Similarity Search
                  in Graph Databases},
  booktitle    = {{ACM} {KDD}},
  pages        = {2062--2072},
  @publisher    = {{ACM}},
  year         = {2020},
  @url          = {https://doi.org/10.1145/3394486.3403257},
  @doi          = {10.1145/3394486.3403257},
  @timestamp    = {Sun, 19 Jan 2025 13:22:15 +0100},
  @biburl       = {https://dblp.org/rec/conf/kdd/QinBS20.bib},
  @bibsource    = {dblp computer science bibliography, https://dblp.org}
}

@inproceedings{DBLP:conf/icde/PengCCX22,
  author       = {Yun Peng and
                  Byron Choi and
                  Tsz Nam Chan and
                  Jianliang Xu},
  title        = {{LAN:} Learning-based Approximate k-Nearest Neighbor Search in Graph
                  Databases},
  booktitle    = {{IEEE} {ICDE}},
  pages        = {2508--2521},
  @publisher    = {{IEEE}},
  year         = {2022},
  @url          = {https://doi.org/10.1109/ICDE53745.2022.00233},
  @doi          = {10.1109/ICDE53745.2022.00233},
  @timestamp    = {Wed, 18 Dec 2024 13:51:13 +0100},
  @biburl       = {https://dblp.org/rec/conf/icde/PengCCX22.bib},
  @bibsource    = {dblp computer science bibliography, https://dblp.org}
}

@article{DBLP:journals/pacmmod/ChengYWHZ25,
  author       = {Qihao Cheng and
                  Da Yan and
                  Tianhao Wu and
                  Zhongyi Huang and
                  Qin Zhang},
  title        = {Computing Approximate Graph Edit Distance via Optimal Transport},
  journal      = {{ACM} {PACMMOD}},
  volume       = {3},
  number       = {1},
  pages        = {23:1--23:26},
  year         = {2025},
  @url          = {https://doi.org/10.1145/3709673},
  @doi          = {10.1145/3709673},
  @timestamp    = {Tue, 08 Apr 2025 20:24:31 +0200},
  @biburl       = {https://dblp.org/rec/journals/pacmmod/ChengYWHZ25.bib},
  @bibsource    = {dblp computer science bibliography, https://dblp.org}
}

@article{DBLP:journals/tsmc/SanfeliuF83,
  author       = {Alberto Sanfeliu and
                  King{-}Sun Fu},
  title        = {A distance measure between attributed relational graphs for pattern
                  recognition},
  journal      = {{IEEE} {TSMC}},
  volume       = {13},
  number       = {3},
  pages        = {353--362},
  year         = {1983},
  @url          = {https://doi.org/10.1109/TSMC.1983.6313167},
  @doi          = {10.1109/TSMC.1983.6313167},
  @timestamp    = {Wed, 20 May 2020 16:24:55 +0200},
  @biburl       = {https://dblp.org/rec/journals/tsmc/SanfeliuF83.bib},
  @bibsource    = {dblp computer science bibliography, https://dblp.org}
}

@inproceedings{DBLP:conf/icde/Cui0WLW20,
  author       = {Ningning Cui and
                  Xiaochun Yang and
                  Bin Wang and
                  Jianxin Li and
                  Guoren Wang},
  title        = {SVkNN: Efficient Secure and Verifiable k-Nearest Neighbor Query on
                  the Cloud Platform},
  booktitle    = {{IEEE} {ICDE}},
  pages        = {253--264},
  @publisher    = {{IEEE}},
  year         = {2020},
  @url          = {https://doi.org/10.1109/ICDE48307.2020.00029},
  @doi          = {10.1109/ICDE48307.2020.00029},
  @timestamp    = {Wed, 19 Aug 2020 13:33:30 +0200},
  @biburl       = {https://dblp.org/rec/conf/icde/Cui0WLW20.bib},
  @bibsource    = {dblp computer science bibliography, https://dblp.org}
}

@inproceedings{DBLP:conf/icde/LeiLLT19,
  author       = {Xinyu Lei and
                  Alex X. Liu and
                  Rui Li and
                  Guan{-}Hua Tu},
  title        = {SecEQP: {A} Secure and Efficient Scheme for SkNN Query Problem Over
                  Encrypted Geodata on Cloud},
  booktitle    = {{IEEE} {ICDE}},
  pages        = {662--673},
  @publisher    = {{IEEE}},
  year         = {2019},
  @url          = {https://doi.org/10.1109/ICDE.2019.00065},
  @doi          = {10.1109/ICDE.2019.00065},
  @timestamp    = {Thu, 14 Oct 2021 16:36:48 +0200},
  @biburl       = {https://dblp.org/rec/conf/icde/LeiLLT19.bib},
  @bibsource    = {dblp computer science bibliography, https://dblp.org}
}

@inproceedings{DBLP:conf/icde/WuLLYYW19,
  author       = {Songrui Wu and
                  Qi Li and
                  Guoliang Li and
                  Dong Yuan and
                  Xingliang Yuan and
                  Cong Wang},
  title        = {ServeDB: Secure, Verifiable, and Efficient Range Queries on Outsourced
                  Database},
  booktitle    = {{IEEE} {ICDE}},
  pages        = {626--637},
  @publisher    = {{IEEE}},
  year         = {2019},
  @url          = {https://doi.org/10.1109/ICDE.2019.00062},
  @doi          = {10.1109/ICDE.2019.00062},
  @timestamp    = {Wed, 26 Feb 2025 11:15:12 +0100},
  @biburl       = {https://dblp.org/rec/conf/icde/WuLLYYW19.bib},
  @bibsource    = {dblp computer science bibliography, https://dblp.org}
}

@article{DBLP:journals/pacmmod/ZhangWXPX24,
  author       = {Xinyi Zhang and
                  Qichen Wang and
                  Cheng Xu and
                  Yun Peng and
                  Jianliang Xu},
  title        = {FedKNN: Secure Federated k-Nearest Neighbor Search},
  journal      = {{ACM} {PACMMOD}},
  volume       = {2},
  number       = {1},
  pages        = {011:1--011:26},
  year         = {2024},
  @url          = {https://doi.org/10.1145/3639266},
  @doi          = {10.1145/3639266},
  @timestamp    = {Sun, 19 Jan 2025 15:06:03 +0100},
  @biburl       = {https://dblp.org/rec/journals/pacmmod/ZhangWXPX24.bib},
  @bibsource    = {dblp computer science bibliography, https://dblp.org}
}

@article{DBLP:journals/pvldb/BaterEEGKR17,
  author       = {Johes Bater and
                  Gregory Elliott and
                  Craig Eggen and
                  Satyender Goel and
                  Abel N. Kho and
                  Jennie Rogers},
  title        = {{SMCQL:} Secure Query Processing for Private Data Networks},
  journal      = {{PVLDB}},
  volume       = {10},
  number       = {6},
  pages        = {673--684},
  year         = {2017},
  @url          = {http://www.vldb.org/pvldb/vol10/p673-rogers.pdf},
  @doi          = {10.14778/3055330.3055334},
  @timestamp    = {Sat, 30 Sep 2023 10:24:08 +0200},
  @biburl       = {https://dblp.org/rec/journals/pvldb/BaterEEGKR17.bib},
  @bibsource    = {dblp computer science bibliography, https://dblp.org}
}

@article{DBLP:journals/vldb/TongZSPFXZZCXXL25,
  author       = {Yongxin Tong and
                  Yuxiang Zeng and
                  Yang Song and
                  Xuchen Pan and
                  Zeheng Fan and
                  Chunbo Xue and
                  Zimu Zhou and
                  Xiaofei Zhang and
                  Lei Chen and
                  Yi Xu and
                  Ke Xu and
                  Weifeng Lv},
  title        = {Hu-Fu: efficient and secure spatial queries over data federation},
  journal      = {{VLDB} J.},
  volume       = {34},
  number       = {2},
  pages        = {19},
  year         = {2025},
  @url          = {https://doi.org/10.1007/s00778-024-00896-3},
  @doi          = {10.1007/S00778-024-00896-3},
  @timestamp    = {Fri, 07 Mar 2025 18:30:43 +0100},
  @biburl       = {https://dblp.org/rec/journals/vldb/TongZSPFXZZCXXL25.bib},
  @bibsource    = {dblp computer science bibliography, https://dblp.org}
}

@article{DBLP:journals/tkde/DuWWCJM22,
  author       = {Minxin Du and
                  Shuangke Wu and
                  Qian Wang and
                  Dian Chen and
                  Peipei Jiang and
                  Aziz Mohaisen},
  title        = {GraphShield: Dynamic Large Graphs for Secure Queries With Forward
                  Privacy},
  journal      = {{IEEE} {TKDE}},
  volume       = {34},
  number       = {7},
  pages        = {3295--3308},
  year         = {2022},
  @url          = {https://doi.org/10.1109/TKDE.2020.3024883},
  @doi          = {10.1109/TKDE.2020.3024883},
  @timestamp    = {Mon, 13 Jun 2022 20:56:58 +0200},
  @biburl       = {https://dblp.org/rec/journals/tkde/DuWWCJM22.bib},
  @bibsource    = {dblp computer science bibliography, https://dblp.org}
}

@inproceedings{DBLP:conf/stoc/Goldreich87,
  author       = {Oded Goldreich},
  @editor       = {Alfred V. Aho},
  title        = {Towards a Theory of Software Protection and Simulation by Oblivious
                  RAMs},
  booktitle    = {{ACM} {STOC}},
  pages        = {182--194},
  @publisher    = {{ACM}},
  year         = {1987},
  @url          = {https://doi.org/10.1145/28395.28416},
  @doi          = {10.1145/28395.28416},
  @timestamp    = {Tue, 06 Nov 2018 11:07:04 +0100},
  @biburl       = {https://dblp.org/rec/conf/stoc/Goldreich87.bib},
  @bibsource    = {dblp computer science bibliography, https://dblp.org}
}

@inproceedings{DBLP:conf/sigmod/ShangLZYW10,
  author       = {Haichuan Shang and
                  Xuemin Lin and
                  Ying Zhang and
                  Jeffrey Xu Yu and
                  Wei Wang},
  @editor       = {Ahmed K. Elmagarmid and
                  Divyakant Agrawal},
  title        = {Connected substructure similarity search},
  booktitle    = {{ACM} {SIGMOD}},
  pages        = {903--914},
  publisher    = {{ACM}},
  year         = {2010},
  @url          = {https://doi.org/10.1145/1807167.1807264},
  @doi          = {10.1145/1807167.1807264},
  @timestamp    = {Sun, 19 Jan 2025 13:27:19 +0100},
  @biburl       = {https://dblp.org/rec/conf/sigmod/ShangLZYW10.bib},
  @bibsource    = {dblp computer science bibliography, https://dblp.org}
}

@article{DBLP:journals/pvldb/ZhangYJ10,
  author       = {Shijie Zhang and
                  Jiong Yang and
                  Wei Jin},
  title        = {{SAPPER:} Subgraph Indexing and Approximate Matching in Large Graphs},
  journal      = {{PVLDB}},
  volume       = {3},
  number       = {1},
  pages        = {1185--1194},
  year         = {2010},
  @url          = {https://vldb.org/pvldb/vol3/R105.pdf},
  @doi          = {10.14778/1920841.1920988},
  @timestamp    = {Mon, 03 Mar 2025 10:23:19 +0100},
  @biburl       = {https://dblp.org/rec/journals/pvldb/ZhangYJ10.bib},
  @bibsource    = {dblp computer science bibliography, https://dblp.org}
}

@article{DBLP:journals/prl/BunkeS98,
  author       = {Horst Bunke and
                  Kim Shearer},
  title        = {A graph distance metric based on the maximal common subgraph},
  journal      = {Pattern Recognit. Lett.},
  volume       = {19},
  number       = {3-4},
  pages        = {255--259},
  year         = {1998},
  @url          = {https://doi.org/10.1016/S0167-8655(97)00179-7},
  @doi          = {10.1016/S0167-8655(97)00179-7},
  @timestamp    = {Sat, 22 Feb 2020 19:31:32 +0100},
  @biburl       = {https://dblp.org/rec/journals/prl/BunkeS98.bib},
  @bibsource    = {dblp computer science bibliography, https://dblp.org}
}

@article{DBLP:journals/bioinformatics/TianMSSP07,
  author       = {Yuanyuan Tian and
                  Richard C. McEachin and
                  Carlos Santos and
                  David J. States and
                  Jignesh M. Patel},
  title        = {{SAGA:} a subgraph matching tool for biological graphs},
  journal      = {Bioinform.},
  volume       = {23},
  number       = {2},
  pages        = {232--239},
  year         = {2007},
  @url          = {https://doi.org/10.1093/bioinformatics/btl571},
  @doi          = {10.1093/BIOINFORMATICS/BTL571},
  @timestamp    = {Mon, 03 Feb 2025 10:38:33 +0100},
  @biburl       = {https://dblp.org/rec/journals/bioinformatics/TianMSSP07.bib},
  @bibsource    = {dblp computer science bibliography, https://dblp.org}
}

@inproceedings{DBLP:conf/sigmod/YanYH05,
  author       = {Xifeng Yan and
                  Philip S. Yu and
                  Jiawei Han},
  @editor       = {Fatma {\"{O}}zcan},
  title        = {Substructure Similarity Search in Graph Databases},
  booktitle    = {{ACM} {SIGMOD}},
  pages        = {766--777},
  @publisher    = {{ACM}},
  year         = {2005},
  @url          = {https://doi.org/10.1145/1066157.1066244},
  @doi          = {10.1145/1066157.1066244},
  @timestamp    = {Thu, 11 Mar 2021 15:20:15 +0100},
  @biburl       = {https://dblp.org/rec/conf/sigmod/YanYH05.bib},
  @bibsource    = {dblp computer science bibliography, https://dblp.org}
}

@article{DBLP:journals/tkde/ZhuQYC20,
  author       = {Yuanyuan Zhu and
                  Lu Qin and
                  Jeffrey Xu Yu and
                  Hong Cheng},
  title        = {Answering Top-{\it k} Graph Similarity Queries
                  in Graph Databases},
  journal      = {{IEEE} {TKDE}},
  volume       = {32},
  number       = {8},
  pages        = {1459--1474},
  year         = {2020},
  @url          = {https://doi.org/10.1109/TKDE.2019.2906608},
  @doi          = {10.1109/TKDE.2019.2906608},
  @timestamp    = {Wed, 19 Mar 2025 21:16:32 +0100},
  @biburl       = {https://dblp.org/rec/journals/tkde/ZhuQYC20.bib},
  @bibsource    = {dblp computer science bibliography, https://dblp.org}
}

@inproceedings{DBLP:conf/sigmod/ZhuLZZY12,
  author       = {Gaoping Zhu and
                  Xuemin Lin and
                  Ke Zhu and
                  Wenjie Zhang and
                  Jeffrey Xu Yu},
  @editor       = {K. Sel{\c{c}}uk Candan and
                  Yi Chen and
                  Richard T. Snodgrass and
                  Luis Gravano and
                  Ariel Fuxman},
  title        = {TreeSpan: efficiently computing similarity all-matching},
  booktitle    = {{ACM} {SIGMOD}},
  pages        = {529--540},
  @publisher    = {{ACM}},
  year         = {2012},
  @url          = {https://doi.org/10.1145/2213836.2213896},
  @doi          = {10.1145/2213836.2213896},
  @timestamp    = {Sun, 19 Jan 2025 13:27:19 +0100},
  @biburl       = {https://dblp.org/rec/conf/sigmod/ZhuLZZY12.bib},
  @bibsource    = {dblp computer science bibliography, https://dblp.org}
}

@article{DBLP:journals/prl/BunkeA83,
  author       = {Horst Bunke and
                  G. Allermann},
  title        = {Inexact graph matching for structural pattern recognition},
  journal      = {Pattern Recognit. Lett.},
  volume       = {1},
  number       = {4},
  pages        = {245--253},
  year         = {1983},
  @url          = {https://doi.org/10.1016/0167-8655(83)90033-8},
  @doi          = {10.1016/0167-8655(83)90033-8},
  @timestamp    = {Sat, 22 Feb 2020 19:32:50 +0100},
  @biburl       = {https://dblp.org/rec/journals/prl/BunkeA83.bib},
  @bibsource    = {dblp computer science bibliography, https://dblp.org}
}

@inproceedings{DBLP:conf/icdcs/CaoYWRL11,
  author       = {Ning Cao and
                  Zhenyu Yang and
                  Cong Wang and
                  Kui Ren and
                  Wenjing Lou},
  title        = {Privacy-Preserving Query over Encrypted Graph-Structured Data in Cloud
                  Computing},
  booktitle    = {{IEEE} {ICDCS}},
  pages        = {393--402},
  @publisher    = {{IEEE} Computer Society},
  year         = {2011},
  @url          = {https://doi.org/10.1109/ICDCS.2011.84},
  @doi          = {10.1109/ICDCS.2011.84},
  @timestamp    = {Fri, 24 Mar 2023 00:04:18 +0100},
  @biburl       = {https://dblp.org/rec/conf/icdcs/CaoYWRL11.bib},
  @bibsource    = {dblp computer science bibliography, https://dblp.org}
}

@article{DBLP:journals/tdsc/GeYH24,
  author       = {Xinrui Ge and
                  Jia Yu and
                  Rong Hao},
  title        = {Privacy-Preserving Graph Matching Query Supporting Quick Subgraph
                  Extraction},
  journal      = {{IEEE} {TDSC}},
  volume       = {21},
  number       = {3},
  pages        = {1286--1300},
  year         = {2024},
  @url          = {https://doi.org/10.1109/TDSC.2023.3276360},
  @doi          = {10.1109/TDSC.2023.3276360},
  @timestamp    = {Sat, 31 Aug 2024 20:43:29 +0200},
  @biburl       = {https://dblp.org/rec/journals/tdsc/GeYH24.bib},
  @bibsource    = {dblp computer science bibliography, https://dblp.org}
}

@article{DBLP:journals/tifs/WangZJHW22,
  author       = {Songlei Wang and
                  Yifeng Zheng and
                  Xiaohua Jia and
                  Hejiao Huang and
                  Cong Wang},
  title        = {OblivGM: Oblivious Attributed Subgraph Matching as a Cloud Service},
  journal      = {{IEEE} {TIFS}},
  volume       = {17},
  pages        = {3582--3596},
  year         = {2022},
  @url          = {https://doi.org/10.1109/TIFS.2022.3210881},
  @doi          = {10.1109/TIFS.2022.3210881},
  @timestamp    = {Sun, 13 Nov 2022 17:52:33 +0100},
  @biburl       = {https://dblp.org/rec/journals/tifs/WangZJHW22.bib},
  @bibsource    = {dblp computer science bibliography, https://dblp.org}
}

@article{DBLP:journals/tifs/WangZJW24,
  author       = {Songlei Wang and
                  Yifeng Zheng and
                  Xiaohua Jia and
                  Cong Wang},
  title        = {eGrass: An Encrypted Attributed Subgraph Matching System With Malicious
                  Security},
  journal      = {{IEEE} {TIFS}},
  volume       = {19},
  pages        = {5999--6014},
  year         = {2024},
  @url          = {https://doi.org/10.1109/TIFS.2024.3409089},
  @doi          = {10.1109/TIFS.2024.3409089},
  @timestamp    = {Thu, 04 Jul 2024 22:02:59 +0200},
  @biburl       = {https://dblp.org/rec/journals/tifs/WangZJW24.bib},
  @bibsource    = {dblp computer science bibliography, https://dblp.org}
}

@article{DBLP:journals/vldb/HuangHZGYZ22,
  author       = {Kai Huang and
                  Haibo Hu and
                  Shuigeng Zhou and
                  Jihong Guan and
                  Qingqing Ye and
                  Xiaofang Zhou},
  title        = {Privacy and efficiency guaranteed social subgraph matching},
  journal      = {{VLDB} J.},
  volume       = {31},
  number       = {3},
  pages        = {581--602},
  year         = {2022},
  @url          = {https://doi.org/10.1007/s00778-021-00706-0},
  @doi          = {10.1007/S00778-021-00706-0},
  @timestamp    = {Thu, 07 Nov 2024 07:51:12 +0100},
  @biburl       = {https://dblp.org/rec/journals/vldb/HuangHZGYZ22.bib},
  @bibsource    = {dblp computer science bibliography, https://dblp.org}
}

@inproceedings{DBLP:conf/icde/HuangL0TZC0024,
  author       = {Kai Huang and
                  Yunqi Li and
                  Qingqing Ye and
                  Yao Tian and
                  Xi Zhao and
                  Yue Cui and
                  Haibo Hu and
                  Xiaofang Zhou},
  title        = {{FRESH:} Towards Efficient Graph Queries in an Outsourced Graph},
  booktitle    = {{IEEE} {ICDE}},
  pages        = {4545--4557},
  @publisher    = {{IEEE}},
  year         = {2024},
  @url          = {https://doi.org/10.1109/ICDE60146.2024.00346},
  @doi          = {10.1109/ICDE60146.2024.00346},
  @timestamp    = {Mon, 03 Mar 2025 21:09:06 +0100},
  @biburl       = {https://dblp.org/rec/conf/icde/HuangL0TZC0024.bib},
  @bibsource    = {dblp computer science bibliography, https://dblp.org}
}

@inproceedings{DBLP:conf/dasfaa/GaoXLCY018,
  author       = {Jiuru Gao and
                  Jiajie Xu and
                  Guanfeng Liu and
                  Wei Chen and
                  Hongzhi Yin and
                  Lei Zhao},
  @editor       = {Jian Pei and
                  Yannis Manolopoulos and
                  Shazia Sadiq and
                  Jianxin Li},
  title        = {A Privacy-Preserving Framework for Subgraph Pattern Matching in Cloud},
  booktitle    = {{DASFAA}},
  @series       = {Lecture Notes in Computer Science},
  volume       = {10827},
  pages        = {307--322},
  @publisher    = {Springer},
  year         = {2018},
  @url          = {https://doi.org/10.1007/978-3-319-91452-7\_20},
  @doi          = {10.1007/978-3-319-91452-7\_20},
  @timestamp    = {Fri, 28 Feb 2025 17:23:41 +0100},
  @biburl       = {https://dblp.org/rec/conf/dasfaa/GaoXLCY018.bib},
  @bibsource    = {dblp computer science bibliography, https://dblp.org}
}

@inproceedings{DBLP:conf/uss/WangZJ24,
  author       = {Songlei Wang and
                  Yifeng Zheng and
                  Xiaohua Jia},
  @editor       = {Davide Balzarotti and
                  Wenyuan Xu},
  title        = {GraphGuard: Private Time-Constrained Pattern Detection Over Streaming
                  Graphs in the Cloud},
  booktitle    = {{USENIX} {Security}},
  @publisher    = {{USENIX} Association},
  year         = {2024},
  @url          = {https://www.usenix.org/conference/usenixsecurity24/presentation/wang-songlei},
  @timestamp    = {Mon, 22 Jul 2024 17:10:49 +0200},
  @biburl       = {https://dblp.org/rec/conf/uss/WangZJ24.bib},
  @bibsource    = {dblp computer science bibliography, https://dblp.org}
}

@inproceedings{DBLP:conf/icde/FanCXB15,
  author       = {Zhe Fan and
                  Byron Choi and
                  Jianliang Xu and
                  Sourav S. Bhowmick},
  @editor       = {Johannes Gehrke and
                  Wolfgang Lehner and
                  Kyuseok Shim and
                  Sang Kyun Cha and
                  Guy M. Lohman},
  title        = {Asymmetric structure-preserving subgraph queries for large graphs},
  booktitle    = {{IEEE} {ICDE}},
  pages        = {339--350},
  @publisher    = {{IEEE} Computer Society},
  year         = {2015},
  @url          = {https://doi.org/10.1109/ICDE.2015.7113296},
  @doi          = {10.1109/ICDE.2015.7113296},
  @timestamp    = {Fri, 24 Mar 2023 00:00:01 +0100},
  @biburl       = {https://dblp.org/rec/conf/icde/FanCXB15.bib},
  @bibsource    = {dblp computer science bibliography, https://dblp.org}
}

@inproceedings{DBLP:conf/icde/XuJCXB21,
  author       = {Lyu Xu and
                  Jiaxin Jiang and
                  Byron Choi and
                  Jianliang Xu and
                  Sourav S. Bhowmick},
  title        = {Privacy Preserving Strong Simulation Queries on Large Graphs},
  booktitle    = {{IEEE} {ICDE}},
  pages        = {1500--1511},
  @publisher    = {{IEEE}},
  year         = {2021},
  @url          = {https://doi.org/10.1109/ICDE51399.2021.00133},
  @doi          = {10.1109/ICDE51399.2021.00133},
  @timestamp    = {Tue, 07 May 2024 20:05:36 +0200},
  @biburl       = {https://dblp.org/rec/conf/icde/XuJCXB21.bib},
  @bibsource    = {dblp computer science bibliography, https://dblp.org}
}

@article{DBLP:journals/pacmmod/XuCPXB23,
  author       = {Lyu Xu and
                  Byron Choi and
                  Yun Peng and
                  Jianliang Xu and
                  Sourav S. Bhowmick},
  title        = {A Framework for Privacy Preserving Localized Graph Pattern Query Processing},
  journal      = {{ACM} {PACMMOD}},
  volume       = {1},
  number       = {2},
  pages        = {129:1--129:27},
  year         = {2023},
  @url          = {https://doi.org/10.1145/3589274},
  @doi          = {10.1145/3589274},
  @timestamp    = {Wed, 18 Dec 2024 13:51:13 +0100},
  @biburl       = {https://dblp.org/rec/journals/pacmmod/XuCPXB23.bib},
  @bibsource    = {dblp computer science bibliography, https://dblp.org}
}

@inproceedings{DBLP:conf/soda/ChanCMS19,
  author       = {T.{-}H. Hubert Chan and
                  Kai{-}Min Chung and
                  Bruce M. Maggs and
                  Elaine Shi},
  editor       = {Timothy M. Chan},
  title        = {Foundations of Differentially Oblivious Algorithms},
  booktitle    = {{ACM-SIAM} {SODA}},
  pages        = {2448--2467},
  @publisher    = {{SIAM}},
  year         = {2019},
  @url          = {https://doi.org/10.1137/1.9781611975482.150},
  @doi          = {10.1137/1.9781611975482.150},
  @timestamp    = {Thu, 15 Jul 2021 13:49:01 +0200},
  @biburl       = {https://dblp.org/rec/conf/soda/ChanCMS19.bib},
  @bibsource    = {dblp computer science bibliography, https://dblp.org}
}

@article{DBLP:journals/tdsc/WuLNHW22,
  author       = {Pengfei Wu and
                  Qi Li and
                  Jianting Ning and
                  Xinyi Huang and
                  Wei Wu},
  title        = {Differentially Oblivious Data Analysis With Intel {SGX:} Design, Optimization,
                  and Evaluation},
  journal      = {{IEEE} {TDSC}},
  volume       = {19},
  number       = {6},
  pages        = {3741--3758},
  year         = {2022},
  @url          = {https://doi.org/10.1109/TDSC.2021.3106317},
  @doi          = {10.1109/TDSC.2021.3106317},
  @timestamp    = {Mon, 04 Nov 2024 22:21:21 +0100},
  @biburl       = {https://dblp.org/rec/journals/tdsc/WuLNHW22.bib},
  @bibsource    = {dblp computer science bibliography, https://dblp.org}
}

@article{DBLP:journals/tifs/ZhengZLGZWSL23,
  author       = {Yandong Zheng and
                  Hui Zhu and
                  Rongxing Lu and
                  Yunguo Guan and
                  Songnian Zhang and
                  Fengwei Wang and
                  Jun Shao and
                  Hui Li},
  title        = {PGSim: Efficient and Privacy-Preserving Graph Similarity Query Over
                  Encrypted Data in Cloud},
  journal      = {{IEEE} {TIFS}},
  volume       = {18},
  pages        = {2030--2045},
  year         = {2023},
  @url          = {https://doi.org/10.1109/TIFS.2023.3262147},
  @doi          = {10.1109/TIFS.2023.3262147},
  @timestamp    = {Sun, 22 Oct 2023 11:15:08 +0200},
  @biburl       = {https://dblp.org/rec/journals/tifs/ZhengZLGZWSL23.bib},
  @bibsource    = {dblp computer science bibliography, https://dblp.org}
}

@article{DBLP:journals/jacm/ChorKGS98,
  author       = {Benny Chor and
                  Eyal Kushilevitz and
                  Oded Goldreich and
                  Madhu Sudan},
  title        = {Private Information Retrieval},
  journal      = {J. {ACM}},
  volume       = {45},
  number       = {6},
  pages        = {965--981},
  year         = {1998},
  @url          = {https://doi.org/10.1145/293347.293350},
  @doi          = {10.1145/293347.293350},
  @timestamp    = {Tue, 14 Jun 2022 13:12:43 +0200},
  @biburl       = {https://dblp.org/rec/journals/jacm/ChorKGS98.bib},
  @bibsource    = {dblp computer science bibliography, https://dblp.org}
}

@inproceedings{DBLP:conf/dac/OhlrichEGS93,
  author       = {Miles Ohlrich and
                  Carl Ebeling and
                  Eka Ginting and
                  Lisa Sather},
  @editor       = {Alfred E. Dunlop},
  title        = {SubGemini: Identifying SubCircuits using a Fast Subgraph Isomorphism
                  Algorithm},
  booktitle    = {{ACM} {DAC}},
  pages        = {31--37},
  @publisher    = {{ACM} Press},
  year         = {1993},
  @url          = {https://doi.org/10.1145/157485.164556},
  @doi          = {10.1145/157485.164556},
  @timestamp    = {Wed, 14 Nov 2018 10:58:37 +0100},
  @biburl       = {https://dblp.org/rec/conf/dac/OhlrichEGS93.bib},
  @bibsource    = {dblp computer science bibliography, https://dblp.org}
}

@article{sprinzak2003reliable,
  title={How reliable are experimental protein--protein interaction data?},
  author={Sprinzak, Einat and Sattath, Shmuel and Margalit, Hanah},
  journal={Journal of Molecular Biology (JMB)},
  volume={327},
  number={5},
  pages={919--923},
  year={2003},
  @publisher={Elsevier}
}

@article{DBLP:journals/pvldb/ZouCO09,
  author       = {Lei Zou and
                  Lei Chen and
                  M. Tamer {\"{O}}zsu},
  title        = {DistanceJoin: Pattern Match Query In a Large Graph Database},
  journal      = {{PVLDB}},
  volume       = {2},
  number       = {1},
  pages        = {886--897},
  year         = {2009},
  @url          = {http://www.vldb.org/pvldb/vol2/vldb09-876.pdf},
  @doi          = {10.14778/1687627.1687727},
  @timestamp    = {Tue, 21 Mar 2023 21:05:57 +0100},
  @biburl       = {https://dblp.org/rec/journals/pvldb/ZouCO09.bib},
  @bibsource    = {dblp computer science bibliography, https://dblp.org}
}

@inproceedings{DBLP:conf/icde/TianP08,
  author       = {Yuanyuan Tian and
                  Jignesh M. Patel},
  @editor       = {Gustavo Alonso and
                  Jos{\'{e}} A. Blakeley and
                  Arbee L. P. Chen},
  title        = {{TALE:} {A} Tool for Approximate Large Graph Matching},
  booktitle    = {{IEEE} {ICDE}},
  pages        = {963--972},
  @publisher    = {{IEEE} Computer Society},
  year         = {2008},
  @url          = {https://doi.org/10.1109/ICDE.2008.4497505},
  @doi          = {10.1109/ICDE.2008.4497505},
  @timestamp    = {Mon, 03 Feb 2025 10:38:33 +0100},
  @biburl       = {https://dblp.org/rec/conf/icde/TianP08.bib},
  @bibsource    = {dblp computer science bibliography, https://dblp.org}
}

@article{DBLP:journals/tods/MaCFHW14,
  author       = {Shuai Ma and
                  Yang Cao and
                  Wenfei Fan and
                  Jinpeng Huai and
                  Tianyu Wo},
  title        = {Strong simulation: Capturing topology in graph pattern matching},
  journal      = {{ACM} {TODS}},
  volume       = {39},
  number       = {1},
  pages        = {4:1--4:46},
  year         = {2014},
  @url          = {https://doi.org/10.1145/2528937},
  @doi          = {10.1145/2528937},
  @timestamp    = {Tue, 10 Dec 2019 14:58:47 +0100},
  @biburl       = {https://dblp.org/rec/journals/tods/MaCFHW14.bib},
  @bibsource    = {dblp computer science bibliography, https://dblp.org}
}

@inproceedings{DBLP:conf/stoc/Cook71,
  author       = {Stephen A. Cook},
  @editor       = {Michael A. Harrison and
                  Ranan B. Banerji and
                  Jeffrey D. Ullman},
  title        = {The Complexity of Theorem-Proving Procedures},
  booktitle    = {{ACM} {STOC}},
  pages        = {151--158},
  @publisher    = {{ACM}},
  year         = {1971},
  @url          = {https://doi.org/10.1145/800157.805047},
  @doi          = {10.1145/800157.805047},
  @timestamp    = {Mon, 26 Nov 2018 15:05:57 +0100},
  @biburl       = {https://dblp.org/rec/conf/stoc/Cook71.bib},
  @bibsource    = {dblp computer science bibliography, https://dblp.org}
}

@article{DBLP:journals/pvldb/QiuKMNK23,
  author       = {Lina Qiu and
                  Georgios Kellaris and
                  Nikos Mamoulis and
                  Kobbi Nissim and
                  George Kollios},
  title        = {Doquet: Differentially Oblivious Range and Join Queries with Private
                  Data Structures},
  journal      = {{PVLDB}},
  volume       = {16},
  number       = {13},
  pages        = {4160--4173},
  year         = {2023},
  @url          = {https://www.vldb.org/pvldb/vol16/p4160-qiu.pdf},
  @doi          = {10.14778/3625054.3625055},
  @timestamp    = {Tue, 14 Oct 2025 19:47:07 +0200},
  @biburl       = {https://dblp.org/rec/journals/pvldb/QiuKMNK23.bib},
  @bibsource    = {dblp computer science bibliography, https://dblp.org}
}

@techreport{AMDMemoryEncryption2021,
  title        = {AMD Memory Encryption},
  author       = {David Kaplan and Jeremy Powell and Tom Woller},
  @institution  = {Advanced Micro Devices, Inc.},
  year         = {2021},
  @month        = {October},
  @number       = {70363},
  type         = {White Paper},
  @url          = {https://docs.amd.com/v/u/en-US/memory-encryption-white-paper},
  @note         = {Document ID 70363}
}

@techreport{IntelTDX2022,
  title        = {Intel{\textregistered} Trust Domain Extensions (Intel TDX)},
  author       = {Intel Corporation},
  @institution  = {Intel Corporation},
  year         = {2025},
  @month        = {February},
  type         = {White Paper},
  @url          = {https://cdrdv2-public.intel.com/690419/TDX-Whitepaper-February2022.pdf},
  @note         = {Intel Document}
}

@misc{NVIDIAConfidentialComputing2023,
  author       = {NVIDIA Corporation},
  title        = {AI Security with Confidential Computing},
  year         = {2023},
  @howpublished = {\url{https://www.nvidia.com/en-us/data-center/solutions/confidential-computing/}},
  @note         = {Accessed: January 15, 2026}
}

@inproceedings{DBLP:conf/tcc/DworkMNS06,
  author       = {Cynthia Dwork and
                  Frank McSherry and
                  Kobbi Nissim and
                  Adam D. Smith},
  @editor       = {Shai Halevi and
                  Tal Rabin},
  title        = {Calibrating Noise to Sensitivity in Private Data Analysis},
  booktitle    = {{Springer} {TCC}},
  @series       = {Lecture Notes in Computer Science},
  volume       = {3876},
  pages        = {265--284},
  @publisher    = {Springer},
  year         = {2006},
  @url          = {https://doi.org/10.1007/11681878\_14},
  @doi          = {10.1007/11681878\_14},
  @timestamp    = {Tue, 14 Oct 2025 19:39:32 +0200},
  @biburl       = {https://dblp.org/rec/conf/tcc/DworkMNS06.bib},
  @bibsource    = {dblp computer science bibliography, https://dblp.org}
}

@inproceedings{DBLP:conf/icde/Liu0MLY24,
  author       = {Shang Liu and
                  Yang Cao and
                  Takao Murakami and
                  Jinfei Liu and
                  Masatoshi Yoshikawa},
  title        = {{CARGO:} Crypto-Assisted Differentially Private Triangle Counting
                  Without Trusted Servers},
  booktitle    = {{IEEE} {ICDE}},
  pages        = {1671--1684},
  @publisher    = {{IEEE}},
  year         = {2024},
  @url          = {https://doi.org/10.1109/ICDE60146.2024.00136},
  @doi          = {10.1109/ICDE60146.2024.00136},
  @timestamp    = {Tue, 08 Apr 2025 16:48:08 +0200},
  @biburl       = {https://dblp.org/rec/conf/icde/Liu0MLY24.bib},
  @bibsource    = {dblp computer science bibliography, https://dblp.org}
}

@article{DBLP:journals/tkde/LiuWLCL24,
  author       = {Yuhan Liu and
                  Tianhao Wang and
                  Yixuan Liu and
                  Hong Chen and
                  Cuiping Li},
  title        = {Edge-Protected Triangle Count Estimation Under Relationship Local
                  Differential Privacy},
  journal      = {{IEEE} {TKDE}},
  volume       = {36},
  number       = {10},
  pages        = {5138--5152},
  year         = {2024},
  @url          = {https://doi.org/10.1109/TKDE.2024.3381832},
  @doi          = {10.1109/TKDE.2024.3381832},
  @timestamp    = {Thu, 25 Dec 2025 12:46:21 +0100},
  @biburl       = {https://dblp.org/rec/journals/tkde/LiuWLCL24.bib},
  @bibsource    = {dblp computer science bibliography, https://dblp.org}
}

@inproceedings{DBLP:conf/sigmod/ZhangCPSX15,
  author       = {Jun Zhang and
                  Graham Cormode and
                  Cecilia M. Procopiuc and
                  Divesh Srivastava and
                  Xiaokui Xiao},
  @editor       = {Timos K. Sellis and
                  Susan B. Davidson and
                  Zachary G. Ives},
  title        = {Private Release of Graph Statistics using Ladder Functions},
  booktitle    = {{ACM} {SIGMOD}},
  pages        = {731--745},
  @publisher    = {{ACM}},
  year         = {2015},
  @url          = {https://doi.org/10.1145/2723372.2737785},
  @doi          = {10.1145/2723372.2737785},
  @timestamp    = {Sun, 06 Oct 2024 21:14:20 +0200},
  @biburl       = {https://dblp.org/rec/conf/sigmod/ZhangCPSX15.bib},
  @bibsource    = {dblp computer science bibliography, https://dblp.org}
}

@article{DBLP:journals/csur/LiPRSRN23,
  author       = {Yang Li and
                  Michael Purcell and
                  Thierry Rakotoarivelo and
                  David Smith and
                  Thilina Ranbaduge and
                  Kee Siong Ng},
  title        = {Private Graph Data Release: {A} Survey},
  journal      = {{ACM} {CSUR}},
  volume       = {55},
  number       = {11},
  pages        = {226:1--226:39},
  year         = {2023},
  @url          = {https://doi.org/10.1145/3569085},
  @doi          = {10.1145/3569085},
  @timestamp    = {Wed, 07 Jun 2023 14:24:59 +0200},
  @biburl       = {https://dblp.org/rec/journals/csur/LiPRSRN23.bib},
  @bibsource    = {dblp computer science bibliography, https://dblp.org}
}

@article{DBLP:journals/pvldb/QinJS0ZC22,
  author       = {Lianke Qin and
                  Rajesh Jayaram and
                  Elaine Shi and
                  Zhao Song and
                  Danyang Zhuo and
                  Shumo Chu},
  title        = {Differentially Oblivious Relational Database Operators},
  journal      = {{PVLDB}},
  volume       = {16},
  number       = {4},
  pages        = {842--855},
  year         = {2022},
  @url          = {https://www.vldb.org/pvldb/vol16/p842-qin.pdf},
  @doi          = {10.14778/3574245.3574267},
  @timestamp    = {Sun, 19 Jan 2025 13:44:37 +0100},
  @biburl       = {https://dblp.org/rec/journals/pvldb/QinJS0ZC22.bib},
  @bibsource    = {dblp computer science bibliography, https://dblp.org}
}

@article{DBLP:journals/pvldb/KimSHHC15,
  author       = {Jinha Kim and
                  Hyungyu Shin and
                  Wook{-}Shin Han and
                  Sungpack Hong and
                  Hassan Chafi},
  title        = {Taming Subgraph Isomorphism for {RDF} Query Processing},
  journal      = {{PVLDB}},
  volume       = {8},
  number       = {11},
  pages        = {1238--1249},
  year         = {2015},
  @url          = {http://www.vldb.org/pvldb/vol8/p1238-kim.pdf},
  @doi          = {10.14778/2809974.2809985},
  @timestamp    = {Sun, 19 Jan 2025 13:44:39 +0100},
  @biburl       = {https://dblp.org/rec/journals/pvldb/KimSHHC15.bib},
  @bibsource    = {dblp computer science bibliography, https://dblp.org}
}

@inproceedings{DBLP:conf/focs/HenzingerHK95,
  author       = {Monika Rauch Henzinger and
                  Thomas A. Henzinger and
                  Peter W. Kopke},
  title        = {Computing Simulations on Finite and Infinite Graphs},
  booktitle    = {{IEEE} {FOCS}},
  pages        = {453--462},
  @publisher    = {{IEEE} Computer Society},
  year         = {1995},
  @url          = {https://doi.org/10.1109/SFCS.1995.492576},
  @doi          = {10.1109/SFCS.1995.492576},
  @timestamp    = {Tue, 08 Jul 2025 16:43:45 +0200},
  @biburl       = {https://dblp.org/rec/conf/focs/HenzingerHK95.bib},
  @bibsource    = {dblp computer science bibliography, https://dblp.org}
}

@inproceedings{DBLP:conf/sp/XuCP15,
  author       = {Yuanzhong Xu and
                  Weidong Cui and
                  Marcus Peinado},
  title        = {Controlled-Channel Attacks: Deterministic Side Channels for Untrusted
                  Operating Systems},
  booktitle    = {{IEEE} {SP}},
  pages        = {640--656},
  @publisher    = {{IEEE} Computer Society},
  year         = {2015},
  @url          = {https://doi.org/10.1109/SP.2015.45},
  @doi          = {10.1109/SP.2015.45},
  @timestamp    = {Fri, 24 Mar 2023 00:03:11 +0100},
  @biburl       = {https://dblp.org/rec/conf/sp/XuCP15.bib},
  @bibsource    = {dblp computer science bibliography, https://dblp.org}
}

@article{DBLP:journals/tkde/Gexinrui26,
  author       = {Xinrui Ge and
                  Jia Yu and
                  Rong Hao},
  title        = {Privacy-Preserving Graph Similarity Matching Query over Encrypted Graph Database},
  journal      = {{IEEE} {TKDE} (Early Access)},
  @volume       = {36},
  @number       = {10},
  @pages        = {5138--5152},
  year         = {2026},
  @url          = {https://doi.org/10.1109/TKDE.2024.3381832},
  @doi          = {10.1109/TKDE.2024.3381832},
  @timestamp    = {Thu, 25 Dec 2025 12:46:21 +0100},
  @biburl       = {https://dblp.org/rec/journals/tkde/LiuWLCL24.bib},
  @bibsource    = {dblp computer science bibliography, https://dblp.org}
}

@inproceedings{DBLP:conf/ccs/KotiKPG24,
  author       = {Nishat Koti and
                  Varsha Bhat Kukkala and
                  Arpita Patra and
                  Bhavish Raj Gopal},
  @editor       = {Bo Luo and
                  Xiaojing Liao and
                  Jun Xu and
                  Engin Kirda and
                  David Lie},
  title        = {Graphiti: Secure Graph Computation Made More Scalable},
  booktitle    = {{ACM} {CCS}},
  pages        = {4017--4031},
  @publisher    = {{ACM}},
  year         = {2024},
  @url          = {https://doi.org/10.1145/3658644.3670393},
  @doi          = {10.1145/3658644.3670393},
  @timestamp    = {Sun, 22 Dec 2024 15:48:20 +0100},
  @biburl       = {https://dblp.org/rec/conf/ccs/KotiKPG24.bib},
  @bibsource    = {dblp computer science bibliography, https://dblp.org}
}

@inproceedings{DBLP:conf/ccs/LaiYSLLL19,
  author       = {Shangqi Lai and
                  Xingliang Yuan and
                  Shifeng Sun and
                  Joseph K. Liu and
                  Yuhong Liu and
                  Dongxi Liu},
  @editor       = {Steven D. Galbraith and
                  Giovanni Russello and
                  Willy Susilo and
                  Dieter Gollmann and
                  Engin Kirda and
                  Zhenkai Liang},
  title        = {GraphSE{\({^2}\)}: An Encrypted Graph Database for Privacy-Preserving
                  Social Search},
  booktitle    = {{ACM} {AsiaCCS}},
  pages        = {41--54},
  @publisher    = {{ACM}},
  year         = {2019},
  @url          = {https://doi.org/10.1145/3321705.3329803},
  @doi          = {10.1145/3321705.3329803},
  @timestamp    = {Sun, 19 Jan 2025 13:26:44 +0100},
  @biburl       = {https://dblp.org/rec/conf/ccs/LaiYSLLL19.bib},
  @bibsource    = {dblp computer science bibliography, https://dblp.org}
}

@inproceedings{DBLP:conf/crypto/ApplebaumDI0Z17,
  author       = {Benny Applebaum and
                  Ivan Damg{\aa}rd and
                  Yuval Ishai and
                  Michael Nielsen and
                  Lior Zichron},
  @editor       = {Jonathan Katz and
                  Hovav Shacham},
  title        = {Secure Arithmetic Computation with Constant Computational Overhead},
  booktitle    = {{Springer} {CRYPTO}},
  @series       = {Lecture Notes in Computer Science},
  volume       = {10401},
  pages        = {223--254},
  @publisher    = {Springer},
  year         = {2017},
  @url          = {https://doi.org/10.1007/978-3-319-63688-7\_8},
  @doi          = {10.1007/978-3-319-63688-7\_8},
  @timestamp    = {Tue, 14 May 2019 10:00:48 +0200},
  @biburl       = {https://dblp.org/rec/conf/crypto/ApplebaumDI0Z17.bib},
  @bibsource    = {dblp computer science bibliography, https://dblp.org}
}

@inproceedings{DBLP:conf/ccs/KellerOS16,
  author       = {Marcel Keller and
                  Emmanuela Orsini and
                  Peter Scholl},
  @editor       = {Edgar R. Weippl and
                  Stefan Katzenbeisser and
                  Christopher Kruegel and
                  Andrew C. Myers and
                  Shai Halevi},
  title        = {{MASCOT:} Faster Malicious Arithmetic Secure Computation with Oblivious
                  Transfer},
  booktitle    = {{ACM} {CCS}},
  pages        = {830--842},
  @publisher    = {{ACM}},
  year         = {2016},
  @url          = {https://doi.org/10.1145/2976749.2978357},
  @doi          = {10.1145/2976749.2978357},
  @timestamp    = {Sun, 19 Jan 2025 13:26:54 +0100},
  @biburl       = {https://dblp.org/rec/conf/ccs/KellerOS16.bib},
  @bibsource    = {dblp computer science bibliography, https://dblp.org}
}

@inproceedings{DBLP:conf/ndss/LeeJFG21,
  author       = {Hyun Bin Lee and
                  Tushar M. Jois and
                  Christopher W. Fletcher and
                  Carl A. Gunter},
  title        = {{DOVE:} {A} Data-Oblivious Virtual Environment},
  booktitle    = {{NDSS}},
  @publisher    = {The Internet Society},
  year         = {2021},
  @url          = {https://www.ndss-symposium.org/ndss-paper/dove-a-data-oblivious-virtual-environment/},
  @timestamp    = {Thu, 17 Jun 2021 16:04:48 +0200},
  @biburl       = {https://dblp.org/rec/conf/ndss/LeeJFG21.bib},
  @bibsource    = {dblp computer science bibliography, https://dblp.org}
}

@misc{repo:dos,
  @author       = {Owner, Name},
  title        = {ObliGraph},
  howpublished = {\url{https://github.com/cslyuxu/ObliGraph}},
  year         = {2026},
  @note         = {GitHub repository, commit \texttt{<COMMIT_SHA>} (accessed 2026-04-30)},
  note         = {Accessed: Oct. 24, 2026}
}

@article{DBLP:journals/pami/CordellaFSV04,
  author       = {Luigi P. Cordella and
                  Pasquale Foggia and
                  Carlo Sansone and
                  Mario Vento},
  title        = {A (Sub)Graph Isomorphism Algorithm for Matching Large Graphs},
  journal      = {{IEEE} {TPAMI}},
  volume       = {26},
  number       = {10},
  pages        = {1367--1372},
  year         = {2004},
  @url          = {https://doi.org/10.1109/TPAMI.2004.75},
  @doi          = {10.1109/TPAMI.2004.75},
  @timestamp    = {Mon, 13 Jun 2022 12:03:26 +0200},
  @biburl       = {https://dblp.org/rec/journals/pami/CordellaFSV04.bib},
  @bibsource    = {dblp computer science bibliography, https://dblp.org}
}

@article{DBLP:journals/pvldb/ChamaniD0PJ23,
  author       = {Javad Ghareh Chamani and
                  Ioannis Demertzis and
                  Dimitrios Papadopoulos and
                  Charalampos Papamanthou and
                  Rasool Jalili},
  title        = {GraphOS: Towards Oblivious Graph Processing},
  journal      = {{PVLDB}},
  volume       = {16},
  number       = {13},
  pages        = {4324--4338},
  year         = {2023},
  @url          = {https://www.vldb.org/pvldb/vol16/p4324-chamani.pdf},
  @doi          = {10.14778/3625054.3625067},
  @timestamp    = {Thu, 19 Dec 2024 16:40:20 +0100},
  @biburl       = {https://dblp.org/rec/journals/pvldb/ChamaniD0PJ23.bib},
  @bibsource    = {dblp computer science bibliography, https://dblp.org}
}

@inproceedings{DBLP:conf/sigmod/HanKGPH19,
  author       = {Myoungji Han and
                  Hyunjoon Kim and
                  Geonmo Gu and
                  Kunsoo Park and
                  Wook{-}Shin Han},
  @editor       = {Peter Boncz and
                  Stefan Manegold and
                  Anastasia Ailamaki and
                  Amol Deshpande and
                  Tim Kraska},
  title        = {Efficient Subgraph Matching: Harmonizing Dynamic Programming, Adaptive
                  Matching Order, and Failing Set Together},
  booktitle    = {{ACM} {SIGMOD}},
  pages        = {1429--1446},
  @publisher    = {{ACM}},
  year         = {2019},
  @url          = {https://doi.org/10.1145/3299869.3319880},
  @doi          = {10.1145/3299869.3319880},
  @timestamp    = {Wed, 03 Jun 2026 20:25:42 +0200},
  @biburl       = {https://dblp.org/rec/conf/sigmod/HanKGPH19.bib},
  @bibsource    = {dblp computer science bibliography, https://dblp.org}
}

@misc{data:aid,
  @author       = {{National Cancer Institute}},
  title        = {{AIDS} Antiviral Screen Data of October 1999},
  howpublished = {\url{https://cactus.nci.nih.gov/download/nci/}},
  year         = {2022},
  note         = {Accessed: Oct. 24, 2026}
}

@misc{data:pubchem,
  @author       = {{National Center for Biotechnology Information}},
  title        = {{PubChem}},
  howpublished = {\url{https://pubchem.ncbi.nlm.nih.gov/source/}},
  year         = {2026},
  note         = {Accessed: Oct. 24, 2026}
}

@misc{data:graphgen,
    @author       = {Owner, Name},
  @key          = {GraphGen},
  @title        = {{GraphGen}: {A} Synthetic Graph Data Generator},
  @note = {2007},
  title        = {GraphGen},
  howpublished = {\url{https://cse.hkust.edu.hk/graphgen/}},
  year         = {2007},
  note         = {Accessed: Oct. 24, 2026}
}

@misc{data:snap,
  author       = {Jure Leskovec and Andrej Krevl},
  title        = {{SNAP} Datasets: {Stanford} Large Network Dataset Collection},
  howpublished = {\url{https://snap.stanford.edu/data/}},
  year         = {2014}
}

@article{DBLP:journals/pacmmod/WuDH26,
  author       = {Zhiang Wu and
                  Wei Dong and
                  Xiao Hu},
  title        = {Differentially Oblivious Multi-way Join},
  journal      = {{PACMMOD}},
  volume       = {4},
  number       = {3},
  pages        = {162:1--162:26},
  year         = {2026},
  @url          = {https://doi.org/10.1145/3802039},
  @doi          = {10.1145/3802039},
  @timestamp    = {Wed, 09 Sep 2026 18:15:33 +0200},
  @biburl       = {https://dblp.org/rec/journals/pacmmod/WuDH26.bib},
  @bibsource    = {dblp computer science bibliography, https://dblp.org}
}

\clearpage

\balance
\appendix

\section{Summary of Notations}
\label{sec:apx:notation}

This appendix presents a table of notations and their meanings for easy reference.

\begin{table}[ht]
\caption{Frequently used notations}
\vspace{-3.5ex}
	\label{tab:notations}
		\begin{center}
  \resizebox{1.0\linewidth}{!}{
            \begin{tabular}{|m{2.6cm}<{\centering}|m{7.5cm}|}\hline 
  \textbf{Notation} & \multicolumn{1}{c|}{\textbf{Description}}  \\  \hline \hline
        $G=(V_G,E_G, \Sigma_G, \ell_G)$ & A graph, where (i) $V_G$ is the node set; (ii) $E_G: V_G\times V_G \to \{0, 1\}$ is the edge function; (iii) $\Sigma_G$ is the label set; and (iv) $\ell_G$: $ V_G \to \Sigma_G$ is the node-labeling function \\
		\hline
    $Q$;~$G$ & query graph; data graph \\ \hline
    $u/v$; $e_G(u, v)$ & node $u/v$; $e_G(u, v)$ $=$ $1$ indicates that there exists a directed edge from $u$ to $v$ in graph $G$. Otherwise, $e_G(u, v)$~$=$~$0$. \\ \hline
    $\dis{G_1}{G_2}$ & The graph edit distance between graphs $G_1$ and~$G_2$ \\ \hline
    $\mathcal{D}$ $=$ $\{G_1, ..., G_n\}$ & a graph database consists of $n$ data graphs $G_1, ..., G_n$ \\ \hline 
    $(Q, \mathcal{F}, \tau)$ & a graph pattern query, where $Q$ is the query graph, $\mathcal{F}$ is the query semantics, and $\tau$ is the similarity parameter \\ \hline 
    $M$; $\mathcal{M}_i$ & candidate mapping ($\cm$) for mapping a data graph to a query graph;  the $\cm$ set at depth $i$  \\ \hline
    $\cseq_Q^{\mathcal{F}}$; $\cseq_G^{\mathcal{F}}$ & the candidate sequence of $Q$ and $G$ for the query semantics $\mathcal{F}$, which is used for the $\cm$ enumeration \\ \hline
    $\mathsf{dp}[\cdot]$ & the dynamic programming table used in $\base$ and $\ppdo$ to store the cost of each evaluated candidate mapping \\ \hline
    $\epsilon$; $\lambda$ & privacy budget; maximum error probability  \\ \hline
		\end{tabular}
}
		\end{center}
\end{table}

\section{Proof}\label{sec:proof}

\begin{manuallemma}{\ref{lemma:base}}
The $\base$ search is $\FO$ if $\mathsf{Cost}_\mathcal{F}(\cdot)$ is $\FO$. 
\end{manuallemma}

\begin{proof}
Consider a database $\mathcal{D}$ and any two queries $(Q_1, \mathcal{F}, \tau)$ and $(Q_2, \mathcal{F}, \tau)$ such that $Q_1$ and $Q_2$ have the same number of nodes. Assume that $\mathsf{Cost}_\mathcal{F}(\cdot)$ is $\FO$. Then, with simulations on Algorithm~\ref{alg:base} for querying $Q_1$ and $Q_2$ on $\mathcal{D}$, it is easy to know that: (i)  $\base$ accesses data in all phases in a deterministic order, depending only on the node set size of the query graph, the query semantics $\mathcal{F}$, the similarity parameter $\tau$, and the size of the candidate sequence $\cseq_Q^{\mathcal{F}}$ for the query graph. As these parameters are the same for both queries, the memory access patterns are the same for both queries; and (ii) $\base$ performs simple $\FO$ arithmetic and logical operations, and $\FO$ cost computation as assumed. 
Hence, Equation~(\ref{eqa:fo}) holds that $\base$ is $\FO$ \wrt query graph. Similarly, 
if $\mathsf{Cost}_\mathcal{F}(\cdot)$ is $\FO$, $\base$ is also $\FO$ \wrt data graph,  yielding that $\base$ is $\FO$. Lemma~\ref{lemma:base} holds.
\end{proof}

\begin{manuallemma}{\ref{lemma:depth3}}
The sensitivity of $d^*$ is bounded by $\max{\big\{|V_Q|, |V_G|\big\}}$.
\end{manuallemma}

\begin{proof} 
We prove Lemma~\ref{lemma:depth3} by analyzing the two exhausted cases for neighboring query graph and neighboring databases. 
\begin{itemize}
\item Assume two neighboring queries $(Q, \gsim, \tau)$ and $(Q', \gsim, \tau)$  such that $\dis{Q}{Q'}$ $=$ $1$.\footref{fn:neighbor} If $Q$ and $Q'$ differ by one edge, the pruning depth $d^*_{Q'}$ for $Q'$ equals the pruning depth $d^*_Q$ for $Q$ due to the same $V_Q$ and $V_G$ with unchanged labels. It is evident that $|d^*_{Q'} - d^*_{Q}|$ $=$ $0$ $<$ $\max{\big\{|V_Q|, |V_G|\big\}}$; otherwise, $Q$ and $Q'$ have the same topological structure except for the label of one node $v$. W.l.o.g., assume that $\ell_Q(v) = l_1$, $\ell_{Q'}(v) = l_2$, and there are $n_1$ and $n_2$ nodes of $G$ having labels $l_1$ and $l_2$, respectively. Then, we have $|d^*_{Q'} - d^*_{Q}|$ $=$ $|n_1-n_2|$ $\le$ $|V_G|$ $<$ $\max{\big\{|V_Q|, |V_G|\big\}}$. Therefore, Lemma~\ref{lemma:depth3} holds for the case of a neighboring query graph.

\item Assume two neighboring graph databases $\mathcal{D}$ and $\mathcal{D}'$ such that (i) $\mathcal{D}$ and $\mathcal{D}'$ differ by one graph, namely graph $G$ in $\mathcal{D}$ and graph $G'$ in $\mathcal{D}'$; and (ii) $\dis{G}{G'}$ $=$ $1$.\footref{fn:neighbor} 
Similarly, if $G$ and $G'$ differ by one edge, the pruning depth $d^*_{G'}$ for $G'$ equals the pruning depth $d^*_G$ for $G$ due to the same $V_Q$ and $V_G$ with unchanged labels, and hence, $|d^*_{G'} - d^*_{G}|$ $=$ $0$ $<$ $\max{\big\{|V_Q|, |V_G|\big\}}$; 
otherwise, $G$ and $G'$ have the same topological structure except for the label of one node $v$, yielding that $\Big|\max_{u\in V_Q} \big\{|N_{G'}(u)|\big\}$ $-$ $ \max_{u\in V_Q} \big\{|N_G(u)|\big\}\Big|$ $\le$ $1$. Then, we have
\begin{align}
\nonumber
|d^*_{G'} - d^*_{G}| &= |V_Q|\cdot \Big|\max_{u\in V_Q} \big\{|N_{G'}(u)|\big\}- \max_{u\in V_Q} \big\{|N_G(u)|\big\}\Big| \\
\nonumber
   &\le |V_Q| \\
   \nonumber
   &\le \max{\big\{|V_Q|, |V_G|\big\}}.
\end{align}

Therefore, Lemma~\ref{lemma:depth3} also holds for this case.
\end{itemize}
With the analyses above,  Lemma~\ref{lemma:depth3} holds. 
\end{proof}

\begin{manuallemma}{\ref{lemma:laplace}}
For any neighboring inputs $\mathcal{X}_1$ and $\mathcal{X}_2$, the Laplace mechanism satisfies $(\epsilon, 0)$-differential privacy.
\end{manuallemma}

\begin{proof}
For $\mathcal{X}_1$ and $\mathcal{X}_2$, the Laplace mechanism outputs $\mathcal{F}(\mathcal{X}_1, f,\epsilon)$ $=$ $f(\mathcal{X}_1)$ $+$ $\xi_1$ and $\mathcal{F}(\mathcal{X}_2, f,\epsilon)$ $=$ $f(\mathcal{X}_2)$ $+$ $\xi_2$, respectively. 
As $\Delta f$ is the sensitivity of $f$, we have
$\Delta f$ $=$ $\max{\Big\{\big|f(\mathcal{X}_1) - f(\mathcal{X}_2)\big|\Big\}}$.
Let $p_{\mathcal{X}_1}$ and $p_{\mathcal{X}_2}$ denote the probability density functions of $\mathcal{F}(\mathcal{X}_1, f,\epsilon)$ and $\mathcal{F}(\mathcal{X}_2, f,\epsilon)$, respectively.
Then, for any output $o$ $\in$ $\mathbb{R}_{\ge 0}$, we have
\begin{align}
\nonumber
\frac{p_{\mathcal{X}_1}(o)}{p_{\mathcal{X}_2}(o)} &= 
\frac{Pr\big[f(\mathcal{X}_1) + \xi_1 = o\big]}{Pr\big[f(\mathcal{X}_2) + \xi_2 = o\big]}
= \frac{Pr\big[\xi_1 = o - f(\mathcal{X}_1)\big]}{Pr\big[\xi_2 = o - f(\mathcal{X}_2)\big]} \\
\nonumber
&= \frac{\frac{\epsilon}{2\Delta f}\cdot e^{-\frac{\epsilon}{\Delta f}\cdot\Big(\big|o-f(\mathcal{X}_1)\big|\Big)}}{\frac{\epsilon}{2\Delta f}\cdot e^{-\frac{\epsilon}{\Delta f}\cdot\Big(\big|o-f(\mathcal{X}_2)\big|\Big)}} 
= e^{\frac{\epsilon}{\Delta f}\cdot\Big(\big|o-f(\mathcal{X}_2)\big| - \big|o-f(\mathcal{X}_1)\big|\Big)} \\
\nonumber
&\le e^{\frac{\epsilon}{\Delta f}\cdot\Big(\big|f(\mathcal{X}_1)  - f(\mathcal{X}_2)\big|\Big)} \le e^{\frac{\epsilon}{\Delta f}\cdot  \Delta f} = e^\epsilon.
\end{align} 
Therefore, the mechanism satisfies $(\epsilon, 0)$-differential privacy. 
\end{proof}

\end{document}